\documentclass[useAMS, usenatbib, a4paper,fleqn]{mn2e}
\usepackage{aas_macros ,multirow,graphicx, xspace, amsmath, url,acronym,microtype}
\usepackage{hyperref}
\usepackage{listings}
\usepackage{amssymb}
\usepackage{cleveref}
\usepackage[top=0.85in, bottom=0.85in, left=0.8in, right=0.8in]{geometry}
\usepackage{times}
\acrodef{hst}[HST]{Hubble Space Telescope}
\acrodef{cfht}[CFHT]{Canada France Hawaii Telescope}
\acrodef{vlt}[VLT]{Very Large Telescope}
\acrodef{cos}[COS]{Cosmic Origins Spectrograph}
\acrodef{fos}[FOS]{Faint Object Spectrograph}
\acrodef{vimos}[VIMOS]{Visible Multi-Object Spectrograph}
\acrodef{muse}[MUSE]{Multi Unit Spectroscopic Explorer}
\acrodef{deimos}[DEIMOS]{Deep Imaging Multi-Object Spectrograph}
\acrodef{gmos}[GMOS]{Gemini Multi-Object Spectrograph}
\acrodef{vvds}[VVDS]{\ac{vlt} \ac{vimos} Deep Survey}
\acrodef{gdds}[GDDS]{Gemini Deep Deep Survey}
\acrodef{idl}[IDL]{Interactive Data Language}
\acrodef{iraf}[IRAF]{Image Reduction and Analysis Facility}
\acrodef{gto}[GTO]{Guaranteed Observing Time}
\acrodef{go}[GO]{General Observer}
\acrodef{stsci}[STScI]{Space Telescope Science Institute}
\acrodef{eso}[ESO]{European Southern Observatory}
\acrodef{pi}[PI]{principal investigator}
\acrodef{mos}[MOS]{multi-object spectroscopy}
\acrodef{ifu}[IFU]{integral field units}

\acrodef{los}[LOS]{line-of-sight}
\acrodef{losp}[LOS]{lines-of-sight}
\acrodef{lss}[LSS]{large scale structure}
\acrodef{wmap}[WMAP]{Wilkinson Microwave Anisotropy Probe}
\acrodef{hipass}[HIPASS]{H~{\sc i} Parkes All Sky Survey}
\acrodef{whim}[WHIM]{warm-hot intergalactic medium}
\acrodef{icm}[ICM]{intracluster medium}
\acrodef{uv}[UV]{ultra-violet}
\acrodef{lbg}[LBG]{Lyman break galaxies}

\acrodef{ccd}[CCD]{charge-coupled device}
\acrodef{sdss}[SDSS]{Sloan Digital Sky Survey}
\acrodef{dr7}[DR7]{Seventh Data Release}
\acrodef{dr8}[DR8]{Eighth Data Release}
\acrodef{dr10}[DR10]{Tenth Data Release}
\acrodef{gimic}[GIMIC]{Galaxies-Intergalactic Medium Interaction Calculation}
\acrodef{gmbcg}[GMBCG]{Gaussian Mixture Brightest Cluster Galaxy}
\acrodef{redmapper}[redMaPPer]{red-sequence Matched-filter Probabilistic Percolation}

\acrodef{bcg}[BCG]{brightest cluster galaxy}
\acrodef{lrg}[LRG]{luminous red galaxies}
\acrodef{gmm}[GMM]{Gaussian Mixture Model}
\acrodef{galex}[GALEX]{Galaxy Evolution Explorer}

\acrodef{uv}[UV]{ultra-violet}
\acrodef{nuv}[NUV]{near ultra-violet}
\acrodef{fuv}[FUV]{far ultra-violet}
\acrodef{igm}[IGM]{intergalactic medium}
\acrodef{cgm}[CGM]{circumgalactic medium}
\acrodef{ism}[ISM]{interstellar medium}
\acrodef{cmb}[CMB]{cosmic microwave background}

\acrodef{psf}[PSF]{point spread function}
\acrodef{lsf}[LSF]{line spread function}
\acrodef{fwhm}[FWHM]{full width at half maximum}
\acrodef{fppos}[FP-POS]{fixed-pattern noise position}
\acrodef{qso}[QSO]{quasi-stellar object}
\acrodef{sn}[SN]{supernova}
\acrodef{agn}[AGN]{active galactic nuclei}
\acrodef{grb}[GRB]{gamma-ray burst}
\acrodef{snr}[S/N]{signal-to-noise ratio}
\acrodef{bla}[BLA]{broad \lya}

\newcommand{\calcos}{{\sc calcos}\xspace}

\newcommand{\vpfit}{{\sc vpfit}\xspace}

\newcommand{\hi}{H~{\sc i}\xspace}
\newcommand{\ovi}{O~{\sc vi}\xspace}
\newcommand{\ovii}{O~{\sc vii}\xspace}
\newcommand{\oiv}{O~{\sc iv}\xspace}
\newcommand{\oiii}{O~{\sc iii}\xspace}
\newcommand{\oii}{O~{\sc ii}\xspace}
\newcommand{\oi}{O~{\sc i}\xspace}
\newcommand{\cii}{C~{\sc ii}\xspace}
\newcommand{\ciid}{C~{\sc ii}*\xspace}
\newcommand{\ciii}{C~{\sc iii}\xspace}
\newcommand{\civ}{C~{\sc iv}\xspace}
\newcommand{\nI}{N~{\sc i}\xspace}
\newcommand{\nii}{N~{\sc ii}\xspace}
\newcommand{\niii}{N~{\sc iii}\xspace}
\newcommand{\nv}{N~{\sc v}\xspace}
\newcommand{\alii}{Al~{\sc ii}\xspace}
\newcommand{\siii}{Si~{\sc ii}\xspace}
\newcommand{\siiii}{Si~{\sc iii}\xspace}
\newcommand{\siiv}{Si~{\sc iv}\xspace}
\newcommand{\neviii}{Ne~{\sc viii}\xspace}
\newcommand{\feiii}{Fe~{\sc iii}\xspace}
\newcommand{\feii}{Fe~{\sc ii}\xspace}

\newcommand{\mpc}{\,Mpc\xspace}

\newcommand{\kpc}{\,kpc\xspace}
\newcommand{\cm}{\,cm$^{-2}$\xspace}

\newcommand{\msun}{\,M$_{\sun}$}

\newcommand{\sk}{\smallskip}

\title[Towards detecting the WHIM in inter-cluster filaments]{Towards the
statistical detection of the warm-hot intergalactic medium in inter-cluster
filaments of the cosmic web\thanks{Based partly on
observations made with the NASA/ESA Hubble Space Telescope under
program GO 12958.}}

\author[Nicolas Tejos et al.]{
\parbox[t]{\textwidth}{
\vspace{-1.0cm} 
Nicolas Tejos,$^{1}$\thanks{E-mail: \href{mailto:ntejos@gmail.com}{ntejos@gmail.com}}
J. Xavier Prochaska,$^{1}$
Neil H. M. Crighton,$^{2}$ 
Simon L. Morris,$^{3}$
Jessica K. Werk,$^{1}$
Tom Theuns,$^{3,4}$
Nelson Padilla,$^{5}$
Rich M. Bielby$^{3}$ and
Charles W. Finn$^{3,4}$
}
\vspace*{6pt} \\ 
$^{1}$ Department of Astronomy and Astrophysics, UCO/Lick Observatory, University of California, 1156 High Street, Santa Cruz, CA 95064, USA\\
$^{2}$ Centre for Astrophysics and Supercomputing, Swinburne University of Technology, Hawthorn, Victoria 3122, Australia\\
$^{3}$ Department of Physics, Durham University, South Road, Durham, DH1 3LE, UK\\ 
$^{4}$ Institute for Computational Cosmology, Department of Physics, University of Durham, South Road, Durham, DH1 3LE, UK\\
$^{5}$ Instituto de Astrof\'isica, Centro de Astro-Ingenier\'ia, Pontificia Universidad Cat\'olica de Chile, Av. Vicu\~na Mackenna 4860, Santiago, Chile\\
\vspace*{-0.5cm}}

\begin{document}
\date{Accepted version, \today}

\pagerange{\pageref{firstpage}--\pageref{lastpage}} \pubyear{2015}

\maketitle

\label{firstpage}

\begin{abstract}

Modern analyses of structure formation predict a universe tangled in a
`cosmic web' of dark matter and diffuse baryons. These theories
further predict that at low-$z$, a significant fraction of the baryons
will be shock-heated to $T \sim 10^{5}-10^{7}$\,K yielding a
\ac{whim}, but whose actual existence has eluded a firm observational
confirmation. We present a novel experiment to detect the \ac{whim},
by targeting the putative filaments connecting galaxy clusters. We use
HST/COS to observe a remarkable QSO sightline that passes within
$\Delta d = 3$\mpc from the $7$ inter-cluster axes connecting $7$
independent cluster-pairs at redshifts $0.1 \le z \le 0.5$. We find
tentative excesses of total \hi, narrow \hi (NLA; Doppler parameters
$b<50$\kms), broad \hi (BLA; $b \ge 50$\kms) and \ovi absorption lines
within rest-frame velocities of $\Delta v \lesssim 1000$\kms from the
cluster-pairs redshifts, corresponding to $\sim 2$, $\sim 1.7$, $\sim
6$ and $\sim 4$ times their field expectations, respectively. Although
the excess of \ovi likely comes from gas close to individual galaxies,
we conclude that most of the excesses of NLAs and BLAs are truly
intergalactic. We find the covering fractions, $f_c$, of BLAs close to
cluster-pairs are $\sim 4-7$ times higher than the random expectation
(at the $\sim 2 \sigma$ c.l.), whereas the $f_c$ of NLAs and \ovi are
not significantly enhanced. We argue that a larger relative excess of
BLAs compared to those of NLAs close to cluster-pairs may be a
signature of the \ac{whim} in inter-cluster filaments. By extending
the present analysis to tens of sightlines our experiment offers a
promising route to detect the \ac{whim}.

\end{abstract}


%v3.0

\section{Introduction}\label{sec:intro}

\acresetall
%\subsection{Motivation}\label{intro:motivation}

Perhaps the most distinctive feature of the cosmic web is its intricate
pattern of filamentary structures. Cosmological simulations in a
$\Lambda$~cold dark matter ($\Lambda$CDM) paradigm predict that these
filaments account for $\sim 40\%$ of all mass in the Universe at $z=0$
and occupy roughly $\sim 10\%$ of the volume
\citep[e.g.][]{Aragon-Calvo2010}. When gas and hydrodynamical effects
are included in these simulations, a remarkable conclusion is reached:
$\sim 30-50\%$ of baryons at low-$z$ should reside in dense filaments,
primarily in the form of a diffuse gas phase with temperatures $T \sim
10^{5}-10^{7}$\,K, which would be very difficult to detect
\citep[e.g.][]{Cen1999,Dave2001}. This material is usually referred to
as the \ac{whim}, and indeed is currently the best candidate to host a
significant fraction of the so-called `missing baryons' at $z<1$
\citep[][and references therein]{Persic1992,Fukugita1998,Bregman2007,
  Prochaska2009,Shull2012}. According to these models, the physical
origin of the \ac{whim} is through gravitational shocks from the
collapse of matter into the \ac{lss} of the Universe.
%\footnote{Galactic winds might also contribute to the \ac{whim}
%  although at a much lesser extent \citep[e.g.][]{Cen2006}.}

One well-studied example of gravitational shock-heating is the
so-called \ac{icm} of galaxy clusters, where the virial temperatures
typically reach $T\sim 10^{7}-10^{8}$\,K.  A plasma at these
temperatures mostly cools through {\it Bremsstrahlung}
(a.k.a. free-free) thermal radiation, emitting $X$-rays at $\sim$~keV
energies that may be observed with modern satellites \citep[e.g.][and
  references therein]{Kravtsov2012}. $X$-ray spectroscopy has also
revealed the presence of high-ionization state metal emission lines in
the \ac{icm}, consistent with these large temperatures
\citep[e.g.][]{Sanders2008}. Thereby one constrains the density,
chemical abundances and morphology of the \ac{icm}. Several decades of
research have revealed a highly enriched medium ($\sim \frac{1}{3}$
solar) with a total mass consistent with the cosmic ratio of baryons to
dark matter \citep[e.g.][]{Allen2008}.
%Although some galaxy clusters show irregularities
%in their \ac{icm}, most of them have morphology consistent with a
%spherical geometry (refs).

In the $\Lambda$CDM paradigm, galaxy clusters correspond to the nodes
of the cosmic web, i.e., they mark the intersection of several
filamentary threads. These models further predict that matter flows
through the filamentary structures, driving the growth of the galaxy
clusters. Ideally, one would image these filaments in a similar manner
to the \ac{icm} to reveal their structure and physical properties as
tests of the cosmic web paradigm.
%and hence a spherical geometry is justified even if
%equilibrium is not fully reached yet in these structures. Beyond their
%virial radii, a different picture is expected: matter should flow
%through collimated filamentary structures, implying an anysotropic
%distribution (at least over scales $\lesssim 300$ \mpc). 
Unfortunately, once at the outskirts of galaxy clusters, the densities
and temperatures are too low for viable $X$-ray detection in emission
(e.g.\ {\it Bremsstrahlung} radiation is proportional to the
density squared of the emitting gas). To study this dominant component
of the cosmic web and its putative relationship to a \ac{whim}, one
must pursue alternate strategies.

%The presence of a galaxy cluster, however, does afford a special
%opportunity to identify and study filaments.  Namely, the cluster
%[marks] a volume in the universe where such threads intersect
%producing an [enhanced opportunity] This enhancment may not be
%sufficient for analysis as the volume filling factor is expected to be
%small even in this environment.  To further [enhance] the probability
%of isolating cosmic threads, one may consider close cluster pairs.
%[wording stinks above]

In principle, one may scour the volumes surrounding galaxy clusters for
signatures of cosmic filaments. A random search, however, would be
compromised by the fact that their volume filling factor is predicted
to be low, even in this environment. To raise the probability of
isolating a cosmic filament, researchers have turned to pairs of
neighbouring clusters on the expectation that these massive structures
will be preferentially connected.
%Because the volumetric filling factors of these filaments are well
%below unity (refs), the chances of having one of these filaments at a
%random direction from the centre of a cluster is low.  If we aim to
%detect these filaments and the \ac{whim}, we must know {\it where} to
%look for them.
Indeed, cosmological dark matter simulations find high probabilities of
having a coherent filamentary structure between close ($< 20$\mpc) and
massive ($>10^{14}$\msun) galaxy clusters \citep[e.g.][]{Colberg2005a,
  Gonzalez2010, Aragon-Calvo2010}. This probability is mostly a
function of the galaxy cluster masses and the separation between them:
the larger the masses and the shorter the separation, the higher the
probability. Therefore, the volume {\it between} close pairs of galaxy
clusters is a natural place to search for signatures of filaments and
an associated \ac{whim}.

Inter-cluster filaments (i.e. filaments between galaxy cluster pairs)
have been inferred from galaxy distributions, either individually from
spectroscopic galaxy surveys \citep[e.g.][]{Pimbblet2004}, or by
stacking analysis from photometric galaxy surveys
\citep[e.g.][]{Zhang2013}. While these studies confirm the strategy to
focus on cluster pairs, they provide limited information into the
nature of cosmic filaments; these luminous systems represent $\lesssim
10\%$ of the baryonic matter, their distribution and motions need not
trace the majority of the gas, and they offer no insight into the
presence of a \ac{whim}.
%but their geometries and properties should be different than that of
%the bulk of the matter.  (refs) and their presence does not imply a
%\ac{whim} necessarily.

\begin{table*}
  \begin{minipage}{0.65\textwidth}
    \centering
    \caption{Properties of the observed \ac{qso} Q1410.}\label{tab:qso_info}
    \begin{tabular}{@{}lcccccc@{}}
      \hline
      
      \multicolumn{1}{c}{QSO Name} & R.A.          & Dec.         & $z_{\rm QSO}$ & \multicolumn{3}{c}{Magnitudes}\\
                                   &  (hr min sec) & (deg min sec)&              & $r$ & NUV & FUV \\ 
      \multicolumn{1}{c}{(1)}      & (2)           & (3)          & (4)          & (5)           & (6)     & (7)     \\
      \hline
  
      SDSS J141038.39+230447.1          & 14 10 38.39 &  $+$23 04 47.18 & 0.7958    & 17.0 & 17.4 & 18.7 \\

      \hline

    \end{tabular}
    \end{minipage}
    \begin{minipage}{0.65\textwidth}
    (1) Name of the QSO.
    (2) Right ascension (J2000). 
    (3) Declination (J2000). 
    (4) Redshift of the QSO. 
    (5) Apparent $r$ (visual) magnitude from \ac{sdss}.
    (6) Apparent near-\ac{uv} magnitude from \acs{galex}.
    (7) Apparent far-\ac{uv} magnitude from \acs{galex}.\\
    
  \end{minipage}
\end{table*}

Promising results from stacking multiple inter-cluster regions have
found an excess of $X$-ray counts in such regions with respect to the
background \citep{Fraser-McKelvie2011}. In contrast to galaxies, one
would be truly observing the bulk of baryonic matter. Unfortunately, the
geometry of the emission and the actual origin of the detected photons
was not well constrained by this original work. Remarkable detection of
individual inter-cluster filaments have also been reported from
gravitational weak lensing signal \citep{Dietrich2012},\footnote{See
  also \citet{Jauzac2012} for a weak lensing signal of a filament
  connecting to a single galaxy cluster.}  and $X$-ray emission
\citep{Kull1999,Werner2008}. Despite their indisputable potential for
characterizing cosmological filaments, these techniques are currently
limited to the most massive systems with geometries maximizing the
observed surface densities, i.e. filaments almost aligned with the
\ac{los}.\\
%but it is still uncertain if one can extrapolate these observational
%results to other cluster-pairs.

%A complementary approach to emission searches, is to look for {\it
%  absorption} signatures of the \ac{whim} in the spectrum of
%  background sources.

%\subsection{Our strategy}

To complement these and other relevant studies to address the
`missing baryons' problem
\citep[e.g.][]{Nevalainen2015,PlanckXXXVII2015,Hernandez2015}, we
have designed a program to detect the putative filaments
connecting cluster pairs in {\it absorption}. This technique has
several advantages over attempts to detect the gas in emission. First,
absorption-line spectroscopy is linearly proportional to the density
of the absorbing gas, offering much greater sensitivity to a diffuse
medium. Second, the absorption lines encode the kinematic
characteristics of the gas, including constraints on the temperature,
turbulence, and \ac{los} velocity. Third, one may assess the chemical
enrichment and ionization state of the gas through the analysis of
multiple ions. The obvious drawback to this technique is that one
requires the fortuitous alignment of a bright background source with
these rare cluster pairs, to probe a greatly reduced spatial volume:
in essence a single pinprick through a given filament. However, with a
large enough survey one may also statistically map the
geometry/morphology of the filaments.

Here we focus on \ac{fuv} spectroscopy leveraging the unprecedented
sensitivity of the \ac{cos} onboard the \ac{hst}, to greatly increase
the sample of inter-cluster filaments probed.\footnote{We note that
  $X$-ray spectroscopy could also be used to trace the \ac{whim} in
  absorption, mostly through \ovii absorption lines
  \citep[e.g.][]{Nicastro2005,Fang2010,Zappacosta2010,Nicastro2010}. However,
  the poor sensitivities of current $X$-ray spectrographs considerably
  limits the sample sizes for these studies. Furthermore, such poor
  sensitivities and poor spectral resolutions make the interpretation
  of signals particularly challenging \citep[e.g.][]{Yao2012}.} With
such \ac{uv} capabilities we can directly access \hi~\lya---the
strongest and most common transition for probing the \ac{igm}. Having
direct coverage of \hi independent of the presence of metals is of
great value for detecting the \ac{whim}
\citep[e.g.][]{Richter2006,Danforth2010}, because this medium may
remain metal poor. Neutral hydrogen generally traces cool and
photoionized gas, but it may also trace collisionally ionized gas in
the \ac{whim} through broad (Doppler parameters $b \gtrsim 50$\kms)
lines \citep[e.g.][]{Tepper-Garcia2012}. Although the \ac{cgm}
surrounding galaxies is responsible for producing \hi\ absorption lines
\citep[especially at column densities $\gtrsim 10^{15}$\cm;
  e.g.][]{Tumlinson2013}, the majority of them must arise in the
diffuse \ac{igm}
\citep[e.g.][]{Prochaska2011b,Tejos2012,Tejos2014}. \ac{fuv}
spectroscopy also allows the detection of the \ovi doublet, a common
highly ionized species. The physical origin of \ovi absorption lines is
controversial, including scenarios of photoionized and/or collisionally
ionized gas in the \ac{cgm} of individual galaxies and/or galaxy groups
\citep[e.g.][]{Tripp2008,Thom2008,Wakker2009,Stocke2014, Savage2014}. Thus, a
collisionally ionized component could well be present, some of which
may come from a \ac{whim} \citep[although
  see][]{Oppenheimer2009,Tepper-Garcia2011}.

In a more general context, \hi and \ovi offer an optimal approach to
study filamentary gas in absorption. As mentioned, this pair of
diagnostics correspond to the most common transitions observed in the
low-$z$ Universe
\citep[e.g.][]{Danforth2008,Tripp2008,Danforth2014},
allowing a good characterization of the background signal against
which one may search for signatures of \ac{whim} in filamentary gas.
Such signatures could include an elevated/suppressed incidence,
covering fractions, and/or unique distributions in the strengths or
widths of the absorption features. In contrast to studies where
absorption systems could be associated with filaments on an individual
basis \citep[e.g.][]{Aracil2006,Narayanan2010},\footnote{We note that
even in individual cases where such absorption does coincide with
known   structures traced by galaxies, it is still unclear whether the
gas is actually produced by a \ac{whim} or individual galaxy halos
\citep[e.g.][]{Stocke2006,Prochaska2011b,Tumlinson2011,Williams2013}.}
our methodology is statistical in nature and a large sample of
independent structures must be collected.

The current advent of big extragalactic surveys makes our approach
feasible. For instance, the \ac{sdss} \citep[][]{Ahn2014} provides
large samples of \ac{lss} traced by galaxies and known \acp{qso} in the
same volume. In particular, by using the galaxy cluster catalog of
\citet{Rykoff2014} we have constructed a cluster-pair sample and found
that, on average, a random sightline extending between $0.1 \le z \le
0.5$ intersects $1 \pm 1$ independent cluster-pairs with projected
separations of $\le 3$\mpc to the inter-cluster axis (defined as the
line segment joining the centers of the two galaxy clusters of a pair;
see \Cref{fig:diagram} for an illustration, and \Cref{sec:field} for
further details), with a very skewed distribution towards zero (see
\Cref{sec:how_unusual}). In order to enhance the efficiency however,
we have cross-matched such cluster-pair sample with known \ac{fuv}
luminous \acp{qso} from the \citet{Schneider2010} catalog, and
identified particular sightlines intersecting more than one of these
structures. Our approach is highly complementary to that of
\citet{Wakker2015}, where a single galaxy filament is targeted with
multiple \ac{qso} sightlines.

In this paper we present \ac{hst}/\ac{cos} \ac{fuv} observations of a
single bright \ac{qso} at $z\approx 0.8$ (namely SDSS
J$141038.39$+$230447.1$, hereafter referred to as Q1410; see
\Cref{tab:qso_info}), whose unique sightline intersects $7$
independent cluster-pairs within $3$\mpc from their inter-cluster
axes. This sightline is highly exceptional; the random expectation of
finding such a number of cluster-pairs is $\lesssim 0.01\%$ (see
\Cref{sec:how_unusual}). With this one dataset, we offer a first
statistical assessment of the presence of diffuse gas close to
cluster-pairs. Although we are only reporting tentative results
($\sim 1-2\sigma$ c.l.), the primary focus of this manuscript is to
establish the experimental design and methodology. Future work will
extend the present study to tens of sightlines, eventually leading {\it towards the statistical detection of the
\ac{whim} in inter-cluster filaments of the cosmic web.}\\

Our paper is structured as follows. In \Cref{sec:data} we present both
the galaxy cluster catalog used to create our cluster-pair sample and
our \ac{hst}/\ac{cos} observations of Q1410. In \Cref{sec:field} we
characterize the volume around the Q1410 sightline in terms of known
clusters and cluster-pairs, quantifying how unusual the Q1410 field
is. In \Cref{sec:abslines} we provide a full characterization of the
Q1410 \ac{hst}/\ac{cos} \ac{fuv} spectrum in terms of absorption line
systems, regardless of the presence of known intervening structures. In
\Cref{sec:dndz} we present our methodology to cross-match the
information provided by the cluster-pair sample and absorption line
systems, while in \Cref{sec:results} we present our observational
results for the Q1410 field. A discussion of these results is presented
in \Cref{sec:discussion}, and a summary of the paper is presented in
\Cref{sec:summary}. Supplementary material is presented in the
Appendix. All distances are in co-moving coordinates assuming
$H_0=67.3$\kms Mpc$^{-1}$, $\Omega_{\rm m}=0.315$, $\Omega_{\rm
  \Lambda}=0.685$, $k=0$ (unless otherwise stated), where $H_0$,
$\Omega_{\rm m}$, $\Omega_{\rm \Lambda}$ and $k$ are the Hubble
constant, mass energy density, `dark energy' density and spatial
curvature, respectively \citep[][]{Planck2013}.

\section{Data}\label{sec:data}

\subsection{Galaxy clusters}\label{data:clusters}

In this section we briefly describe the cluster catalog used in the
present paper. We used \ac{redmapper} \citep{Rykoff2014} applied to the
\ac{sdss} \ac{dr8} \citep{Aihara2011}. This is one of the largest
galaxy cluster catalogs currently available, containing $\sim 25\,000$
rich galaxy clusters ($>20$ galaxies having luminosities $L \ge 0.2
L^*$)\footnote{According to their calibration, this richness limit
  corresponds to a mass of $M\sim 1.8 \ 10^{14}$\msun \citep[uncertain
    up to $\sim 0.25$ in $\ln(M)$;][see also
    \Cref{sec:clusters}]{Rykoff2012}.} at $0.08 \le z \le 0.55$.

The \ac{redmapper} catalog is very well suited for statistical
analysis: it defines clusters properties in terms of probabilities
(e.g. position, richness, redshift, galaxy members), with a well
understood selection function; it adopts an optimal mass-richness
relationship \citep{Rykoff2012}; and it has high completeness and
purity levels compared to others cluster catalogs
\citep[][]{Rykoff2014,Rozo2014a,Rozo2014b}.

In this paper we used an extension of the published \ac{redmapper}
catalog, including galaxy clusters with richness below $20$ but larger
than $10$.\footnote{Kindly provided by E. Rykoff and E. Rozo (private
  communication).} The mass--richness relation relevant to the
\ac{redmapper} catalog is,

\begin{equation}
\ln \left(\frac{M_{200}}{h^{-1}_{70}\,10^{14}M_{\odot}}\right) = 1.72 +
1.08 \ln \left( \frac{\lambda}{60} \right) \ \rm{,}
\label{eq:mass}
\end{equation}

\begin{figure*}
\begin{minipage}{1\textwidth}
    \centering
    \includegraphics[width=1\textwidth]{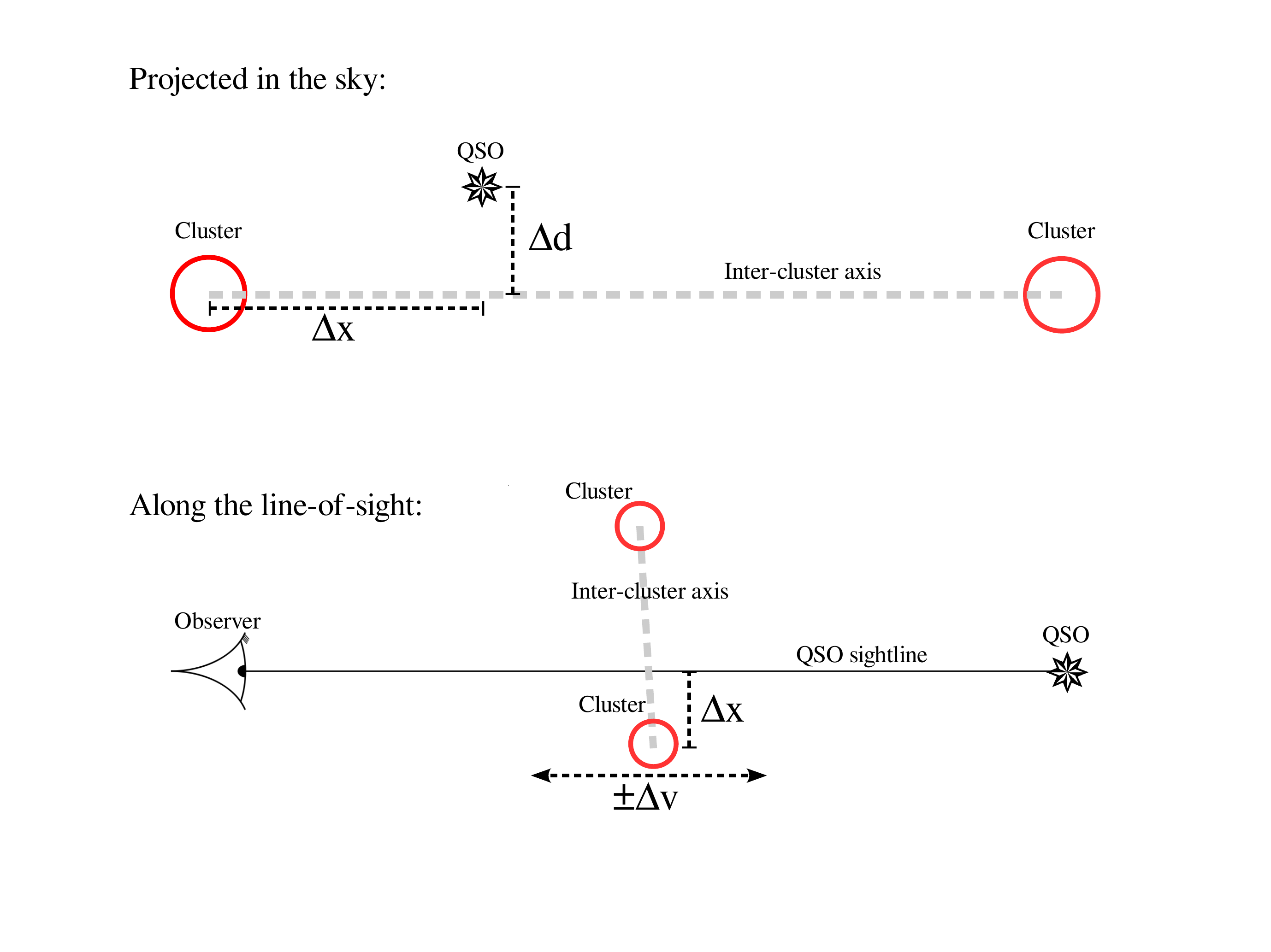}
\end{minipage}
\vspace{-10ex}
\caption{Schematic representation of our survey geometry projected in
  the sky (top) and along the line-of-sight (bottom), for a single
  cluster-pair. Galaxy clusters are represented by red circles, while
  the inter-cluster axis is represented by the grey dashed lines. The
  \ac{qso} itself is represented by a eight-pointed star. The impact
  parameter between the \ac{qso} sightline and the inter-cluster axis
  is defined as $\Delta d$, while the distance to the {\it closest}
  cluster of the pair along the projected inter-cluster axis is defined
  as $\Delta x$. A rest-frame velocity window around the position of
  the cluster-pair is defined as $\pm \Delta v$.}\label{fig:diagram}

\end{figure*}

\noindent with a typical scatter of $\sim 0.25$ in $\ln(M)$
\citep{Rykoff2012}, where $M_{200}$ is the total mass enclosed within
an overdensity of $200$ times the critical density of the
Universe,$\rho_{\rm crit}$; $h_{70}$ is the dimensionless Hubble
parameter $h_{70} \equiv H_0 / (70$\, km s$^{-1}$ Mpc$^{-1})$;
$M_{\odot}$ is the mass of the sun; and $\lambda$ is the richness of
galaxies with luminosities $L\ge 0.2 L^*$ (corrected for
incompleteness). Extrapolating this relation to $\lambda= 10$ we get a
minimum mass $M_{200}(\lambda=10) \approx 0.8 \times 10^{14} M_{\odot}$
in our assumed cosmology (see end of \Cref{sec:intro}). Therefore, our
adopted limit should still ensure a reasonable large minimum mass
limit. We also computed an estimate of the virial radius, $R_{200}$,
simply defined as the radius at which the $M_{200}$ is enclosed,

\begin{equation}
R_{200} \equiv \left( \frac{M_{200}}{\frac{4}{3}\pi200\rho_{\rm crit}}
\right)^{1/3} \ \rm{.}
\label{eq:virial_radii}
\end{equation}

We note that because we are only using these clusters as tracers of
high density regions in the cosmic web, the {\it exact} mass of the
clusters will not be particularly relevant, making potential systematic
uncertainties in the mass-richness calibration and its extrapolation
to lower values not a critical issue. Moreover, low richness cluster
samples suffer more from impurity and incompleteness
\citep[][]{Rykoff2014}, but such issues should not create a fake signal
in differentiating the properties of absorption line systems in
environments traced by these cluster-pairs with respect to the field,
in the presence of real inter-cluster filaments. On the contrary, our
approach is conservative in the sense that impurity and incompleteness
would dilute such a signal, if present.

For the purposes of this paper, we also required the clusters to lie
between $0.1 \le z \le 0.5$, yielding a total of $162\,144$ clusters
(see \Cref{sec:clusters} and \Cref{sec:cluster_control} for further
details).

Although the \ac{redmapper} catalog is mostly based on photometry, the
spectroscopic redshift of the most likely centre is also given when
available (typically from its \ac{bcg}).\footnote{We note that the
  typical photometric redshift uncertainties for the \ac{redmapper}
  clusters are of the order of $\delta z \sim 0.006$.} This is
advantageous for our experiment; a high precision in the cluster
redshifts is needed for a reliable association with the gas observed in
absorption with inter-cluster filaments (if any). About $44\%$
($71\,701/162\,144$) of the \ac{redmapper} clusters have spectroscopic
redshifts, yielding velocity precision of $\approx 30$\kms~in those
clusters' rest frames. We note that for the subsample of clusters most
relevant to our paper (i.e. within $20$\mpc of the Q1410), such
fraction increases to $\sim 70\%$ ($40/57$; see \Cref{sec:clusters} for
further details).

\subsection{Q1410 absorption lines}\label{data:obs}

\subsubsection{Specific selection of Q1410}\label{sec:selection}

In this section we describe in detail the selection criteria of our
targeted \ac{qso}, Q1410. We emphasize that the original selection of
Q1410 was done using the galaxy cluster catalog published by
\citet{Hao2010} instead of the \ac{redmapper} catalog
\citep{Rykoff2014} used here.

The \ac{gmbcg} catalog \citep[][]{Hao2010} is based on data from the
\ac{dr7} of the \ac{sdss} \citep{Abazajian2009}, and similar to
\ac{redmapper}, it is mainly based on photometry. From the \ac{gmbcg}
catalog we searched for cluster-pairs where at least one member has a
spectroscopic redshift and where the redshift difference between them
is less than $3\times$ the combined redshift uncertainty, and over
clusters having \ac{gmbcg} richness $>15$. We then measured the
transverse co-moving separation between clusters at the redshift of the
cluster with spectroscopic identification (if both clusters had
spectroscopic redshifts we used the average redshift), and kept the
ones separated by $<25$\mpc.

\begin{table*}
\centering
\scriptsize
\begin{minipage}{0.75\textwidth}
\centering
\caption{Properties of the \ac{redmapper} clusters at $0.1 \le z \le 0.5$ and within $20$\mpc from the Q1410 sightline.}\label{tab:clusters}
\begin{tabular}{@{}cccccccccrr@{}}

\hline                            

Cluster ID & R.A. & Dec. & $z_{\rm spec}$ & $z_{\rm photo}$  &Richness & Mass                      &  $R_{200}$ &\multicolumn{3}{c}{Impact parameter} \\ 

       & (degrees)& (degrees) &           &                    &         & ($10^{14} \ M_{\sun}$)   &    (\mpc) &(degrees)       &      (Mpc)&   ($R_{200}$)      \\ 

(1)    &(2)   &(3)   &(4)                 &(5)                 &(6)      &(7)                            & (8)        &    (9)         &  (10)     &   (11) \\ 

\hline 
1 & 212.994666 & 21.418861 & 0.1335 & 0.150 $\pm$ 0.006 & 11.0 & 0.92 &0.9 &1.6895 & 16.9 &18.5\\ 
2 & 213.481691 & 22.808351 & 0.1376 & 0.147 $\pm$ 0.005 & 20.4 & 1.80 &1.1 &0.8039 & 8.3 &7.3\\ 
3 & 214.114059 & 23.256258 & 0.1381 & 0.138 $\pm$ 0.005 & 33.0 & 3.03 &1.4 &1.3484 & 13.9 &10.3\\ 
4 & 213.817562 & 24.020988 & 0.1386 & 0.152 $\pm$ 0.006 & 16.5 & 1.43 &1.1 &1.4184 & 14.7 &13.9\\ 
5 & 213.014763 & 22.125043 & 0.1413 & 0.141 $\pm$ 0.006 & 18.3 & 1.60 &1.1 &1.0093 & 10.6 &9.7\\ 
6 & 213.223623 & 22.319874 & 0.1417 & 0.150 $\pm$ 0.005 & 20.3 & 1.79 &1.1 &0.9208 & 9.7 &8.6\\ 
7 & 213.330826 & 22.510559 & \dots & 0.143 $\pm$ 0.005 & 17.9 & 1.56 &1.1 &0.8405 & 9.0 &8.3\\ 
8 & 213.664909 & 22.216942 & 0.1535 & 0.171 $\pm$ 0.007 & 10.9 & 0.91 &0.9 &1.2667 & 14.5 &16.0\\ 
9 & 212.712357 & 23.086999 & 0.1580 & 0.157 $\pm$ 0.006 & 13.0 & 1.11 &1.0 &0.0487 & 0.6 &0.6\\ 
10 & 212.162442 & 21.946327 & 0.1596 & 0.145 $\pm$ 0.006 & 11.3 & 0.95 &0.9 &1.2231 & 14.5 &15.9\\ 
11 & 212.999309 & 24.435343 & 0.1612 & 0.171 $\pm$ 0.006 & 16.1 & 1.39 &1.0 &1.3907 & 16.6 &16.1\\ 
12 & 212.674015 & 22.495683 & 0.1733 & 0.179 $\pm$ 0.006 & 23.0 & 2.05 &1.2 &0.5842 & 7.5 &6.4\\ 
13 & 212.895979 & 23.605206 & 0.1787 & 0.167 $\pm$ 0.007 & 11.4 & 0.96 &0.9 &0.5684 & 7.5 &8.3\\ 
14 & 213.557346 & 22.914918 & 0.1918 & 0.195 $\pm$ 0.007 & 12.4 & 1.05 &0.9 &0.8423 & 11.9 &12.8\\ 
15 & 212.123899 & 22.560692 & 0.2220 & 0.241 $\pm$ 0.012 & 11.2 & 0.94 &0.9 &0.7167 & 11.6 &13.1\\ 
16 & 211.805856 & 23.845513 & 0.2359 & 0.233 $\pm$ 0.008 & 21.4 & 1.90 &1.1 &1.0955 & 18.8 &16.8\\ 
17 & 213.277561 & 24.079481 & 0.2392 & 0.234 $\pm$ 0.012 & 10.6 & 0.89 &0.9 &1.1488 & 20.0 &23.1\\ 
18 & 213.029395 & 23.923686 & 0.2424 & 0.254 $\pm$ 0.010 & 24.9 & 2.23 &1.2 &0.9094 & 16.0 &13.6\\ 
19 & 212.319446 & 22.314272 & 0.2914 & 0.482 $\pm$ 0.021 & 11.4 & 0.96 &0.9 &0.8275 & 17.3 &19.8\\ 
20 & 212.415280 & 22.627699 & 0.3127 & 0.306 $\pm$ 0.020 & 11.4 & 0.96 &0.9 &0.5052 & 11.3 &13.0\\ 
21 & 213.395595 & 23.114057 & \dots & 0.327 $\pm$ 0.020 & 12.2 & 1.03 &0.9 &0.6775 & 15.8 &17.9\\ 
22 & 212.831535 & 22.621344 & 0.3382 & 0.368 $\pm$ 0.017 & 30.0 & 2.73 &1.2 &0.4849 & 11.6 &9.6\\ 
23 & 212.723980 & 22.929555 & 0.3412 & 0.357 $\pm$ 0.017 & 28.1 & 2.54 &1.2 &0.1614 & 3.9 &3.3\\ 
24 & 213.000928 & 22.437411 & \dots & 0.345 $\pm$ 0.018 & 23.6 & 2.10 &1.1 &0.7152 & 17.5 &15.7\\ 
25 & 212.293470 & 23.265859 & 0.3465 & 0.349 $\pm$ 0.019 & 21.2 & 1.88 &1.1 &0.3849 & 9.4 &8.8\\ 
26 & 212.684410 & 23.402739 & 0.3506 & 0.358 $\pm$ 0.024 & 11.4 & 0.96 &0.9 &0.3237 & 8.0 &9.4\\ 
27 & 212.770942 & 23.232412 & 0.3508 & 0.357 $\pm$ 0.022 & 10.4 & 0.87 &0.8 &0.1836 & 4.6 &5.5\\ 
28 & 213.066601 & 23.075991 & 0.3511 & 0.378 $\pm$ 0.021 & 20.6 & 1.82 &1.1 &0.3741 & 9.3 &8.8\\ 
29 & 212.600828 & 23.822434 & 0.3520 & 0.358 $\pm$ 0.020 & 15.0 & 1.29 &0.9 &0.7446 & 18.5 &19.7\\ 
30 & 212.854438 & 22.531418 & \dots & 0.365 $\pm$ 0.021 & 13.5 & 1.15 &0.9 &0.5769 & 14.8 &16.5\\ 
31 & 212.861147 & 22.341353 & 0.3718 & 0.385 $\pm$ 0.016 & 22.8 & 2.03 &1.1 &0.7614 & 19.9 &18.3\\ 
32 & 213.011303 & 23.110255 & 0.3722 & 0.395 $\pm$ 0.019 & 13.6 & 1.16 &0.9 &0.3246 & 8.5 &9.4\\ 
33 & 212.737954 & 23.677308 & 0.3725 & 0.379 $\pm$ 0.017 & 18.4 & 1.61 &1.0 &0.6018 & 15.8 &15.7\\ 
34 & 213.115594 & 22.812719 & 0.3725 & 0.400 $\pm$ 0.022 & 16.9 & 1.46 &1.0 &0.4973 & 13.0 &13.4\\ 
35 & 212.604626 & 23.277634 & \dots & 0.376 $\pm$ 0.018 & 22.2 & 1.98 &1.1 &0.2043 & 5.4 &5.0\\ 
36 & 212.538462 & 23.376505 & \dots & 0.384 $\pm$ 0.019 & 17.6 & 1.53 &1.0 &0.3170 & 8.5 &8.7\\ 
37 & 212.800638 & 23.052477 & 0.4138 & 0.426 $\pm$ 0.015 & 20.0 & 1.76 &1.0 &0.1323 & 3.8 &3.7\\ 
38 & 212.609281 & 22.997888 & 0.4159 & 0.425 $\pm$ 0.015 & 10.6 & 0.89 &0.8 &0.0942 & 2.7 &3.4\\ 
39 & 212.177450 & 22.994366 & 0.4188 & 0.424 $\pm$ 0.015 & 25.4 & 2.28 &1.1 &0.4522 & 13.1 &11.9\\ 
40 & 213.120494 & 22.548543 & \dots & 0.420 $\pm$ 0.014 & 38.4 & 3.57 &1.3 &0.6800 & 19.8 &15.4\\ 
41 & 211.935365 & 23.227769 & 0.4199 & 0.416 $\pm$ 0.019 & 10.7 & 0.90 &0.8 &0.6825 & 19.9 &24.5\\ 
42 & 212.374306 & 22.818205 & \dots & 0.420 $\pm$ 0.019 & 17.3 & 1.50 &1.0 &0.3710 & 10.8 &11.2\\ 
43 & 212.420269 & 23.093511 & \dots & 0.428 $\pm$ 0.018 & 14.5 & 1.24 &0.9 &0.2209 & 6.6 &7.3\\ 
44 & 212.593164 & 22.699892 & \dots & 0.430 $\pm$ 0.017 & 12.1 & 1.02 &0.8 &0.3848 & 11.5 &13.6\\ 
45 & 211.957129 & 23.106715 & \dots & 0.434 $\pm$ 0.014 & 28.6 & 2.59 &1.1 &0.6471 & 19.4 &16.9\\ 
46 & 212.999005 & 22.821469 & 0.4358 & 0.443 $\pm$ 0.019 & 12.5 & 1.06 &0.9 &0.4052 & 12.2 &14.3\\ 
47 & 212.607221 & 22.800250 & \dots & 0.440 $\pm$ 0.018 & 13.0 & 1.10 &0.9 &0.2837 & 8.6 &10.0\\ 
48 & 212.115194 & 23.041513 & 0.4397 & 0.442 $\pm$ 0.016 & 11.1 & 0.94 &0.8 &0.5027 & 15.3 &18.7\\ 
49 & 212.968630 & 22.717594 & \dots & 0.441 $\pm$ 0.021 & 11.0 & 0.92 &0.8 &0.4604 & 14.0 &17.3\\ 
50 & 212.980494 & 22.615241 & \dots & 0.441 $\pm$ 0.018 & 15.6 & 1.34 &0.9 &0.5505 & 16.7 &18.2\\ 
51 & 212.172529 & 23.471857 & \dots & 0.445 $\pm$ 0.018 & 15.6 & 1.35 &0.9 &0.5952 & 18.3 &19.8\\ 
52 & 212.710891 & 23.162554 & \dots & 0.452 $\pm$ 0.017 & 11.4 & 0.96 &0.8 &0.0951 & 3.0 &3.6\\ 
53 & 212.336805 & 22.647284 & \dots & 0.454 $\pm$ 0.019 & 15.5 & 1.34 &0.9 &0.5251 & 16.4 &18.0\\ 
54 & 212.658634 & 23.039911 & 0.4582 & 0.437 $\pm$ 0.014 & 37.2 & 3.44 &1.3 &0.0399 & 1.3 &1.0\\ 
55 & 212.346270 & 23.242090 & 0.4585 & 0.449 $\pm$ 0.016 & 15.6 & 1.35 &0.9 &0.3310 & 10.4 &11.4\\ 
56 & 213.213654 & 22.988852 & 0.4603 & 0.477 $\pm$ 0.013 & 47.9 & 4.52 &1.4 &0.5176 & 16.4 &11.9\\ 
57 & 213.210168 & 23.112030 & 0.4615 & 0.469 $\pm$ 0.013 & 48.9 & 4.62 &1.4 &0.5071 & 16.1 &11.7\\ 

\hline
\end{tabular}
  \end{minipage}
\begin{minipage}{0.75\textwidth}
(1) Cluster ID.
(2) Right ascension (J2000).
(3) Declination (J2000). 
(4) Spectroscopic redshift.
(5) Photometric redshift.
(6) Richness of galaxies having $L\ge 0.2 L^*$, corrected for incompleteness (hence non-integer).
(7) Inferred mass using \Cref{eq:mass}; typical scatter of $\sim 0.25$ in $\ln(M)$ \citep{Rykoff2012}.
(8) Inferred virial radii of the cluster using \Cref{eq:virial_radii}.
(9) Projected separation to the Q1410 sightline in degrees.
(10) Projected separation to the Q1410 sightline in \mpc.
(11) Projected separation to the Q1410 sightline in units of our $R_{200}$ estimation.
%{\bf Notes:} \
%$^a$: Spectroscopic redshift retrieved from \ac{sdss} \ac{dr10}.\
%$^b$: Spectroscopic redshift retrieved from \ac{sdss} \ac{dr10}, from a
%companion object ($< 5$ arcseconds).

\end{minipage}
\end{table*}

We selected our target from the \ac{qso} catalog published by
\citet{Schneider2010}, which is also based on \ac{sdss} \ac{dr7}
data. This catalog comprises $\gtrsim 100\,000$ \acp{qso} with well
known magnitudes and spectroscopic redshifts. We looked for \acp{qso}
having redshifts greater than individual cluster-pairs and located
inside their sky-projected cylinder areas as defined above. We imposed
a magnitude limit of $r<17.5$\,mag to select relatively bright
\acp{qso}. We gave priority to \acp{qso} $z>0.3$, ensuring large
redshift path coverage. We also searched in the \ac{galex}
\citep{Martin2005} database and prioritized those \acp{qso} with high
\ac{fuv} fluxes to ensure no higher-$z$ Lyman Limit Systems (LLS) were
present,\footnote{We believe that biasing against LLS is unimportant
  for our present study.} enabling a \ac{snr} $S/N\sim10$ spectra to be
observed in a relatively short exposure time (no larger than $15$
orbits). For each of these \acp{qso} we counted the number of
independent cluster-pairs (defined as those which were separated by
more than $1000$ km s$^{-1}$ from another, and by more than $5000$ km
s$^{-1}$ from the background \ac{qso} in rest-frame velocity space) at
impact parameters $\Delta d \le 2$\mpc from the \ac{qso} sightline (see
\Cref{fig:diagram} for an illustration). We note that in this paper we
adopted a larger limit of $3$\mpc as the fiducial minimum impact
parameter (see \Cref{sec:field}), motivated by the results obtained in
\Cref{sec:results}. We finally selected the sightline that
maximized the number of independent cluster-pair structures, for the
mininum observing time. \Cref{tab:qso_info} presents a summary of the
properties of the targeted \ac{qso}, Q1410.

\subsubsection{HST/COS observations and data reduction of Q1410}

In this paper we present \ac{fuv} spectroscopic data of Q1410 (see
\Cref{tab:qso_info} for a summary of its main properties) from
\ac{hst}/\ac{cos} \citep{Green2012} taken under program \ac{go} 12958
(PI Tejos).

The \ac{qso} Q1410 was observed in August 2013 using both G130M and
G160M gratings centred at $1318$ and $1611$ \AA\ respectively, using
the four \acp{fppos} available for each configuration. These settings
provided medium-resolution ($R \equiv \frac{\lambda}{\Delta\lambda}
\sim 16\,000-21\,000$ or $FWHM \sim 0.07-0.09$\AA) over the \ac{fuv}
wavelength range of $\sim 1160-1790$ \AA, but having two $\sim 20$
\AA\ gaps around each central wavelength. We chose this approach rather
than a continuous wavelength coverage to increase the \ac{snr} at
spectral regions where we expected to find \hi\ absorption line systems
associated with inter-cluster filaments. Having multiple \ac{fppos} (as
in our observations) is crucial, however, to minimize the effect of
fixed-pattern noise present in \ac{cos} \citep[see][for more details on
  the technical aspects of \ac{cos}]{Osterman2011,Green2012}.

    \begin{figure*}
    \begin{minipage}{1\textwidth}
    \centering
    \includegraphics[width=1\textwidth]{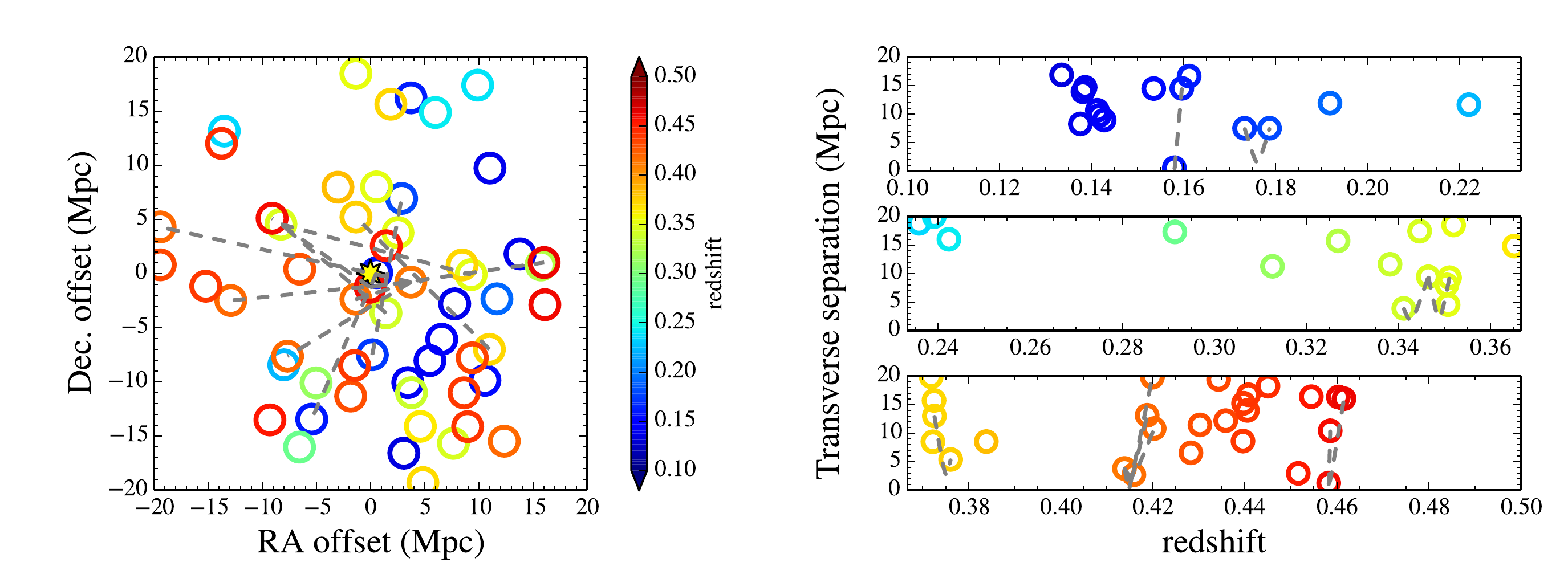}
    \end{minipage}
    
    \caption{Representation of the \acp{lss} within $20$\mpc
    around the Q1410 sightline at redshifts $0.1 \le z \le 0.5$, as
    traced by the $57$ galaxy clusters from the \ac{redmapper}
    catalog (see \Cref{sec:field} and
    \Cref{tab:clusters}). Galaxy clusters are represented by coloured
    circles according to redshift, as given by the colour bar
    scale. Cluster-pairs at impact parameters $\Delta d \le 3$\mpc of
    the Q1410 sightline are represented by dashed grey lines. The left
    panel shows to the projected in the sky distributions in co-moving
    \mpc, where the Q1410 sightline is represented by the yellow star
    at the origin. The right panel shows the distribution along the
    line-of-sight in terms of total transverse separation (in co-moving
    \mpc) as a function of redshift (note that inter-cluster axes
    appear as hyperbolas in these coordinates). See \Cref{sec:field}
    for further details.}  \label{fig:field}
    
    \end{figure*}

\begin{table*}
\centering
\scriptsize
\begin{minipage}{0.65\textwidth}
\centering
\caption{Clusters-pairs around Q1410.}\label{tab:cpairs}
\begin{tabular}{@{}ccccccccc@{}}

\hline                            

Pair ID & Cluster IDs & $z$ &\multicolumn{2}{c}{Separation between clusters}  & $\Delta d$ &  $\Delta x$ & Both spec-$z$?  & Grouped ID\\ 

        &             &     &Transverse (Mpc)  &  LOS (km s$^{-1}$)                & (Mpc)            &          (Mpc)           &                 &       \\ 

    (1) &(2)          & (3) & (4)              & (5)                                 & (6)              &            (7)           & (8)             & (9)    \\

\hline 
1 & 9,10 & 0.1588 & 14.7 & 419 & 0.48 & 0.31 &y&1  \\ 
2 & 12,13 & 0.1760 & 14.7 & 1378 & 1.54 & 7.33 &y&2  \\ 
3 & 23,25 & 0.3439 & 12.7 & 1189 & 1.85 & 3.43 &y&3  \\ 
4 & 25,28 & 0.3488 & 18.1 & 1021 & 2.29 & 9.00 &y&4  \\ 
5 & 34,35 & 0.3726 & 17.3 & 771 & 2.76 & 4.63 &n&5  \\ 
6 & 37,42 & 0.4139 & 13.2 & 1365 & 2.59 & 2.79 &n&6  \\ 
7 & 37,38 & 0.4149 & 5.3 & 435 & 1.86 & 1.99 &y&6  \\ 
8 & 37,39 & 0.4163 & 16.7 & 1043 & 1.16 & 3.62 &y&6  \\ 
9 & 37,41 & 0.4169 & 23.6 & 1284 & 0.03 & 3.80 &y&6  \\ 
10 & 54,55 & 0.4584 & 11.1 & 68 & 1.05 & 0.69 &y&7  \\ 
11 & 54,57 & 0.4599 & 16.2 & 682 & 1.24 & 0.21 &y&7  \\ 

\hline
\end{tabular}
  \end{minipage}
\begin{minipage}{0.65\textwidth}
(1) Cluster-pair ID.
(2) IDs of clusters defining the cluster-pair as given in \Cref{tab:clusters}.
(3) Redshift of the cluster-pair.
(4) Transverse separation between clusters in \mpc.
(5) Along the line-of-sight separation between the clusters in rest-frame \kms.
(6) Impact parameter to the Q1410 sightline.
(7) Projected on the sky distance to the {\it closest} cluster of the pair, along the inter-cluster axis.
(8) Whether both clusters have spectroscopic redshifts.
(9) Grouped ID for independent cluster-pairs.
\end{minipage}
\end{table*}

Data reduction was performed in the same fashion as presented in
\citet{Finn2014} and \citet{Tejos2014}, for the G130M and G160M
\ac{cos} gratings. In summary, we used the \calcos v2.18.5 pipeline
with extraction windows of $25$ and $20$ pixels for the G130M and G160M
gratings, respectively. We applied a customized background estimation
smoothing (boxcar) over $1000$ and $500$ pixels for the \ac{fuv}A and
\ac{fuv}B stripes, respectively, while masking out and linearly
interpolating over regions close to geocoronal emission lines and
pixels flagged as having bad quality. The uncertainty was calculated in
the same manner as in \calcos but using our customized background. The
co-alignment was performed using strong Galactic \ac{ism} features as
reference. We finally binned the original spectra having dispersions of
$\sim 0.010$ and $\sim 0.012$ \AA\, pixel$^{-1}$ for the G130M and
G160M gratings respectively, into a single linear wavelength scale of
$0.0395$ \AA\, pixel$^{-1}$ (roughly corresponding to two pixels per
resolution element). Due to the difficulties in assessing the degree of
geocoronal contamination in the final reduced Q1410 spectrum, we opted
to mask out the spectral regions close to rest-frame \nI, \hi~\lya\ and
\oi\ (namely $1300.0-1307.5$, $1198.5-1201.0$ and $1213.5-1217.8$~\AA,
respectively).

Our pseudo-continuum\footnote{i.e. including intrinsic broad emission
  lines and the Galaxy's damped \lya~system wings.} fit was modelled as
in \citet{Tejos2014}, but also introducing the presence of three
partial LLS breaks at $\approx 1232$, $1401$ and
$1637$~\AA. \Cref{fig:spectrum} shows the reduced Q1410 spectrum (black
line), its corresponding uncertainty (green lines) and our adopted
pseudo-continuum fit (blue dotted line).
%The figure also shows our Voigt profile fit solutions and residuals
%(red lines and grey dots respectively; see \Cref{sec:voigt_fit}), and
%their corresponding IDs and reliability labels (see
%\Cref{sec:reliability} and \Cref{tab:abslines}).

\section{Characterization of large-scale structures around the Q1410 sightline}\label{sec:field}

\subsection{Galaxy clusters}\label{sec:clusters}

From the \ac{redmapper} catalog described in \Cref{data:clusters}, we
define a subsample of clusters according to the following criteria:

\begin{itemize}
\item the redshift has to lie between $0.1 \le z \le 0.5$; and,
\item the impact parameter to the Q1410 sightline has to be no larger
  than $20$\mpc.
\end{itemize}

There are a total of $57$ clusters from the \ac{redmapper} catalog
satisfying the aforementioned criteria, whose relevant information is
presented in \Cref{tab:clusters}.  We also show their distribution
around the Q1410 sightline in \Cref{{fig:field}} (coloured circles).

The redshift range of $0.1 \le z \le 0.5$ was chosen to ensure
simultaneous coverage of both \hi\ and \ovi\ transitions from our
\ac{cos} data, while the impact parameter of $20$\mpc (arbitrary) was
chosen to cover scales expected to be relevant for inter-cluster
filaments \citep[e.g.][]{Colberg2005a, Gonzalez2010}.

In \Cref{sec:cluster_control} we show how our subsample of clusters
compares to appropriate control samples drawn from the full
\ac{redmapper} catalog. We found no statistically significant
differences for the mass (richness) and redshift distributions between
our subsample and the control samples, implying that no noticeable bias
is present in the subsample close to the Q1410 sightline.

\subsection{Cluster-pairs}\label{sec:pairs}

From the subsample of clusters around the Q1410 sightline presented in
\Cref{tab:clusters}, we define a sample of cluster-pairs according to
the following criteria:

\begin{itemize}

\item the rest-frame velocity difference between the clusters redshifts
  has to be $< 2000$\kms;

\item at least one of the two members has to have a spectroscopic
  redshift determination (typically from a \ac{bcg}), and the other has
  to have a redshift uncertainty no larger than
  $0.05$.\footnote{Although a photometric uncertainty of $\delta z =
    0.05$ corresponds to a very large $\delta v \approx
    15\,000/(1+z)$\kms, we note that in most of our cluster-pair sample
    both clusters have spectroscopic redshifts (see
    \Cref{tab:cpairs}).}

\item the transverse separation between the cluster centres has to be
  no larger than $25$\mpc; and,

\item the impact parameter between the inter-cluster axis and the Q1410
  sightline has to be $\Delta d \le 3$\mpc.

\end{itemize}

When these criteria are satisfied, we assign the cluster-pair redshift
to be the average between the two cluster members. There are a total of
$11$ cluster-pairs satisfying these criteria around the Q1410 sightline
(see the grey dashed lines in \Cref{{fig:field}}), and whose relevant
information is presented in \Cref{tab:cpairs}.

We chose $2000$\kms (arbitrary) for the rest-frame velocity difference
limit for the clusters in a cluster-pair, in order to account for the
typical velocity dispersion of galaxy clusters ($\sim 600$\kms) and a
contribution from a cosmological redshift difference. We note however
that the majority of the clusters in a given cluster-pair have
rest-frame velocity differences $<1000$\kms and that {\it all} of them
have $<1400$\kms (see the fifth column of \Cref{tab:cpairs}). The need
for relatively small redshift uncertainties for the clusters is
necessary to minimize the dilution of a real signal due to
unconstrained positions for the cluster-pairs along the
line-of-sight. The $25$\mpc (arbitrary) maximum separation between
clusters in a cluster-pair was motivated by theoretical results from
$N$-body simulations in $\Lambda$CDM universes. These studies indicate
that at $< 25$\mpc there is relatively high probability of having
coherent filamentary structures between galaxy clusters
\citep[e.g.][]{Colberg2005a, Gonzalez2010}. We stress that the majority
of cluster-pairs in our sample have projected separations $\sim
10-15$\mpc (see the fourth column of \Cref{tab:cpairs}). The choice for
the maximum impact parameter between the cluster-pair inter-cluster
axis and the Q1410 sightline of $\Delta d = 3$\mpc was directly
motivated by one of our observational results (see \Cref{sec:results}),
and is in good agreement with the typical scales for the radii of
inter-cluster filaments inferred from $N$-body simulations
\citep[e.g.][]{Colberg2005a, Gonzalez2010,Aragon-Calvo2010}.

\subsection{Independent cluster-pairs}\label{sec:grouped}

As expected from the clustering of galaxy clusters, many cluster-pairs
are grouped together and hence might not be tracing independent
structures. We therefore have grouped cluster-pairs if they are within
$1000$\kms from one another and we treat them as independent. There are
a total of $7$ independent cluster-pairs; a unique identifier is given
for each of these in the last column of \Cref{tab:cpairs}. We can
clearly see these structures in the right panel of \Cref{fig:field} at
$z \sim 0.16, 0.18, 0.34, 0.35, 0.37, 0.41$ and $0.46$.

Again, this velocity limit of $1000$\kms is arbitrary and chosen to
account for the typical velocity dispersion of galaxy clusters and a
contribution from a cosmological redshift difference. As reference, if
$3000$\kms is used instead, then there are $6$ independent structures
rather than $7$ (i.e. the structures at $z\sim 0.34-0.35$ are joined
together). We note however that in our subsequent analysis of
associating \ac{igm} absorption lines with cluster-pairs, we will only
use the impact parameter to the {\it closest} cluster-pair
independently of the group it belongs to (see
\Cref{sec:results}). Therefore, the velocity limit for grouping
cluster-pairs is irrelevant for the main results of this paper. Still,
this definition allows us to quantify how many independent
cluster-pairs the Q1410 sightline is probing, making sure that our
results are not dominated by a single coherent structure spanning a
large redshift range.

    \begin{figure*}
    \begin{minipage}{1\textwidth}
    \centering
    \includegraphics[width=1\textwidth]{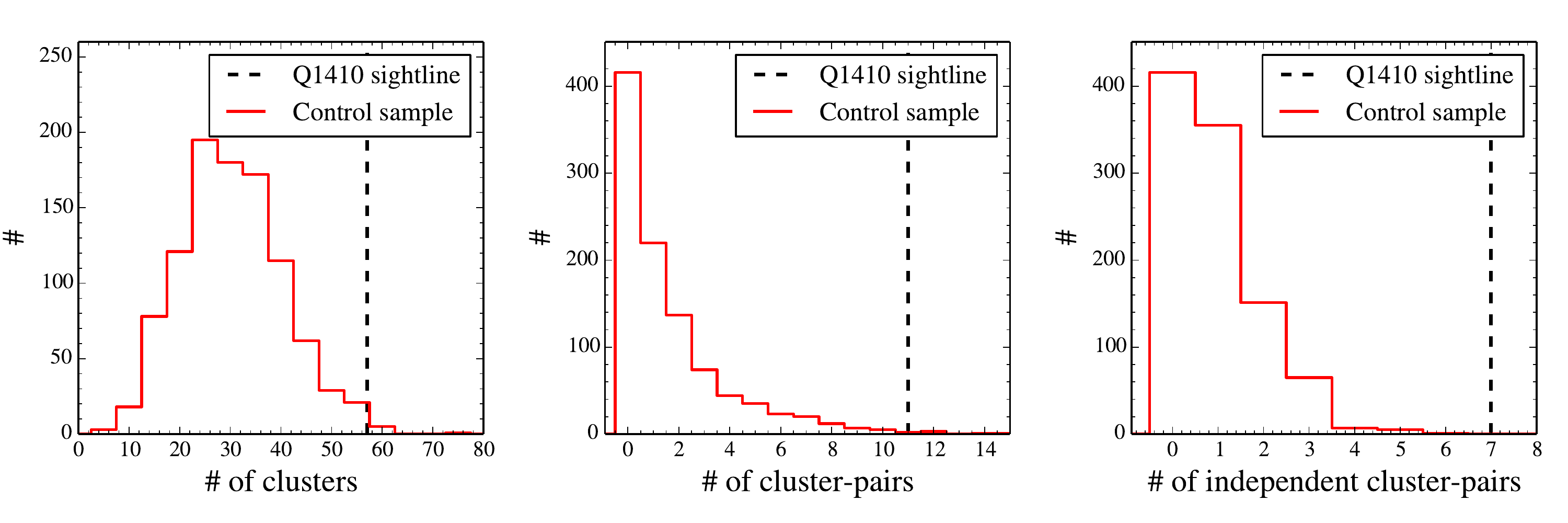}
    \end{minipage}

    \caption{Comparison between our observed (dashed black vertical
    lines) number of \ac{redmapper} clusters (left panel),
    cluster-pairs (middle panel) and independent cluster-pairs (right
    panel) in the Q1410 sightline satisfying our criteria (see
    \Cref{sec:clusters}, \Cref{sec:pairs} and
    \Cref{sec:grouped}, respectively), and the distributions from
    control samples (solid red histograms; see
    \Cref{sec:how_unusual}). The Q1410 sightline is highly
    exceptional in terms of number of LSS traced by galaxy clusters
    close to it.}\label{fig:how_unusual}
    
    \end{figure*}

\subsection{How unusual is the Q1410 field?}\label{sec:how_unusual}

As explained in \Cref{sec:selection}, the field around Q1410 was
selected to maximize the presence of cluster-pairs close to the
\ac{qso} sightline. Therefore, it is by no means a randomly selected
sightline. To quantify how unusual the sightline is, we have performed
the same search for clusters, cluster-pairs and independent cluster-pairs
in $1000$ randomly selected sightlines having coordinates R.A. $\in
[140,222]$ degrees and Dec. $\in [4,56]$ degrees (i.e. well within the
\ac{sdss} footprint), using the same set of criteria used to
characterize the Q1410 field (see \Cref{sec:clusters}, \Cref{sec:pairs}
and \Cref{sec:grouped}).

In \Cref{fig:how_unusual} we compare our observed (dashed black
vertical lines) number of \ac{redmapper} clusters (left panel),
cluster-pairs (middle panel) and independent cluster-pairs (right
panel), to the distributions from our control samples (solid red
histograms). There are a total of $57$ clusters at redshifts $0.1 \le z
\le 0.5$ satisfying the condition of being at impact parameter of
$<20$\mpc from the Q1410 sightline, whereas the average random
expectation is $32 \pm 10$ with median of $32$. Likewise, the actual
number of cluster-pairs and independent cluster-pair within our
constraints are $11$ and $7$ respectively, whereas the average random
expectations are $1.6 \pm 2.2$ and $1 \pm 1$, with medians of $1$ and
$1$, respectively. These last two distributions are very skewed towards
zero.

Although the number of clusters around Q1410 exceeds that from the
random expectation at only the $2-3\sigma$ confidence level (c.l.),
the excesses of total and independent cluster-pairs are highly
significant ($> 5\sigma$), making Q1410 a very exceptional sightline.
We take this fact into account when comparing the incidences of
absorption line systems close to cluster-pairs and the field as
estimated from the Q1410 sightline itself (see
\Cref{sec:dndz_field}).

\section{Characterization of absorption lines in the Q1410 spectrum}\label{sec:abslines}
We performed a {\it full} characterization of absorption lines in the
\ac{hst}/\ac{cos} \ac{fuv} spectrum of Q1410. This approach is more
time consuming than just restricting ourselves to spectral regions
associated with the redshifts where known structures exist
(e.g. clusters, cluster-pairs; see \Cref{sec:field}), but is necessary
to avoid potential biases and systematic effects. In particular, our
approach allows us to: (i) identify absorption lines independently of
the presence of known structures; (ii) quantify how the rest of the
redshift path unassociated with these known structures compares to the
field expectation in terms of absorption features (see
\Cref{sec:dndz_field}); and (iii) assess the extent of contamination by
blended unassociated lines in a given redshift. This last point is
crucial to minimize misidentification of lines, but some ambiguous
cases are unavoidable. In this section we present our methodology for
the identification and characterization of absorption lines in the
Q1410 spectrum, and how we handled ambiguity.

\subsection{Absorption line identification}\label{sec:identification}
We searched for individual absorption line components\footnote{In this
  paper a `component' is defined as an individual absorption line in a
  given ion; and a `system' is defined as the union of components lying
  within a given velocity window (usually arbitrary) from a given
  redshift.}  in the continuum normalized \ac{qso} spectrum manually
(i.e. eyeballing), based on an iterative algorithm described as
follows:

\begin{table*}
\centering
\scriptsize
\begin{minipage}{0.77\textwidth}
\centering
\caption{Absorption lines in the Q1410 sightline.}\label{tab:abslines}
\begin{tabular}{@{}clccccccc@{}}

%\hline                            
%Component ID  & Ion & $z$ & $\log (N/$cm$^{-2})$& $b$            & $W_r$  & $\langle S/N \rangle_{\rm res}$ & Label & $\log(\langle S/N \rangle_{\rm res} \times N/b)$& $\Delta v|_{<{}\ {\rm Mpc}}$& $\Delta d|_{<{}\ {\rm km/s}}$ &     Comments  \\ 

%    &     &     &                        & (km s$^{-1}$)& (\AA)  &                                        &       &                         &   (\kms)                &    (\mpc)                &    \\ 

%(1) &(2)  & (3) & (4)                    & (5)            & (6)    &  (7)                                   &  (8)  &        (9)              &       (10)               &       (11)                &  (12)\\ 

%\hline 

\hline                            
Component ID  & Ion& Obs. wavelength & $z$ & $\log (N/$cm$^{-2})$& $b$            & $W_r$  & $\langle S/N \rangle_{\rm res}$   & Label \\ 
 %      &  $\Delta v|_{<3\ {\rm Mpc}}$& $\Delta d|_{<1000\ {\rm km/s}}$ \\ 

    &     &    (\AA)       &      &         & (km s$^{-1}$)          & (\AA)  &                                        &                    \\ 
 %      &    (\kms)                         &    (\mpc)                \\ 

(1) &(2)  & (3) & (4)      & (5)                      & (6)    &  (7)                                   &  (8)  &        (9)                   \\ 
 %      &       (10)                         & (11)                      \\ 

\hline 

1&\civ&1547.8&-0.00024 $\pm$ 0.00013&13.98 $\pm$ 0.82&18 $\pm$ 35&0.205 $\pm$ 0.383&14 $\pm$ 1&a \\ 
2&\siiv&1393.4&-0.00024 $\pm$ 0.00003&13.38 $\pm$ 0.82&15 $\pm$ 22&0.134 $\pm$ 0.231&12 $\pm$ 1&a \\ 
3&\alii&1670.5&-0.00018 $\pm$ 0.00100&14.13 $\pm$ 40.72&11 $\pm$ 131&0.268 $\pm$ 2.683&11 $\pm$ 1&a \\ 
4&\nv&1238.6&-0.00018 $\pm$ 0.00011&13.55 $\pm$ 0.53&23 $\pm$ 55&0.064 $\pm$ 0.105&15 $\pm$ 1&a \\ 
5&\siiv&1393.6&-0.00014 $\pm$ 0.00008&13.73 $\pm$ 0.38&35 $\pm$ 14&0.304 $\pm$ 0.203&12 $\pm$ 1&a \\ 
6&\siiii&1206.4&-0.00012 $\pm$ 0.00003&16.24 $\pm$ 1.53&16 $\pm$ 10&0.534 $\pm$ 1.077&14 $\pm$ 1&a \\ 
7&\civ&1548.0&-0.00011 $\pm$ 0.00011&14.09 $\pm$ 0.66&18 $\pm$ 29&0.227 $\pm$ 0.340&15 $\pm$ 1&a \\ 
8&\siii&1260.3&-0.00010 $\pm$ 0.00001&16.58 $\pm$ 0.08&17 $\pm$ 1&0.700 $\pm$ 0.057&16 $\pm$ 1&a \\ 
9&\cii&1334.4&-0.00010 $\pm$ 0.00003&18.32 $\pm$ 1.65&23 $\pm$ 9&0.824 $\pm$ 1.500&11 $\pm$ 1&a \\ 
10&\alii&1670.7&-0.00006 $\pm$ 0.00107&14.07 $\pm$ 66.61&9 $\pm$ 204&0.222 $\pm$ 2.217&11 $\pm$ 1&a \\ 
11&\ciid&1335.6&-0.00006 $\pm$ 0.00007&14.11 $\pm$ 0.38&24 $\pm$ 32&0.164 $\pm$ 0.188&12 $\pm$ 1&a \\ 
12&\siiii&1206.7&0.00019 $\pm$ 0.00004&13.28 $\pm$ 3.99&11 $\pm$ 41&0.121 $\pm$ 0.894&14 $\pm$ 1&a \\ 
13&\siiii&1207.2&0.00056 $\pm$ 0.00016&12.37 $\pm$ 0.92&17 $\pm$ 88&0.043 $\pm$ 0.169&13 $\pm$ 1&a \\ 
14&\hi&1222.7&0.00579 $\pm$ 0.00003&12.86 $\pm$ 0.17&19 $\pm$ 17&0.036 $\pm$ 0.019&13 $\pm$ 1&c \\ 
15&\hi&1224.7&0.00744 $\pm$ 0.00002&12.90 $\pm$ 0.12&12 $\pm$ 11&0.037 $\pm$ 0.020&14 $\pm$ 1&c \\ 
16&\hi&1225.0&0.00771 $\pm$ 0.00002&12.92 $\pm$ 0.30&7 $\pm$ 14&0.034 $\pm$ 0.032&14 $\pm$ 1&c \\ 
17&\hi&1250.5&0.02865 $\pm$ 0.00001&13.49 $\pm$ 0.06&18 $\pm$ 5&0.115 $\pm$ 0.022&16 $\pm$ 1&b \\ 
18&\hi&1250.9&0.02894 $\pm$ 0.00001&13.57 $\pm$ 0.05&32 $\pm$ 7&0.155 $\pm$ 0.023&16 $\pm$ 1&b \\ 
19&\hi&1259.4&0.03594 $\pm$ 0.00001&13.74 $\pm$ 0.05&27 $\pm$ 4&0.197 $\pm$ 0.025&16 $\pm$ 1&b \\ 
20&\hi&1268.9&0.04375 $\pm$ 0.00001&13.49 $\pm$ 9.56&4 $\pm$ 26&0.044 $\pm$ 0.441&15 $\pm$ 1&c \\ 
21&\hi&1269.4&0.04418 $\pm$ 0.00004&12.93 $\pm$ 0.16&30 $\pm$ 19&0.043 $\pm$ 0.017&15 $\pm$ 1&c \\ 
22&\hi&1292.0&0.06277 $\pm$ 0.00001&13.32 $\pm$ 0.08&21 $\pm$ 7&0.089 $\pm$ 0.019&14 $\pm$ 1&b \\ 
23&\hi&1295.9&0.06599 $\pm$ 0.00004&12.86 $\pm$ 0.17&21 $\pm$ 16&0.036 $\pm$ 0.017&14 $\pm$ 1&c \\ 
24&\hi&1298.3&0.06800 $\pm$ 0.00077&14.32 $\pm$ 6.43&25 $\pm$ 76&0.300 $\pm$ 3.003&14 $\pm$ 1&c \\ 
25&\hi&1324.8&0.08981 $\pm$ 0.00002&13.29 $\pm$ 0.10&22 $\pm$ 10&0.087 $\pm$ 0.025&10 $\pm$ 1&b \\ 
26&\hi&1372.7&0.12920 $\pm$ 0.00002&13.55 $\pm$ 0.06&44 $\pm$ 8&0.159 $\pm$ 0.023&12 $\pm$ 1&b \\ 
27&\hi&1410.9&0.16058 $\pm$ 0.00005&13.40 $\pm$ 0.11&59 $\pm$ 22&0.123 $\pm$ 0.033&14 $\pm$ 1&b \\ 
28&\hi&1416.4&0.16515 $\pm$ 0.00006&12.80 $\pm$ 0.28&23 $\pm$ 26&0.032 $\pm$ 0.026&14 $\pm$ 1&c \\ 
29&\hi&1419.3&0.16748 $\pm$ 0.00005&12.76 $\pm$ 0.17&30 $\pm$ 19&0.030 $\pm$ 0.013&16 $\pm$ 2&c \\ 
30&\hi&1424.7&0.17192 $\pm$ 0.00004&12.88 $\pm$ 0.13&34 $\pm$ 16&0.039 $\pm$ 0.013&19 $\pm$ 2&b \\ 
31&\hi&1429.4&0.17580 $\pm$ 0.00004&12.86 $\pm$ 0.15&31 $\pm$ 17&0.037 $\pm$ 0.014&22 $\pm$ 2&c \\ 
32&\hi&1433.2&0.17894 $\pm$ 0.00002&13.07 $\pm$ 0.09&23 $\pm$ 8&0.057 $\pm$ 0.013&21 $\pm$ 4&b \\ 
33&\hi&1441.1&0.18543 $\pm$ 0.00002&13.00 $\pm$ 0.07&22 $\pm$ 6&0.048 $\pm$ 0.009&24 $\pm$ 1&b \\ 
34&\hi&1444.1&0.18794 $\pm$ 0.00007&13.41 $\pm$ 0.08&112 $\pm$ 28&0.133 $\pm$ 0.026&23 $\pm$ 2&b \\ 
35&\hi&1449.9&0.19267 $\pm$ 0.00002&12.91 $\pm$ 0.09&24 $\pm$ 9&0.041 $\pm$ 0.009&23 $\pm$ 2&b \\ 
36&\cii&1613.5&0.20904 $\pm$ 0.00001&14.53 $\pm$ 1.73&6 $\pm$ 6&0.096 $\pm$ 0.135&8 $\pm$ 2&a \\ 
37&\siii&1523.9&0.20905 $\pm$ 0.00002&12.54 $\pm$ 0.10&11 $\pm$ 8&0.046 $\pm$ 0.016&16 $\pm$ 1&a \\ 
38&\hi&1469.8&0.20908 $\pm$ 0.00001&16.15 $\pm$ 1.03&11 $\pm$ 3&0.257 $\pm$ 0.155&18 $\pm$ 1&a \\ 
39&\siiii&1458.8&0.20909 $\pm$ 0.00001&12.63 $\pm$ 0.06&13 $\pm$ 4&0.066 $\pm$ 0.011&20 $\pm$ 1&a \\ 
40&\ciii&1181.4&0.20915 $\pm$ 0.00001&14.60 $\pm$ 13.57&8 $\pm$ 35&0.113 $\pm$ 1.134&12 $\pm$ 1&a \\ 
41&\niii&1196.8&0.20916 $\pm$ 0.00002&13.84 $\pm$ 0.10&19 $\pm$ 10&0.060 $\pm$ 0.017&13 $\pm$ 1&a \\ 
42&\hi&1470.0&0.20920 $\pm$ 0.00030&13.37 $\pm$ 2.72&14 $\pm$ 47&0.088 $\pm$ 0.571&18 $\pm$ 1&a \\ 
43&\siii&1524.5&0.20955 $\pm$ 0.00001&13.05 $\pm$ 0.15&13 $\pm$ 5&0.109 $\pm$ 0.043&15 $\pm$ 2&a \\ 
44&\cii&1614.2&0.20955 $\pm$ 0.00001&14.96 $\pm$ 1.42&10 $\pm$ 7&0.174 $\pm$ 0.184&8 $\pm$ 2&a \\ 
45&\siiv&1685.8&0.20956 $\pm$ 0.00001&13.67 $\pm$ 0.08&14 $\pm$ 3&0.170 $\pm$ 0.033&10 $\pm$ 1&a \\ 
46&\hi&1470.5&0.20958 $\pm$ 0.00015&16.29 $\pm$ 3.61&14 $\pm$ 7&0.330 $\pm$ 3.272&18 $\pm$ 1&a \\ 
47&\niii&1197.3&0.20962 $\pm$ 0.00001&14.94 $\pm$ 1.48&14 $\pm$ 12&0.151 $\pm$ 0.192&13 $\pm$ 1&a \\ 
48&\siiii&1459.4&0.20962 $\pm$ 0.00001&13.56 $\pm$ 0.07&23 $\pm$ 3&0.256 $\pm$ 0.032&20 $\pm$ 1&a \\ 
49&\ciii&1181.8&0.20964 $\pm$ 0.00236&15.03 $\pm$ 59.37&13 $\pm$ 98&0.188 $\pm$ 1.884&12 $\pm$ 1&a \\ 
50&\nv&1498.5&0.20964 $\pm$ 0.00003&13.57 $\pm$ 0.09&31 $\pm$ 10&0.071 $\pm$ 0.015&17 $\pm$ 1&a \\ 
51&\ovi&1248.3&0.20967 $\pm$ 0.00005&14.31 $\pm$ 0.21&26 $\pm$ 10&0.164 $\pm$ 0.073&16 $\pm$ 1&a \\ 
52&\siii&1524.8&0.20973 $\pm$ 0.00005&13.06 $\pm$ 0.16&34 $\pm$ 13&0.150 $\pm$ 0.058&15 $\pm$ 2&a \\ 
53&\cii&1614.5&0.20976 $\pm$ 0.00001&14.48 $\pm$ 0.04&20 $\pm$ 3&0.233 $\pm$ 0.029&8 $\pm$ 2&a \\ 
54&\hi&1470.7&0.20980 $\pm$ 0.00024&15.93 $\pm$ 2.00&27 $\pm$ 15&0.534 $\pm$ 0.567&18 $\pm$ 1&a \\ 
55&\siiv&1686.2&0.20981 $\pm$ 0.00002&13.27 $\pm$ 0.07&26 $\pm$ 7&0.128 $\pm$ 0.025&10 $\pm$ 1&a \\ 
56&\siiii&1459.7&0.20986 $\pm$ 0.00001&13.17 $\pm$ 0.05&21 $\pm$ 4&0.178 $\pm$ 0.025&20 $\pm$ 1&a \\ 
57&\ciii&1182.1&0.20986 $\pm$ 0.00021&17.15 $\pm$ 0.78&14 $\pm$ 18&0.488 $\pm$ 0.515&12 $\pm$ 1&a \\ 
58&\ovi&1248.5&0.20987 $\pm$ 0.00006&14.16 $\pm$ 0.31&26 $\pm$ 16&0.129 $\pm$ 0.095&16 $\pm$ 1&a \\ 
59&\niii&1197.5&0.20987 $\pm$ 0.00003&14.09 $\pm$ 0.11&25 $\pm$ 12&0.100 $\pm$ 0.031&13 $\pm$ 1&a \\ 
60&\hi&1471.3&0.21028 $\pm$ 0.00001&13.70 $\pm$ 0.05&19 $\pm$ 3&0.157 $\pm$ 0.020&18 $\pm$ 1&a \\ 
61&\ovi&1248.9&0.21028 $\pm$ 0.00001&14.50 $\pm$ 0.03&50 $\pm$ 6&0.270 $\pm$ 0.022&16 $\pm$ 1&a \\ 
62&\nv&1499.3&0.21029 $\pm$ 0.00006&13.53 $\pm$ 0.15&50 $\pm$ 24&0.068 $\pm$ 0.024&16 $\pm$ 1&a \\ 
63&\ciii&1182.5&0.21033 $\pm$ 0.00002&13.09 $\pm$ 0.18&11 $\pm$ 11&0.055 $\pm$ 0.040&12 $\pm$ 1&a \\ 
64&\hi&1478.4&0.21615 $\pm$ 0.00002&13.30 $\pm$ 0.05&42 $\pm$ 7&0.096 $\pm$ 0.012&17 $\pm$ 1&b \\ 
65&\hi&1481.9&0.21900 $\pm$ 0.00002&13.10 $\pm$ 0.07&36 $\pm$ 9&0.063 $\pm$ 0.011&17 $\pm$ 1&b \\ 
66&\hi&1486.8&0.22300 $\pm$ 0.00002&13.32 $\pm$ 0.05&52 $\pm$ 8&0.104 $\pm$ 0.012&17 $\pm$ 1&b \\ 

\hline
\end{tabular}
    
  \end{minipage}
\begin{minipage}{0.77\textwidth}
(1) Absorption component ID.
(2) Ion (see \Cref{sec:identification} for details on the line identification process).
(3) Observed wavelength of the strongest transition of the ion in the HST/COS spectrum.
(4) Redshift from the Voigt profile fitting (see \Cref{sec:voigt_fit}).
(5) Column density from the Voigt profile fitting (see \Cref{sec:voigt_fit}). 
(6) Doppler parameter from the Voigt profile fitting (see \Cref{sec:voigt_fit}).
(7) Inferred rest-frame equivalent width from fitted values (note that uncertainties are greatly overestimated for saturated lines or unconstrained fits; see \Cref{sec:ew}).
(8) Averaged local \ac{snr} (see \Cref{sec:snr}).
(9) Line reliability flag (`a' secure, `b' possible and `c' uncertain; see \Cref{sec:snr}). \\ 
%(10) Velocity difference to the closest cluster-pair within $\Delta d =3$\mpc.
%(11): Projected distance to the closest cluster-pair within $\Delta v=1000$\kms. \\ 
% $^{\dagger}$: Could be a very broad \hi~\lyb at redshift $z = 0.53531$
% instead, but we cannot confirm it with our current data.
\end{minipage}
\end{table*}

\begin{table*}
\centering
\scriptsize
\begin{minipage}{0.77\textwidth}
\centering
\contcaption{}
\begin{tabular}{@{}clccccccccc@{}}

%\hline                            
%Component ID  & Ion & $z$ & $\log (N/$cm$^{-2})$& $b$            & $W_r$  & $\langle S/N \rangle_{\rm res}$ & Label & $\log(\langle S/N \rangle_{\rm res} \times N/b)$& $\Delta v|_{<\ {\rm Mpc}}$& $\Delta d|_{<\ {\rm km/s}}$ &     Comments  \\ 

%    &     &     &                        & (km s$^{-1}$)& (\AA)  &                                        &       &                         &   (\kms)                &    (\mpc)                &    \\ 

%(1) &(2)  & (3) & (4)                    & (5)            & (6)    &  (7)                                   &  (8)  &        (9)              &       (10)               &       (11)                &  (12)\\ 

%\hline 
\hline                            
Component ID  & Ion& Obs. wavelength & $z$ & $\log (N/$cm$^{-2})$& $b$            & $W_r$  & $\langle S/N \rangle_{\rm res}$   & Label \\ 
 %      &  $\Delta v|_{<3\ {\rm Mpc}}$& $\Delta d|_{<1000\ {\rm km/s}}$ \\ 

    &     &    (\AA)       &      &         & (km s$^{-1}$)          & (\AA)  &                                        &                    \\ 
 %      &    (\kms)                         &    (\mpc)                \\ 

(1) &(2)  & (3) & (4)      & (5)                      & (6)    &  (7)                                   &  (8)  &        (9)                   \\ 
 %      &       (10)                         & (11)                      \\ 

\hline

67&\hi&1497.1&0.23150 $\pm$ 0.00001&13.45 $\pm$ 0.04&32 $\pm$ 4&0.125 $\pm$ 0.012&17 $\pm$ 1&b \\ 
68&\ovi&1274.5&0.23508 $\pm$ 0.00001&14.08 $\pm$ 0.05&24 $\pm$ 4&0.111 $\pm$ 0.014&15 $\pm$ 1&a \\ 
69&\hi&1501.5&0.23511 $\pm$ 0.00001&14.48 $\pm$ 0.04&32 $\pm$ 1&0.395 $\pm$ 0.021&17 $\pm$ 1&a \\ 
70&\hi&1509.0&0.24126 $\pm$ 0.00003&13.29 $\pm$ 0.06&54 $\pm$ 10&0.097 $\pm$ 0.014&16 $\pm$ 1&b \\ 
71&\hi&1538.8&0.26584 $\pm$ 0.00005&14.91 $\pm$ 0.41&28 $\pm$ 4&0.419 $\pm$ 0.111&15 $\pm$ 1&a \\ 
72&\ciii&1236.8&0.26587 $\pm$ 0.00001&13.70 $\pm$ 40.19&3 $\pm$ 66&0.041 $\pm$ 0.406&15 $\pm$ 1&a \\ 
73&\hi&1539.0&0.26593 $\pm$ 0.00001&15.88 $\pm$ 0.06&16 $\pm$ 2&0.328 $\pm$ 0.041&15 $\pm$ 1&a \\ 
74&\hi&1541.8&0.26827 $\pm$ 0.00010&12.78 $\pm$ 0.22&50 $\pm$ 37&0.032 $\pm$ 0.017&15 $\pm$ 1&c \\ 
75&\hi&1545.5&0.27132 $\pm$ 0.00004&12.98 $\pm$ 0.12&36 $\pm$ 16&0.049 $\pm$ 0.015&14 $\pm$ 2&b \\ 
76&\hi&1550.9&0.27574 $\pm$ 0.00004&13.87 $\pm$ 0.32&21 $\pm$ 17&0.200 $\pm$ 0.166&14 $\pm$ 1&c \\ 
77&\hi&1566.7&0.28874 $\pm$ 0.00007&13.06 $\pm$ 0.15&49 $\pm$ 25&0.059 $\pm$ 0.022&13 $\pm$ 1&c \\ 
78&\hi$^{\dagger}$&1573.8&0.29459 $\pm$ 0.00017&13.58 $\pm$ 0.13&157 $\pm$ 57&0.193 $\pm$ 0.058&14 $\pm$ 1&b \\ 
79&\hi&1576.7&0.29696 $\pm$ 0.00010&12.75 $\pm$ 0.23&45 $\pm$ 37&0.030 $\pm$ 0.017&13 $\pm$ 1&c \\ 
80&\hi&1578.8&0.29874 $\pm$ 0.00001&13.68 $\pm$ 0.03&36 $\pm$ 4&0.191 $\pm$ 0.015&13 $\pm$ 1&b \\ 
81&\hi&1582.7&0.30188 $\pm$ 0.00002&13.25 $\pm$ 0.07&32 $\pm$ 8&0.084 $\pm$ 0.014&13 $\pm$ 1&b \\ 
82&\hi&1584.2&0.30314 $\pm$ 0.00003&13.32 $\pm$ 0.07&47 $\pm$ 11&0.102 $\pm$ 0.017&13 $\pm$ 1&b \\ 
83&\hi&1613.9&0.32756 $\pm$ 0.00003&13.24 $\pm$ 0.11&24 $\pm$ 11&0.079 $\pm$ 0.024&8 $\pm$ 2&b \\ 
84&\hi&1617.4&0.33044 $\pm$ 0.00004&13.38 $\pm$ 0.09&45 $\pm$ 13&0.113 $\pm$ 0.024&10 $\pm$ 1&b \\ 
85&\hi&1631.6&0.34217 $\pm$ 0.00006&13.75 $\pm$ 0.05&153 $\pm$ 19&0.276 $\pm$ 0.030&10 $\pm$ 1&b \\ 
86&\ovi&1392.6&0.34954 $\pm$ 0.00002&13.77 $\pm$ 0.09&21 $\pm$ 8&0.061 $\pm$ 0.015&11 $\pm$ 1&a \\ 
87&\siiii&1628.6&0.34986 $\pm$ 0.00001&13.50 $\pm$ 0.03&45 $\pm$ 4&0.378 $\pm$ 0.032&10 $\pm$ 1&a \\ 
88&\hi&1641.0&0.34986 $\pm$ 0.00001&15.88 $\pm$ 0.03&17 $\pm$ 1&0.344 $\pm$ 0.015&11 $\pm$ 1&a \\ 
89&\ovi&1393.0&0.34989 $\pm$ 0.00001&14.07 $\pm$ 0.06&19 $\pm$ 5&0.101 $\pm$ 0.018&11 $\pm$ 1&a \\ 
90&\cii&1801.5&0.34989 $\pm$ 0.00001&13.31 $\pm$ 0.36&5 $\pm$ 5&0.030 $\pm$ 0.028&11 $\pm$ 1&a \\ 
91&\niii&1336.2&0.34998 $\pm$ 0.00006&14.04 $\pm$ 0.11&50 $\pm$ 20&0.102 $\pm$ 0.028&12 $\pm$ 1&a \\ 
92&\hi&1641.4&0.35020 $\pm$ 0.00004&14.29 $\pm$ 0.09&97 $\pm$ 10&0.701 $\pm$ 0.116&11 $\pm$ 1&b \\ 
93&\ovi&1393.4&0.35029 $\pm$ 0.00001&14.48 $\pm$ 0.05&35 $\pm$ 5&0.233 $\pm$ 0.029&12 $\pm$ 1&a \\ 
94&\hi&1641.6&0.35035 $\pm$ 0.00001&14.57 $\pm$ 0.06&26 $\pm$ 3&0.346 $\pm$ 0.047&11 $\pm$ 1&a \\ 
95&\hi&1642.4&0.35106 $\pm$ 0.00001&15.43 $\pm$ 0.03&25 $\pm$ 1&0.440 $\pm$ 0.015&11 $\pm$ 1&a \\ 
96&\hi&1652.3&0.35918 $\pm$ 0.00002&13.61 $\pm$ 0.05&31 $\pm$ 5&0.163 $\pm$ 0.022&11 $\pm$ 1&b \\ 
97&\hi&1658.1&0.36397 $\pm$ 0.00003&13.52 $\pm$ 0.06&58 $\pm$ 10&0.156 $\pm$ 0.021&11 $\pm$ 1&b \\ 
98&\hi&1664.1&0.36886 $\pm$ 0.00005&13.25 $\pm$ 0.11&50 $\pm$ 18&0.089 $\pm$ 0.023&11 $\pm$ 1&b \\ 
99&\hi&1666.8&0.37106 $\pm$ 0.00001&14.06 $\pm$ 0.03&36 $\pm$ 2&0.329 $\pm$ 0.022&11 $\pm$ 1&a \\ 
100&\hi&1667.6&0.37176 $\pm$ 0.00003&13.24 $\pm$ 0.08&29 $\pm$ 9&0.081 $\pm$ 0.016&11 $\pm$ 1&a \\ 
101&\hi&1673.8&0.37686 $\pm$ 0.00001&13.59 $\pm$ 0.04&31 $\pm$ 4&0.159 $\pm$ 0.018&11 $\pm$ 1&a \\ 
102&\ovi&1426.5&0.38238 $\pm$ 0.00002&13.43 $\pm$ 0.14&7 $\pm$ 12&0.026 $\pm$ 0.016&21 $\pm$ 2&c \\ 
103&\hi&1680.6&0.38242 $\pm$ 0.00005&12.91 $\pm$ 0.16&27 $\pm$ 17&0.041 $\pm$ 0.017&10 $\pm$ 1&c \\ 
104&\hi&1683.8&0.38508 $\pm$ 0.00001&14.25 $\pm$ 0.03&32 $\pm$ 2&0.345 $\pm$ 0.019&11 $\pm$ 1&a \\ 
105&\hi&1691.0&0.39097 $\pm$ 0.00007&13.19 $\pm$ 0.14&53 $\pm$ 25&0.078 $\pm$ 0.027&10 $\pm$ 1&c \\ 
106&\hi&1704.6&0.40219 $\pm$ 0.00012&13.28 $\pm$ 0.20&79 $\pm$ 49&0.097 $\pm$ 0.048&10 $\pm$ 1&c \\ 
107&\hi&1704.7&0.40224 $\pm$ 0.00003&13.07 $\pm$ 0.30&15 $\pm$ 14&0.053 $\pm$ 0.046&10 $\pm$ 1&c \\ 
108&\hi&1716.3&0.41182 $\pm$ 0.00005&13.47 $\pm$ 0.09&62 $\pm$ 18&0.142 $\pm$ 0.031&9 $\pm$ 1&b \\ 
109&\hi&1723.3&0.41758 $\pm$ 0.00001&14.38 $\pm$ 0.03&19 $\pm$ 1&0.254 $\pm$ 0.014&9 $\pm$ 1&a \\ 
110&\hi&1726.5&0.42022 $\pm$ 0.00006&13.25 $\pm$ 0.12&56 $\pm$ 20&0.090 $\pm$ 0.025&10 $\pm$ 1&b \\ 
111&\hi&1737.3&0.42911 $\pm$ 0.00008&13.08 $\pm$ 0.15&55 $\pm$ 27&0.062 $\pm$ 0.023&9 $\pm$ 1&c \\ 
112&\hi&1740.4&0.43167 $\pm$ 0.00005&13.73 $\pm$ 0.83&23 $\pm$ 14&0.179 $\pm$ 0.213&10 $\pm$ 1&a \\ 
113&\hi&1740.6&0.43183 $\pm$ 0.00083&13.21 $\pm$ 2.74&37 $\pm$ 102&0.079 $\pm$ 0.793&9 $\pm$ 1&c \\ 
114&\hi&1768.4&0.45466 $\pm$ 0.00006&13.46 $\pm$ 0.08&81 $\pm$ 18&0.143 $\pm$ 0.026&9 $\pm$ 1&b \\ 
115&\hi&1775.9&0.46085 $\pm$ 0.00035&13.55 $\pm$ 0.25&192 $\pm$ 141&0.185 $\pm$ 0.115&9 $\pm$ 1&c \\ 
116&\hi&1777.7&0.46235 $\pm$ 0.00002&13.77 $\pm$ 0.07&26 $\pm$ 5&0.199 $\pm$ 0.037&9 $\pm$ 1&a \\ 
117&\oii&1281.0&0.53509 $\pm$ 0.00001&13.85 $\pm$ 0.25&10 $\pm$ 10&0.041 $\pm$ 0.033&15 $\pm$ 1&a \\ 
118&\hi&1574.6&0.53510 $\pm$ 0.00001&16.52 $\pm$ 0.02&16 $\pm$ 1&0.274 $\pm$ 0.010&14 $\pm$ 1&a \\ 
119&\cii&2048.6&0.53510 $\pm$ 0.00001&13.30 $\pm$ 0.56&10 $\pm$ 10&0.034 $\pm$ 0.051&14 $\pm$ 1&a \\ 
120&\oiii&1078.2&0.53512 $\pm$ 0.00001&14.51 $\pm$ 0.25&13 $\pm$ 5&0.078 $\pm$ 0.036&14 $\pm$ 1&a \\ 
121&\ciii&1499.9&0.53520 $\pm$ 0.00001&13.98 $\pm$ 0.25&19 $\pm$ 4&0.177 $\pm$ 0.054&16 $\pm$ 1&a \\ 
122&\oiv&1209.3&0.53524 $\pm$ 0.00003&14.22 $\pm$ 0.11&23 $\pm$ 10&0.076 $\pm$ 0.023&13 $\pm$ 1&a \\ 
123&\hi&1574.8&0.53531 $\pm$ 0.00003&14.99 $\pm$ 0.15&17 $\pm$ 4&0.180 $\pm$ 0.050&14 $\pm$ 1&a \\ 
124&\hi&1717.1&0.67402 $\pm$ 0.00001&14.80 $\pm$ 0.04&31 $\pm$ 3&0.238 $\pm$ 0.024&9 $\pm$ 1&a \\ 
125&\ciii&1752.5&0.79372 $\pm$ 0.00003&14.84 $\pm$ 7.41&4 $\pm$ 9&0.065 $\pm$ 0.653&9 $\pm$ 1&a \\ 
126&\oiv&1412.9&0.79373 $\pm$ 0.00002&14.68 $\pm$ 0.09&20 $\pm$ 5&0.128 $\pm$ 0.032&14 $\pm$ 1&a \\ 
127&\oiii&1259.8&0.79376 $\pm$ 0.00001&14.41 $\pm$ 14.69&3 $\pm$ 22&0.023 $\pm$ 0.234&16 $\pm$ 1&a \\ 
128&\hi&1839.9&0.79378 $\pm$ 0.00005&15.55 $\pm$ 0.10&37 $\pm$ 6&0.426 $\pm$ 0.077&16 $\pm$ 1&a \\ 
129&\ciii&1752.7&0.79396 $\pm$ 0.00002&15.14 $\pm$ 8.93&10 $\pm$ 22&0.151 $\pm$ 1.514&9 $\pm$ 1&a \\ 
130&\oiv&1413.2&0.79401 $\pm$ 0.00002&15.51 $\pm$ 4.99&9 $\pm$ 15&0.097 $\pm$ 0.831&14 $\pm$ 1&a \\ 
131&\oiii&1260.0&0.79404 $\pm$ 0.00001&14.59 $\pm$ 0.20&12 $\pm$ 3&0.077 $\pm$ 0.024&16 $\pm$ 1&a \\ 
132&\hi&1840.2&0.79404 $\pm$ 0.00005&14.91 $\pm$ 0.42&18 $\pm$ 10&0.176 $\pm$ 0.115&16 $\pm$ 1&a \\ 

\hline
\end{tabular}
    
  \end{minipage}
\begin{minipage}{0.77\textwidth}
(1) Absorption component ID.
(2) Ion (see \Cref{sec:identification} for details on the line identification process).
(3) Observed wavelength of the strongest transition of the ion in the HST/COS spectrum.
(4) Redshift from the Voigt profile fitting (see \Cref{sec:voigt_fit}).
(5) Column density from the Voigt profile fitting (see \Cref{sec:voigt_fit}). 
(6) Doppler parameter from the Voigt profile fitting (see \Cref{sec:voigt_fit}).
(7) Inferred rest-frame equivalent width from fitted values (note that uncertainties are greatly overestimated for saturated lines or unconstrained fits; see \Cref{sec:ew}).
(8) Averaged local \ac{snr} (see \Cref{sec:snr}).
(9) Line reliability flag (`a' secure, `b' possible and `c' uncertain; see \Cref{sec:snr}). \\ 
%(10) Velocity difference to the closest cluster-pair within $\Delta d =3$\mpc.
%(11): Projected distance to the closest cluster-pair within $\Delta v=1000$\kms. \\ 
$^{\dagger}$: Could be a very broad \hi~\lyb at redshift $z = 0.53531$
instead, but we cannot confirm it with our current data.
\end{minipage}
\end{table*}

\begin{enumerate}
\item Identify all possible absorption components (\hi\ and metals)
  within $\pm 500$\kms from redshift $z=0$, and assign them to the
  `reliable' category (label~`a'; see \Cref{sec:reliability}).\sk

\item Identify all possible absorption components (\hi\ and metals)
  within $\pm 500$\kms from redshift $z=z_{\rm QSO}$, and assign them
  to the `reliable' category.\sk

\item Identify \hi\ absorption components, showing at least two
  transitions (e.g. \lya\ and \lyb\ or \lyb\ and \lyc, and so on; i.e.
  strong \hi)\footnote{Note that this condition allow us to identify
    strong \hi\ components at redshifts greater than $z>0.477$ by means
    of higher order Lyman series transitions.}, starting at $z=z_{\rm
    QSO}$ until $z=0$, and assign them to the `reliable' category. This
  identification includes the whole Lyman series covered by the
  spectrum in a given component.\sk

\item Identify all possible metal absorption components within $\pm
  200$\kms from each \hi\ redshift found in the previous step, and
  assign them to the `reliable' category. When the wavelength coverage
  allows the detection of multiple transitions of a single ion, we
  require the relative positions of these to coincide; in the case of
  multiple adjacent components blending with each other, we require
  them to have similar kinematic structure across the multiple
  transitions of the same ion.\sk
  
\item Identify high-ionization transitions (namely: \neviii, \ovi, \nv,
  \civ, \siiv) showing in at least two transitions, independently of
  the presence of \hi, starting at $z=z_{\rm QSO}$ until $z=0$, and
  assign them to the `reliable' category.\sk

\item Identify low-ionization transitions (namely: \cii, \ciii, \nii,
  \niii, \oi, \oii, \siii, \siiii, \feii, \feiii and \alii), showing at
  least two transitions, independently of the presence of \hi, starting
  at $z=z_{\rm QSO}$ until $z=0$, and assign them to the `reliable'
  category.\footnote{No low-ionization transition was found without
    having associated \hi, so this step was redundant.}\sk

\item Assume all the unidentified absorption features to be \hi~\lya
  and repeat step (iv). If metals satisfying the criteria in step (iv)
  exist, assign the component to the `reliable' category; otherwise
  assign the component to the `possible' category (label~`b'; see
  \Cref{sec:reliability}).\sk

\item For complex blended systems we allowed for the presence of extra
  heavily blended (hidden) ions, preferentially \hi~\lya unless a metal
  ion showing at least one unblended transition exist, and update the
  identification accordingly. In cases where the metal ion shows at
  least two unblended transitions, assign them to the `reliable'
  category. In the rest of the cases (including \hi~\lya only) assign
  them to the `possible' category.

\end{enumerate}

\noindent We note that we will then degrade some of the components in
the `possible' category to the `uncertain' category (`c'), based on an
equivalent width significance criterium in \Cref{sec:reliability}.

For all the identified components, we set initial guesses for their
redshifts, column densities ($N$), and Doppler parameters ($b$), which
are used as the inputs of our automatic Voigt profile fitting process
described in the \Cref{sec:voigt_fit}. We based these guesses on the
intensities and widths of the spectral features, keeping the number of
individual components to the minimum: we only added a component when
there is a clear presence of multiple adjacent local minima or
asymmetries. In the case of symmetric and intense \hi~\lya\ absorption
lines showing no corresponding \hi~\lyb\ absorption (when the spectral
coverage and the \ac{snr} would have allowed it), this last condition
would require the components to have relatively large $b$-values
(typically $\gtrsim 40-50$\kms). We warn the reader that this is a
potential source of bias, especially for the broad \lya\ systems ($\ge
50$\kms) expected to trace portions of the \ac{whim} (but see
\Cref{sec:caveat}).

\subsection{Voigt profile fitting}\label{sec:voigt_fit}

We fit Voigt profiles to the identified absorption line components
using
\vpfit\footnote{\url{http://www.ast.cam.ac.uk/~rfc/vpfit.html}}. We
accounted for the non-Gaussian \ac{cos} \ac{lsf}, by interpolating
between the closest \ac{cos} \ac{lsf} tabulated values provided by the
\ac{stsci}\footnote{\url{http://www.stsci.edu/hst/cos/performance/spectral_resolution}}
at a given wavelength. We used the guesses provided by the absorption
line search (see \Cref{sec:identification}) as the initial input of
\vpfit, and modified them when needed to reach converged solutions with
low reduced $\chi^2$.\footnote{Our final reduced $\chi^2$ have in
  average (and median) values of $1.1$. See also residuals in
  \Cref{fig:spectrum}.}

When dealing with \hi\ absorption lines, we used at least two spectral
regions associated with their Lyman series transitions when the
spectral coverage allowed it. This means that for those showing only
the \lya\ transition, we also included their associated \lyb\ regions
(even though they do not show evident absorption) when available. This
last condition provides reliable upper limits to the column density of
these components. For strong \hi\ components, we used regions
associated with as many Lyman series transitions as possible, but
excluding those heavily blended or in spectral regions of poor \ac{snr}
($\lesssim 1$ per pixel). For metal transitions we used all spectral
regions available.

We fitted absorption line systems starting from $z=z_{\rm QSO}$ until
$z=0$. When a given system at redshift $0 < z \le z_{\rm QSO}$ showed
strong blends from {\it lower} redshifts, we fitted them all
simultaneously in a given \vpfit iteration (i.e. including all spectral
regions associated with them). When a given system at redshift $0 < z <
z_{\rm QSO}$ showed weak blending from {\it higher} redshifts, we
allowed \vpfit to modify the `effective' continuum by adding the
previously found absorption line solutions to it. This last condition
accounts for the blending of weak lines (especially from higher order
Lyman series)---whose solutions are already well determined---in a more
efficient manner than fitting {\it all} regions involved
simultaneously. In the whole process, we allowed \vpfit to add lines
automatically when the $\chi^2$ and the Kolmogorov-Smirnov test (K-S)
test probabilities were below $0.01$ (see \vpfit manual for details).

\Cref{tab:abslines} shows our final list of identified absorption line
components, and their corresponding fits. Unique component IDs are
given in the first column to components for each ion (second
column). The observed wavelength associated with the strongest
transition of an ion is shown in the third column (but note that some
ions can show up in multiple wavelengths when having multiple
transitions). The fitted redshifts, column densities and Doppler
parameters are given in the fourth, fifth and sixth columns
respectively. Our final reduced $\chi^2$ have an average (and median)
of $1.1$. In \Cref{fig:spectrum} we show how these fits (red line)
compare to the observed spectrum (black line), by means of their
residuals (grey dots) defined as the difference between the two
(i.e. in the same units as the spectrum). We see how these residuals
are mostly distributed within the spectrum uncertainty level (green
line).

\subsection{Rest-frame equivalent widths}\label{sec:ew}

For each component we estimate the rest-frame equivalent width of the
strongest transition, $W_{\rm r}$, using the approximation given by
\citet[][see his equation 9.27]{Draine2011}, based on their fitted $N$
and $b$ values. The resulting values are given in the seventh column of
\Cref{tab:abslines}. We chose this approach in order to avoid
complications when dealing with blended components. We emphasize that
passing from $W_{\rm r} \rightarrow (N,b)$ is not always robust when on
of the flat part of the curve-of-growth, but passing from $(N,b)
\rightarrow W_{\rm r}$ is robust. We compared the results in $W_{\rm
  r}$ from our adopted approach and that from a direct pixel
integration, in a subsample of $10$ unblended and unsaturated lines,
and obtained consistent results within the uncertainties.

    \begin{figure}
    
    \begin{minipage}{0.45\textwidth}
    \centering
    \includegraphics[width=1\textwidth]{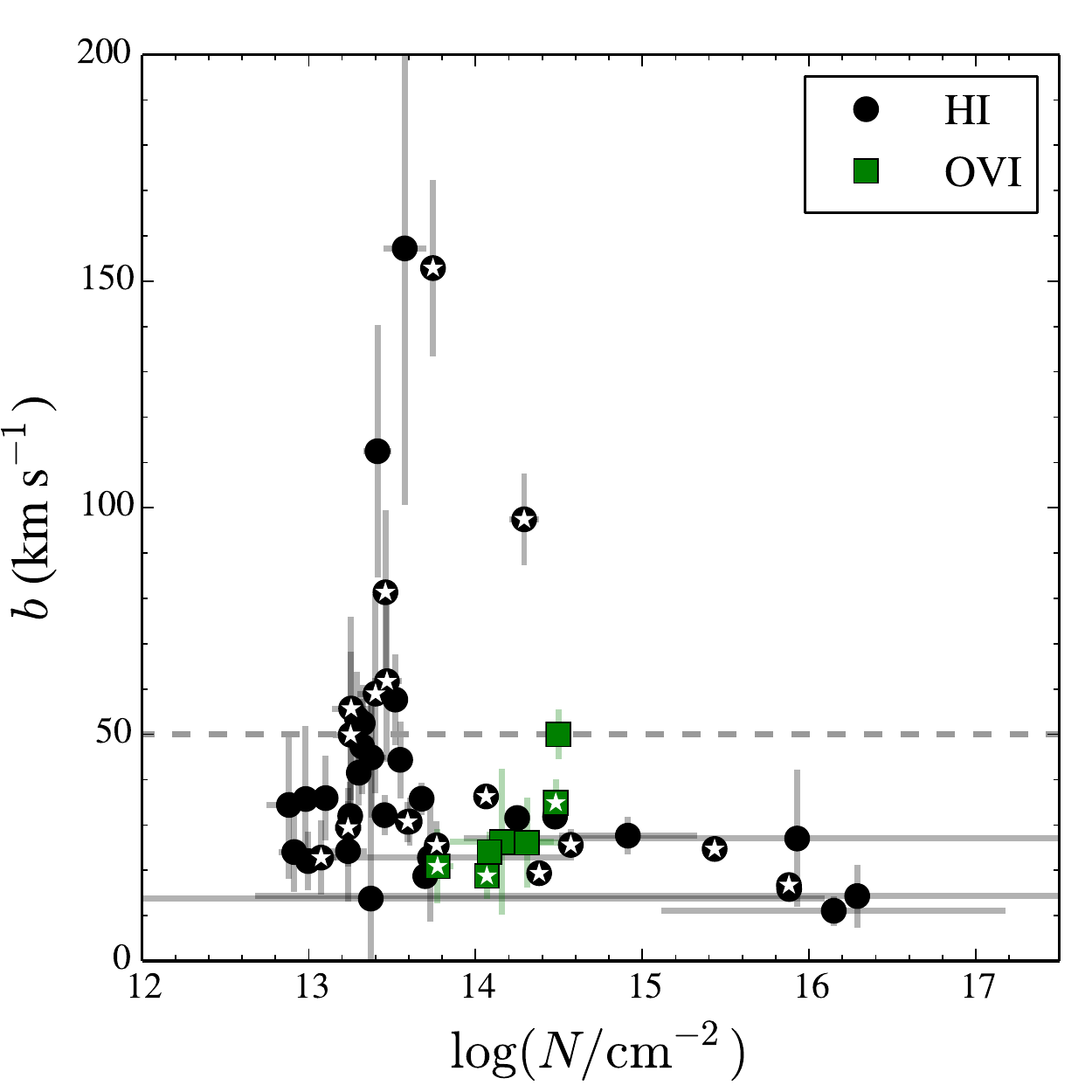}
    \end{minipage}

    \caption{Distribution of Doppler parameters ($b$) as a function
    of column density ($N$) for \hi (black circles) and \ovi (green
    squares) absorption lines found in the \ac{hst}/\ac{cos}
    \ac{fuv} spectrum between $0.1 \le z \le 0.5$ and excluding
    those in category `c' (uncertain; see
    \Cref{sec:reliability}). White stars mark absorbers associated
    to cluster-pairs (see \Cref{sec:dndz}). The horizontal dashed
    grey line corresponds to a Doppler parameter of $b=50$\kms, our
    adopted limit to split \hi lines into broad and narrow.}
    \label{fig:abslines}

\end{figure}

The rest-frame equivalent width uncertainty, $\delta W_{\rm r}$, was
estimated as follows. We first calculated the maximum/minimum
equivalent width, $W_{\rm r}^{\rm max/min}$, still consistent within
$1\sigma$ from the $N$ and $b$ fitted values, i.e. using the
aforementioned approximation for $(N \pm \delta N,b \pm \delta b)
\rightarrow W_{\rm r}^{\rm max/min}$, where $\delta N$ and $\delta b$
are the column density and Doppler parameter uncertainties,
respectively, as given by \vpfit. We then took $\delta W_{\rm r} \equiv
\frac{1}{2}(W_{\rm r}^{\rm max} - W_{\rm r}^{\rm min})$. In
catastrophic cases where the fits are unconstrained (i.e. $\delta
W_{\rm r} \gg W_{\rm r}$), we arbitrarily imposed $\delta W_{\rm r} =
10\times W_{\rm r}$, ensuring a very low significance level.

By using the actual fitted parameters and their corresponding errors,
our $\delta W_{\rm r}$ uncertainty estimation takes into account the
non-Gaussian shape of the \ac{cos} \ac{lsf} (particularly important for
broad absorption lines). For saturated lines, our method will give
unrealistically large $\delta W_{\rm r}$ uncertainties due to a poor
constraint in $N$, which is a conservative choice.
%We note that the real $\delta W_{\rm r}$ is asymmetric, skewed towards
%${\it higher} values. Therefore, our adopted approach is also
%conservative in the sense that we are considering such an
%uncertainty,$ even though what matters for detecting absorption lines
%is the uncertainty towards the {\it smaller} values.

\subsection{Local \acf{snr} estimation}\label{sec:snr}

For each component we estimated the average local spectral \acf{snr}
per pixel, $\langle S/N \rangle_{\rm pixel}$, over the $50$ closest
pixels around its strongest transition, without considering those with
flux values below $90\%$ of the continuum. We then estimated the local
\acf{snr} per resolution element $\langle S/N \rangle_{\rm res}$ as
$\sqrt{2}\langle S/N \rangle_{\rm pixel}$. The resulting values are
given in the eight column of \Cref{tab:abslines}.

\subsection{Absorption line reliability}\label{sec:reliability}

To deal with ambiguity and significance of the absorption lines we have
introduced a reliability flag scheme as follows:

\begin{itemize}
\item {\it Reliable (`a'):} Absorption line components showing at least
  two transitions or showing up in at least two ions, independently of
  the significance of its corresponding $W_{\rm r}$.\sk
\item {\it Probable (`b'):} Absorption line components showing only one
  transition, showing up in only one ion, and having $W_{\rm r}/\delta
  W_{\rm r} \ge 3$.\sk
\item {\it Uncertain (`c'):} Absorption line components showing only
  one transition, showing up in only one ion, and having $W_{\rm
    r}/\delta W_{\rm r} < 3$. Components in this category will be
  excluded from the main scientific analysis presented in this paper.
\end{itemize}

This reliability scheme applied to our absorption line list is shown in
the ninth column of \Cref{tab:abslines}. We also show these flags
together with the ion component ID given (first column of
\Cref{tab:abslines}) in \Cref{fig:spectrum} as vertical labels.
%Although a limit of $W_{\rm r}/\delta W_{\rm r} \ge 3$ to regard
%systems as probable might introduce a small fraction of unreal random
%contamination, such an issue should only dilute any potential signal
%when associating absorption lines to inter-cluster filaments.
 
    \begin{figure*}
    \begin{minipage}{1\textwidth}
    \centering
    \includegraphics[width=1\textwidth]{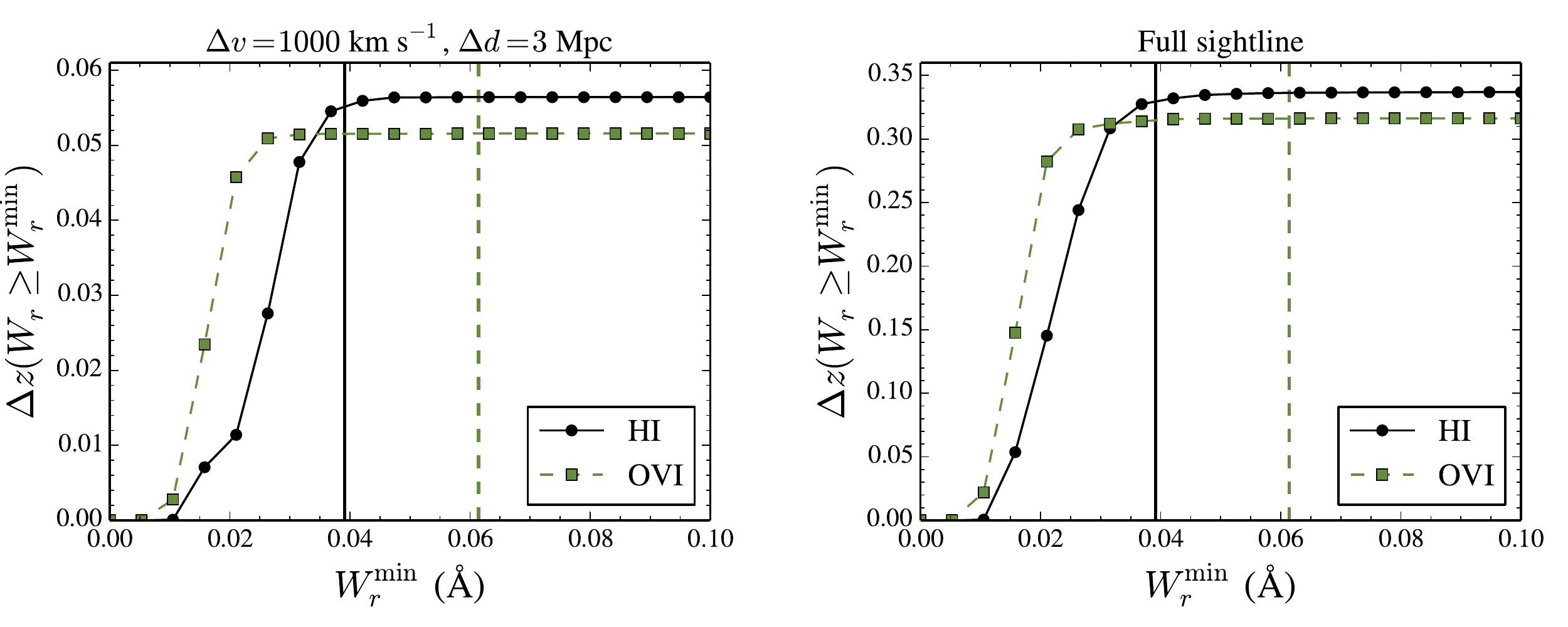}
    \end{minipage}
    
    \caption{Redshift path, $\Delta z$, as a function of minimum
    rest-frame equivalent width ,$W_r^{\rm min}$, for our survey of
    \hi\ (black solid line and circles) and \ovi\ (green dashed
    line and squares) absorption lines. The left panel shows the
    corresponding redshift path associated to regions of our Q1410
    HST/COS spectrum within rest-frame velocity differences $\Delta
    v=1000$\kms from cluster-pairs at impact parameters smaller than
    $\Delta d=3$\mpc. The right panel shows the total corresponding
    redshift path for the full Q1410 \ac{hst}/\ac{cos} spectrum
    between $0.1 \le z \le 0.5$. Vertical lines show the minimum
    rest-frame equivalent width for our {\it detected} \hi\ (solid
    black; $W_r=0.039$\AA) and \ovi\ (dashed green; $W_r=0.061$\AA)
    absorption line samples, excluding those in the category `c'
    (uncertain; see \Cref{sec:reliability}). See
    \Cref{sec:dndz_method} for further details.}  \label{fig:wr_dz}
    \end{figure*}

We note that previous studies on \ac{bla} lines have suggested one to
quantifying the completeness level of these broad lines by means of
$\langle S/N \rangle_{\rm res} \times N / b$ \citep{Richter2006}. The
motivation of this criterion is that the \ac{bla} is sensitive to both
\ac{snr} and the optical depth at the line centres, $\tau_0 \propto N /
b$. Therefore, it is not appropriate to use the commonly adopted
formalism based on a minimum equivalent width threshold for unresolved
lines (these broad lines are usually resolved).  In
\Cref{sec:significance} we compare the proposed approach by
\citet{Richter2006} to ours, and show that imposing a minimum $W_{\rm
  r}/\delta W_{\rm r}$ value (as defined here) is roughly consistent
with imposing a minimum $\langle S/N \rangle_{\rm res} \times N / b$
value for broad lines\footnote{We note that \citet{Danforth2010}
  reached a similar conclusion although using the commonly adopted
  formalism based on a minimum equivalent width threshold for
  unresolved lines.}, but is more conservative when applied to narrow
lines. Moreover, our approach has the advantage of being
straightforwardly applicable to any absorption line irrespective of its
Doppler parameter and ionic transition, hence more appropriate for an
homogeneus analysis.

\Cref{fig:abslines} shows the distribution Doppler parameters $b$ as a
function of column densities $N$, for our sample of \hi (black circles)
and \ovi (green squares) between $0.1 \le z \le 0.5$ and excluding
those in the `c' category (i.e. uncertain). White stars mark components
that lie within $\Delta v \le 1000$\kms and within impact parameters of
$\Delta d \le 3$\mpc from cluster-pairs, our fiducial values for
associating absorption lines with cluster-pairs (see \Cref{sec:dndz}).

\section{Redshift number density of absorption line systems 
around cluster-pairs}\label{sec:dndz}

After having characterized \ac{lss} traced by galaxy cluster-pairs
around the Q1410 sightline (\Cref{sec:field}) and intervening absorption
lines (\Cref{sec:abslines}), we can now provide a cross-match between
the two. Because the completeness level of the \ac{redmapper} clusters
with richness $<20$ ($<40$) at $z\sim 0.1-0.4$ ($z\sim 0.4-0.5$) is
lower than $\sim 50\%$ \citep[see top panel of fig. 22
  of][]{Rykoff2014}, in this paper we will only match absorption lines
close to the position of {\it known} cluster-pairs rather than the
other way around (or both). We also note that the purity of
\ac{redmapper} clusters is fairly constant with richness and redshift
(some trends are present though), but with values above $95\%$ in all
the cases \citep[see bottom panel of fig. 22 of][]{Rykoff2014}; still,
the presence of fake clusters will only dilute any real signal when
associating absorption lines to inter-cluster filaments traced by
cluster-pairs.

In this paper we use the redshift number density of absorption lines,
$dN/dz$, as a function of cluster-pair separation, as the relevant
statistical quantity to characterize inter-cluster filaments (if
any). We have chosen $dN/dz$ as opposed to the number of systems per
absorption distance, $dN/dX$, (or both), only for simplicity. Still, in
\Cref{sec:summary_tables} we provide tables with relevant quantities
and results for both $dN/dz$ and $dN/dX$. We note that our conclusions
are independent of this choice.

At this point, it is also important to emphasize that we do not know a
priori that cluster-pairs in our sample are tracing true inter-cluster
filaments, and that even if they do, we do not know if these could
produce a signal in the observed incidences of \hi\ and
\ovi\ absorption lines at the \ac{snr} level obtained in our Q1410
\ac{hst}/\ac{cos} spectrum. Although cosmological hydrodynamical
simulations suggest that this may be the case, this paper aims to
provide a direct test of such an hypothesis. Therefore, we will explore
the behaviour of $dN/dz$ over a wide range of scales around
cluster-pairs both {along} and {transverse} to the Q1410 \ac{los}. This
means that in the following, we will allow the maximum impact parameter
for clusters and cluster-pair inter-cluster axes to the Q1410 sightline
to be larger than the fiducial values adopted in \Cref{sec:clusters}
and \Cref{sec:pairs} (i.e. larger than $20$ and $3$\mpc, respectively).

\subsection{Measuring the redshift number density of absorption lines around cluster-pairs}\label{sec:dndz_method}

We measure the redshift number density of absorption lines around
cluster-pairs in the following way. Let $\Delta d$ be the maximum
impact parameter between a cluster-pair inter-cluster axis and the
Q1410 sightline in a given calculation. Let $\Delta v$ be the maximum
rest-frame velocity window to a given cluster-pair, at the redshift of
such cluster-pair (see \Cref{fig:diagram} for an schematic
diagram). Then, we define $N(\Delta d,\Delta v, W_{\rm r}^{\rm min})$
as the number of absorption components found within $\Delta v$ from the
closest (in rest-frame velocity space) cluster-pair, from those
cluster-pairs being within $\Delta d$ from the Q1410 sightline, having
rest-frame equivalent widths $W_{\rm r} \ge W_{\rm r}^{\rm
  min}$.\footnote{In our analysis we will also impose the line to be
  detected at $W_{\rm r}/\delta W_{\rm r} \ge 3$ (see
  \Cref{sec:results}), but this is not a requirement of the
  methodology.} Let $\Delta z(\Delta d,\Delta v, W_{\rm r}^{\rm min})$
be the redshift path in which a given absorption line having $W_{\rm r}
\ge W_{\rm r}^{\rm min}$ could have been detected along portions of the
spectrum being within $\Delta v$ to {\it any} cluster-pair, from those
cluster-pairs being within $\Delta d$ from the Q1410 sightline, and
being within our redshift range constraint (i.e. $0.1 \le z \le
0.5$).\footnote{This redshift range condition is not a requirement of
  the methodology in the most general case.} Then, the redshift number
density is calculated as,

\begin{equation}
\frac{dN}{dz}(\Delta d,\Delta v, W_{\rm r}^{\rm min}) = \frac{N(\Delta
  d,\Delta v, W_{\rm r}^{\rm min})}{\Delta z(\Delta d,\Delta v, W_{\rm r}^{\rm min})}
\ \rm{.}
\label{eq:dndz}
\end{equation}

%\noindent and we define the effective total redshift path as,

%\begin{equation} \Delta z(\Delta d,\Delta v) \equiv \frac{N(\Delta
%d,\Delta v)}{dN/dz(\Delta d,\Delta v)}
%\ \rm{.}  \label{eq:dz} \end{equation}

\noindent Our methodology for estimating $\Delta z(\Delta d,\Delta v,
W_{\rm r}^{\rm min})$ is presented in \Cref{sec:dz}.

%Our approach takes into account the slight differences in completeness
%level of $\Delta z_i$ for lines with different $W_{r,i}$. We note that
%in cases where $N(\Delta d,\Delta v)=0$, \Cref{eq:dndz} cannot be
%evaluated and hence $dN/dz$ remains undefined. In such cases, we
%define $dN/dz \equiv 0$, and the redshift path, $\Delta z$, to be the
%same as that of the minimum equivalent width absorption line in the
%given sample.

\Cref{fig:wr_dz} shows the redshift path, $\Delta z$, along the Q1410
sightline as a function of minimum rest-frame equivalent width
,$W_r^{\rm min}$, for our survey of \hi\ (black solid line and circles)
and \ovi\ (green dashed line and squares) absorption lines. The left
panel shows the corresponding redshift path associated with regions of
our Q1410 HST/COS spectrum within rest-frame velocity differences
$\Delta v=1000$\kms from cluster-pairs at impact parameters smaller
than $\Delta d=3$\mpc, while the right panel shows the total
corresponding redshift path for the full Q1410 \ac{hst}/\ac{cos}
spectrum between $0.1 \le z \le 0.5$. We see that the completeness
level is very similar between the portions of the spectrum close to
cluster-pairs and that of the full spectrum. We checked that this is
also the case for multiple choices of $\Delta v$ and $\Delta d$ values,
increasing from our fiducial values to cover the full spectrum (not
shown). The vertical lines in \Cref{fig:wr_dz} show the minimum
rest-frame equivalent width for our \hi\ (black solid; $W_{\rm r}^{\rm
  min}=0.039$\AA) and \ovi\ (dashed green; $W_{\rm r}^{\rm
  min}=0.061$\AA) absorption line samples, excluding those labelled as
`uncertain' (category `c'; see \Cref{sec:reliability}). We see that
these values correspond to high completeness levels and are therefore
adopted as the minimum equivalent widths in the forthcoming analysis
(but note that we could have detected \ovi down to $W_{\rm r} \sim
0.03$\AA\ with a similar completeness).

\subsection{Estimating the field redshift number density from the Q1410 sightline}\label{sec:dndz_field}

In this paper we have introduced slightly different ways to count and
assess the statistical significance of absorption lines compared with
has been done in previous published works. This is so because we opted
to do a uniform analysis for \hi (either total, broad or narrow) and
\ovi\ absorption lines, while previous works have usually focused on
one type at a time. Therefore, we estimate the field redshift number
density of a given species using our own methodology using the Q1410
sightline data alone. This is justified by the fact that cluster-pair
filaments (if any) should only influence specific portions of the
spectrum (in our case about $\sim 1/6$ of it), while the rest should
match the field expectation (i.e. that from a randomly selected
sightline).

As described in section \Cref{sec:how_unusual}, our sightline is
extremely unusual in terms of the number of cluster-pairs close to it
(by construction, see also \Cref{sec:selection}). Therefore, an
estimation of the field value from this sightline alone could be biased
when compared against a representative ensemble of random
sightlines. In order to correct for this potential bias we have
proceeded as follows. Let $N_{\rm tot}$ be the total number of relevant
absorption lines in the full Q1410 sightline between $0.1 \le z \le
0.5$, and $N_{\rm cpairs}$ be the number of such absorption lines
associated with our cluster-pairs assuming fiducial values of $\Delta v
= 1000$\kms and $\Delta d = 3$\mpc (see \Cref{sec:dndz_method} for a
definition of these two quantities). Therefore, our expected number of
absorption lines associated with the field value can be estimated by,

\begin{equation}
N_{\rm field} \approx N_{\rm tot} - N_{\rm cpairs}\left(1 -
\frac{n_{\rm cpair}^{\rm field}}{n_{\rm cpair}^{\rm Q1410}}\right)
\ \rm{,}
\label{eq:nfield}
\end{equation}

\begin{figure*}
  \begin{minipage}{1\textwidth}
    \centering
    \includegraphics[width=1\textwidth]{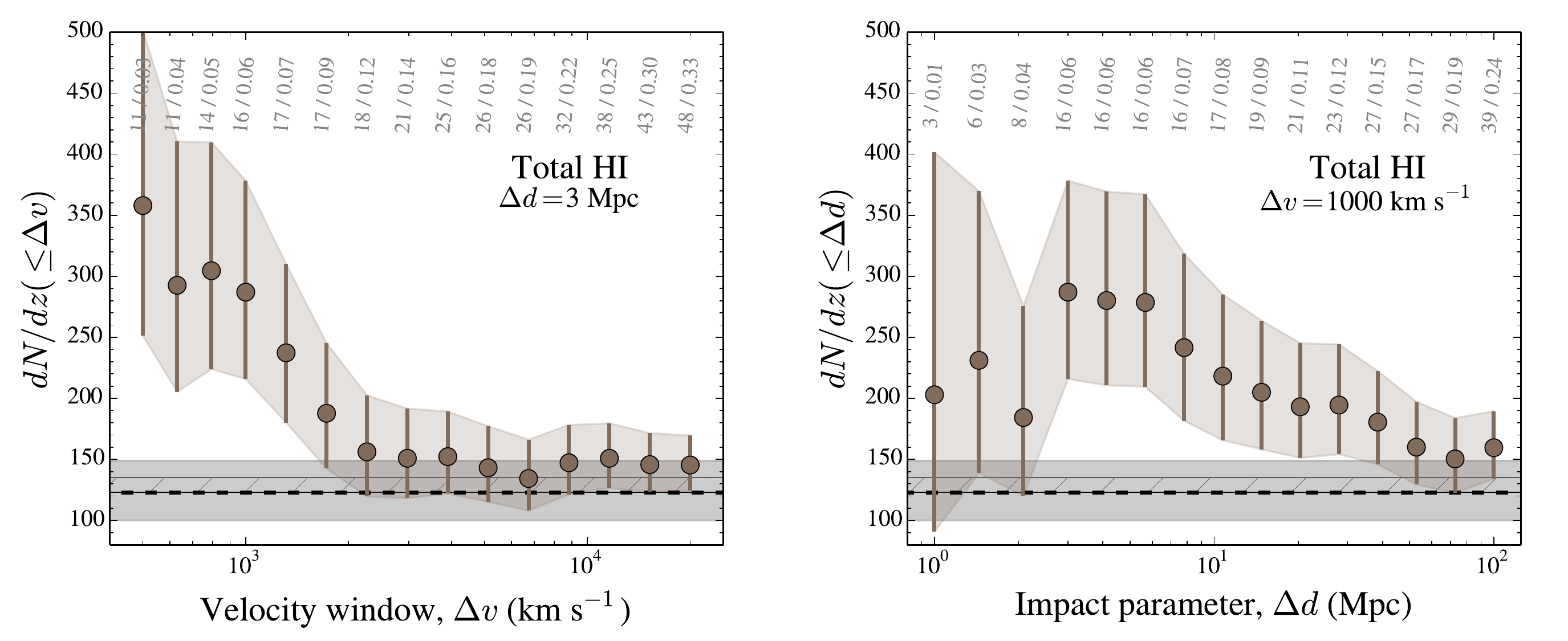}
    \vspace{1ex}
  \end{minipage}
  
  \begin{minipage}{1\textwidth}
    \centering
    \includegraphics[width=1\textwidth]{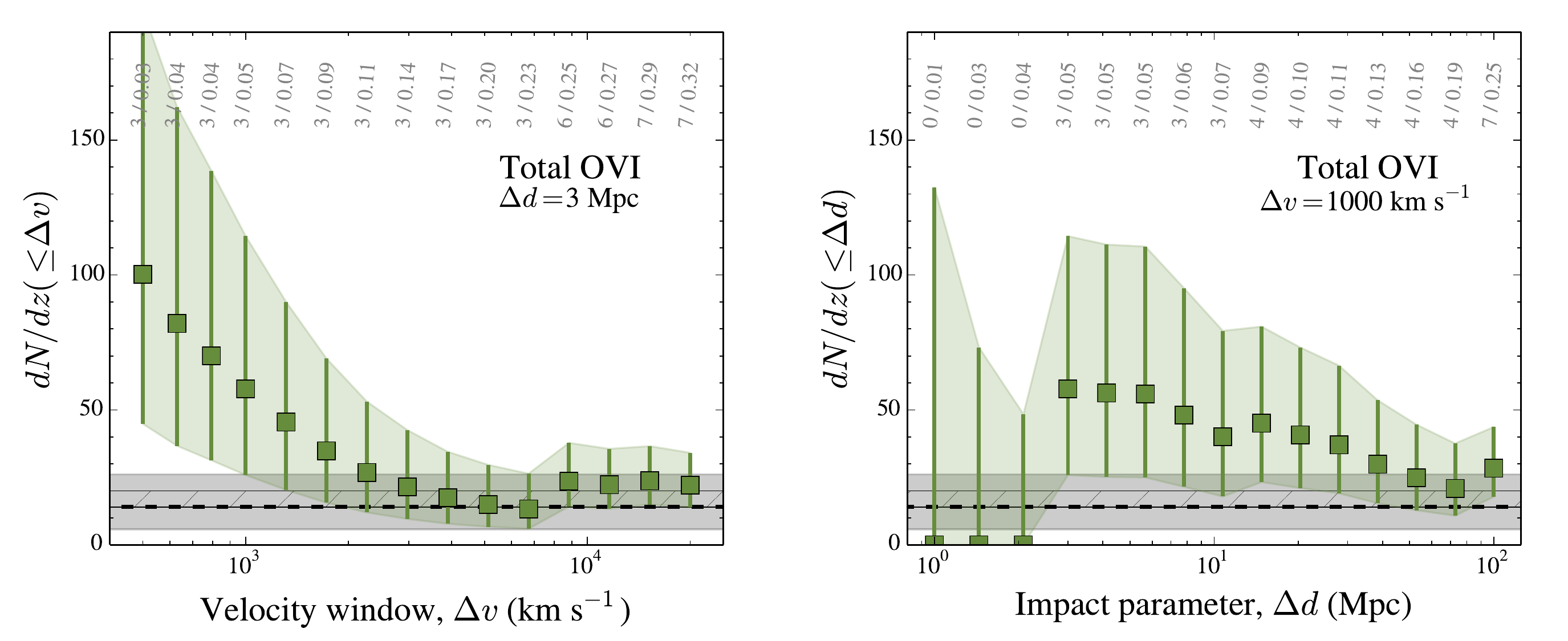}
    
  \end{minipage}
  
  \caption{Redshift number density of total \hi (brown circles; top
    panels) and \ovi (green squares; bottom panels) absorption
    components as a function of rest-frame velocity window ($\Delta v$;
    left panels) and maximum impact parameter to the closest
    cluster-pair axis ($\Delta d$; right panels) for a fixed $\Delta d
    = 3$\mpc and $\Delta v = 1000$\kms, respectively. Note that bins
    are not independent from each other, as emphasized by the coloured
    areas. The total number of lines and redshift paths per bin are
    given in grey numbers on top of the datapoints. The expected field
    value estimated from our Q1410 sightline is represented by the
    horizontal dashed line with its $\pm 1\sigma$ uncertainty
    represented by the darker grey region. The darkest grey hashed
    regions represents the $\pm 1\sigma$ field values from the
    \citet{Danforth2008} survey. See \Cref{sec:dndz_results} for
    further details.} \label{fig:dndz_hi_ovi}

\end{figure*}

\begin{figure*}
    \begin{minipage}{1\textwidth}
    \centering
    \includegraphics[width=1\textwidth]{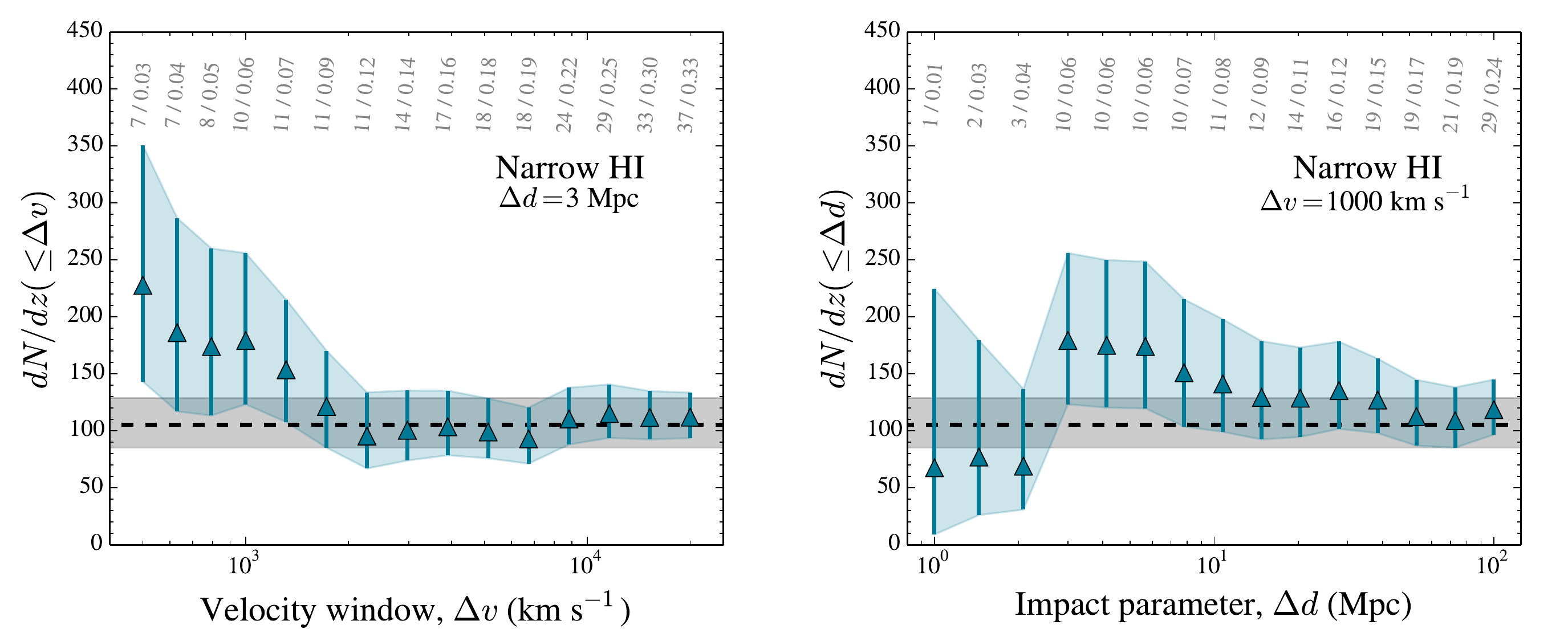}
    \vspace{1ex}
    \end{minipage}
    
    \begin{minipage}{1\textwidth}
    \centering
    \includegraphics[width=1\textwidth]{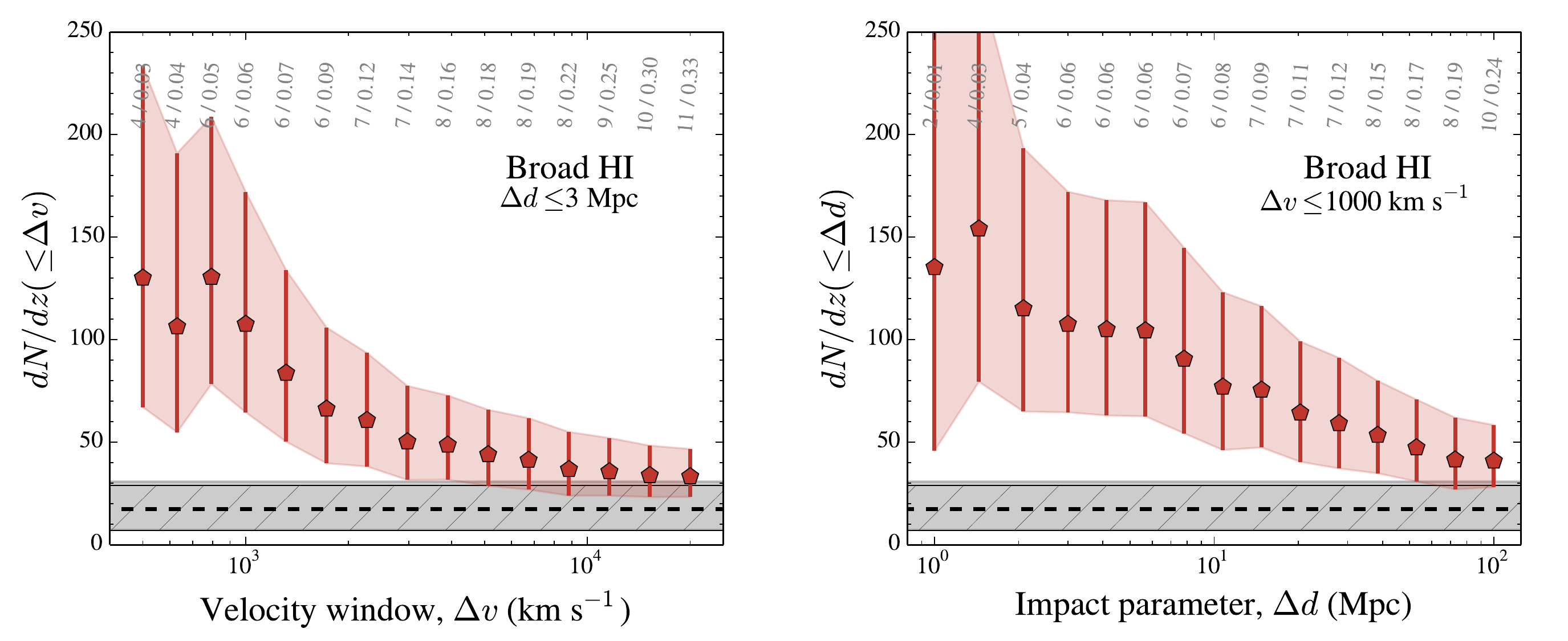}
  \end{minipage}

\caption{Same as \Cref{fig:dndz_hi_ovi} but for narrow ($b < 50$\kms;
  blue triangles; top panels) and broad ($b \ge 50$\kms; red pentagons;
  bottom panels). The darkest grey hashed regions in the bottom panels
  represents the $\pm 1\sigma$ field value from the
  \citet{Danforth2010} BLA survey. See \Cref{sec:dndz_results} for further
  details.} \label{fig:dndz_nla_bla}
\end{figure*}

\noindent where $n_{\rm cpair}^{\rm Q1410}$ and $n_{\rm cpair}^{\rm
  field}$ are the number of independent cluster-pairs found in Q1410
sightline and those randomly expected, respectively. In our case
$n_{\rm cpair}^{\rm Q1410} = 7$ and $n_{\rm cpair}^{\rm field} \approx
1 \pm 1$. Therefore, we take the expected field value to be $N_{\rm
  field} \approx N_{\rm tot} - (0.85 \pm 0.14) \ N_{\rm
  cpairs}$.\footnote{We note that a consistent correction factor of
  $N_{\rm field} = N_{\rm tot} - (0.85 \pm 0.2) \ N_{\rm cpairs}$ is
  found when considering the {\it total} number of cluster-pairs
  instead of independent ones, i.e. $n_{\rm cpair}^{\rm Q1410}=11$ and
  $n_{\rm cpair}^{\rm field}=1.6 \pm 2.2$.}  Finally, we estimate the
relevant redshift number density by using this corrected $N_{\rm
  field}$ as,

\begin{equation}
\frac{dN}{dz}|_{\rm field} = \frac{N_{\rm field}}{\Delta z} \ \rm{,}
\label{eq:dn_dz_field}
\end{equation}

\noindent where $\Delta z$ is the total redshift path associated with
the full Q1410 sightline between $0.1 \le z \le 0.5$. 

This methodology assumes (i) that there is an {\it excess} of
absorption lines in the data compared to the field expectation, and
(ii) that this excess is purely confined within $\Delta v = 1000$\kms
and $\Delta d = 3$\mpc from the {\it known} cluster-pairs. If
assumption (i) is incorrect, then our field expectation estimation will
be underestimated. If assumption (i) is correct, but assumption (ii) is
incorrect, then our field expectation estimation will be
overestimated. In \Cref{sec:results} we show that our field estimations
based on Q1410 alone matches those of comparable previously published
blind surveys, making our assumptions reasonable.

\subsection{Statistical uncertainty estimations}

The statistical uncertainty in our calculations is dominated by the
uncertainty in $N(\Delta d, \Delta v)$, which we assume is Poissonian and
estimate from the analytical approximation given by
\citet{Gehrels1986}: $\sigma^{+}_N \approx \sqrt{N+3/4} + 1$ and
$\sigma^{-}_N \approx \sqrt{N -1/4}$. The statistical uncertainty in
our estimation of $N_{\rm field}$ is taken from the contributions of
both the Poissonian uncertainty of $N_{\rm tot}$, and the statistical
uncertainty of $n_{\rm cpair}^{\rm field}$, which we propagate assuming
independence between these two quantities. Given that the statistical
uncertainties in $\Delta z(\Delta d, \Delta v)$ and $\Delta X(\Delta d,
\Delta v)$ are much smaller, we neglect them.

\section{Results}\label{sec:results}
In this section we report our results on $dN/dz$, for our different
samples of \hi (total, narrow and broad) and \ovi\ absorption lines
observed in the Q1410 sightline (see \Cref{sec:abslines}) applying the
methodology described in \Cref{sec:dndz} to associate them with
cluster-pairs (see \Cref{sec:field}). For simplicity, and in order to
reduce the `shot noise' of the measurements, the following results are
obtained by varying $\Delta v$ and $\Delta d$ for {\it fixed} values of
$\Delta d$ and $\Delta v$, respectively (as opposed to varying both
values at the same time).

A summary of all the results presented in this section (and those of
$dN/dX$, not described here), are given in Tables \ref{tab:hi_summary}
to \ref{tab:bla_summary}.

    \begin{figure*}
    \begin{minipage}{1\textwidth}
    \centering
    \includegraphics[width=1\textwidth]{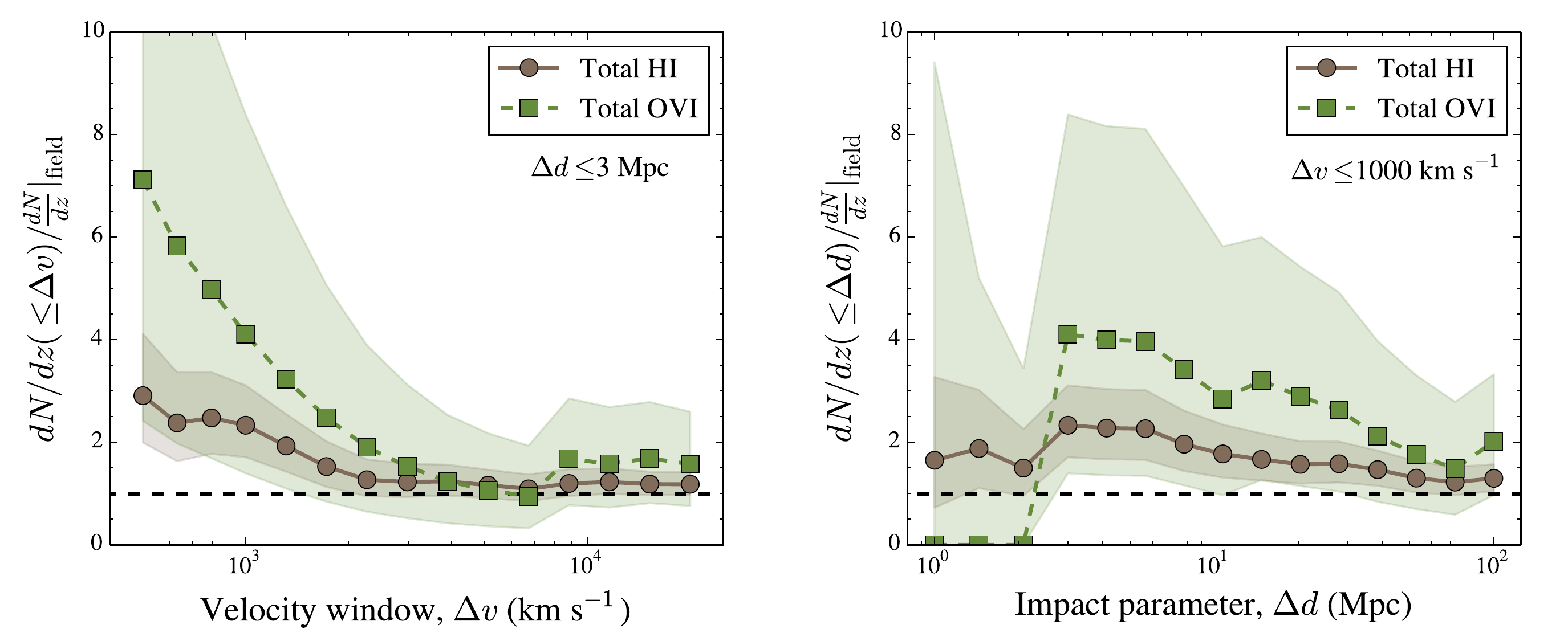}
    \vspace{2ex}
    \end{minipage}

    \begin{minipage}{1\textwidth}
    \centering
    \includegraphics[width=1\textwidth]{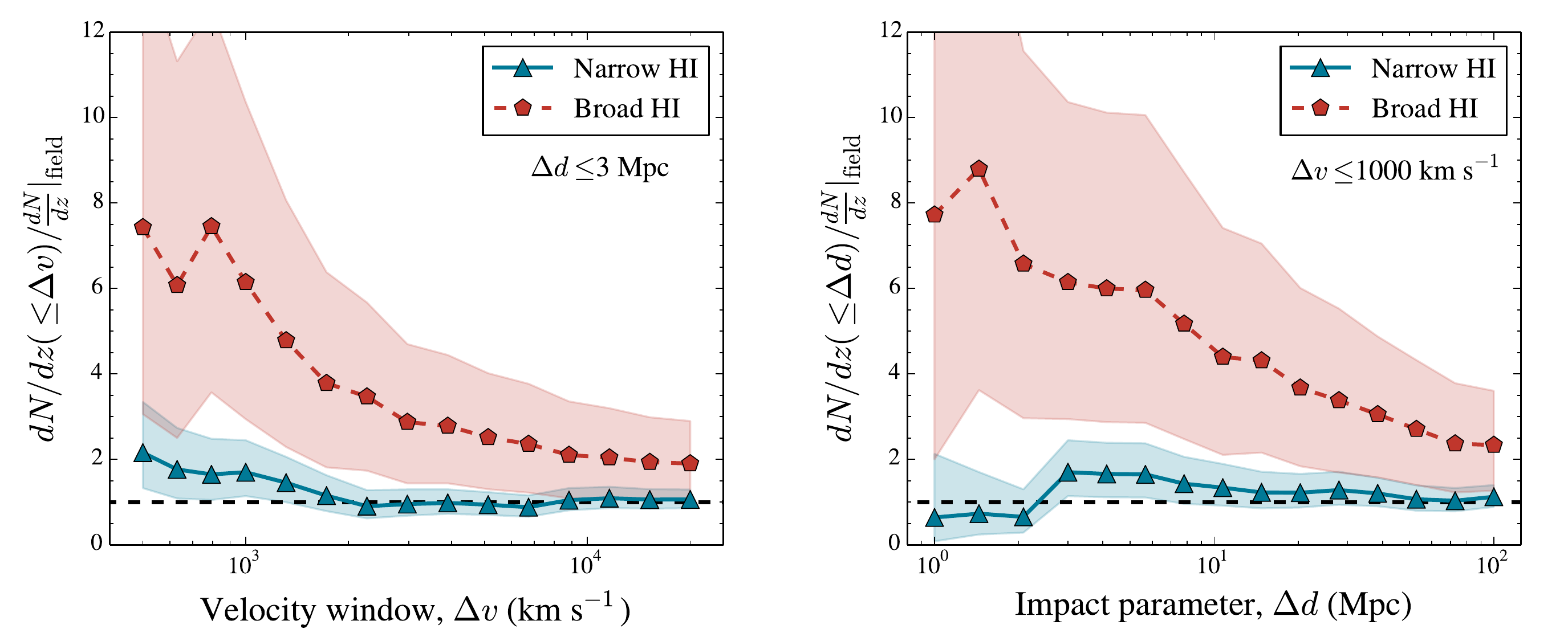}
    \end{minipage}
    
    \caption{Relative excesses of redshift number densities compared
    to the field expectation, as a function of rest-frame velocity
    window ($\Delta v$; left panels) and maximum impact parameter to
    the closest cluster-pair axis ($\Delta d$; right panels), for
    fixed $\Delta d = 3$\mpc and $\Delta v = 1000$\kms values
    respectively. The top panels show those for total \hi (brown
    circles, solid lines) and \ovi (green squares, dashed lines)
    absorption line samples. The bottom panels show those for narrow
    (NLA; $b<50$\kms; blue triangles, solid lines) and broad (BLAs; $b
    \ge 50$\kms; red pentagons, dashed lines) \hi\ absorption line
    samples.  Coloured light shaded areas represent $\pm 1 \sigma$
    statistical uncertainties. See \Cref{sec:excesses} for further
    details.}  \label{fig:excesses} \end{figure*}

\subsection{Redshift number densities}\label{sec:dndz_results}

\Cref{fig:dndz_hi_ovi} shows the $dN/dz$ of total \hi (top panels;
brown circles) and \ovi (bottom panels; green squares) absorption
components as a function of maximum velocity window ($\Delta v$; left
panels) and maximum impact parameter ($\Delta d$; right panels) for a
fixed $\Delta d = 3$\mpc and $\Delta v = 1000$\kms, respectively. The
expected field values following our approach described in
\Cref{sec:dndz_field} are shown by the horizontal dashed line with its
$\pm 1\sigma$ uncertainty represented by the grey region. We also show
the $\pm 1\sigma$ field values from the \citet{Danforth2008} survey as
the darker grey hashed regions, which is consistent with ours.

When we fix $\Delta d = 3$\mpc (left panels), we observe a clear
overall increase in the redshift number density of \hi\ and \ovi
absorption lines with decreasing $\Delta v$. Similarly, when we fix
$\Delta v = 1000$\kms (right panels), we observe an overall increase in
the redshift number density of \hi\ and \ovi absorption lines with
decreasing $\Delta d$, but only down to $\Delta d \sim 3$\mpc; at
$\Delta d \lesssim 3$\mpc a flattening (or even decrease) trend is
observed, which we believe is mostly due to our small sample in such
bins.\footnote{But note that with this limited sample we cannot rule
  out that a real decreasing signal is present either.}  This change of
behaviour motivated our adopted fiducial value of $3$\mpc for the
maximum transverse separation between cluster-pair axis and the Q1410
sightline (see \Cref{sec:pairs}).

%\subsection{Redshift number density for narrow and broad
%\hi}\label{sec:dndz_nla_bla}

% In order to test whether a \ac{whim} signal is present in the \hi~data,
To test whether kinematic trends are present in the
\hi~data, we repeated the $dN/dz$ measurements for both narrow (NLA;
$b < 50$\kms) and broad (BLA; $b \ge 50$\kms) \hi~\lya absorption
lines. Although the canonical value for BLAs tracing the \ac{whim} is
$40$\kms, this limit assumes that the broadening is purely thermal.
Following more recent work \citep[][]{Richter2006,Danforth2010}, it is
acknowledged that non-thermal broadening mechanisms are likely to be
present in absorption line samples. Thus, our adopted value of $b\ge
50$\kms is more conservative (see also \Cref{sec:whim}).

\Cref{fig:dndz_nla_bla} is equivalent to \Cref{fig:dndz_hi_ovi} but for
NLA (top panels) and BLA (bottom panels) absorption line samples. The
hashed darker grey area in the bottom panels represents the field value
obtained by \citet{Danforth2010} for BLAs.

When we fix $\Delta d = 3$\mpc (left panels), we observe a clear
overall increase in the redshift number density of both narrow and
broad \hi absorption lines with decreasing $\Delta v$. When we fix
$\Delta v = 1000$\kms (right panels), we also observe an overall
increase down to $\Delta d \sim 3$\mpc; below this scale a decreasing
trend may be present for narrow \hi lines, while for broad \hi lines
the increasing trend persists.

We also note that our estimation field expectations are fully
consistent with those from previous blind surveys
\citep[][]{Danforth2008, Danforth2010}.\footnote{Note that in the case
  of BLAs, both field values have similar uncertainty. This is because
  \citet{Danforth2010} included a systematic contribution to the error,
  whereas ours is purely statistical.}  This implies that our
characterization of absorption lines (see \Cref{sec:abslines}) and our
methodology for estimating the field expectation from our Q1410 data
alone (see \Cref{sec:dndz_field}) are reasonable. Therefore, we can
conclude that the vast majority (if not all) of the observed excesses
come from scales within $\Delta v =1000$\kms and $\Delta d = 3$\mpc
(see \Cref{sec:discussion} for further discussion).\\

%Similarly to the cases of the total \hi and \ovi samples, our
%estimation of the field expectation for BLAs is fully consistent with
%that of previous blind surveys \citep{Danforth2010}. Following the same
%argument as in \label{sec:dndz_hi_ovi}, we conclude that the vast
%majority (if not all) of the observed excesses come from scales within
%$\Delta v =1000$\kms and $\Delta d = 3$\mpc.

%In the case of the BLA sample, the $dN/dz$ does not overlap with its
%field expectation at large $\Delta v$ or $\Delta d$ values (at least at
%the $1\sigma$ c.l.). This is driven by the relatively large excess of
%BLAs observed close to cluster-pairs compared to the field (i.e. a
%factor of $\sim 6$; see \Cref{sec:excesses}), which makes the last term
%in \Cref{eq:nfield} to be of the order of $\sim 50\% N_{\rm tot}$.

\subsection{Relative excesses with respect to the field}\label{sec:excesses}

    \begin{figure}
    \begin{minipage}{0.5\textwidth}
    \centering
    \includegraphics[width=1\textwidth]{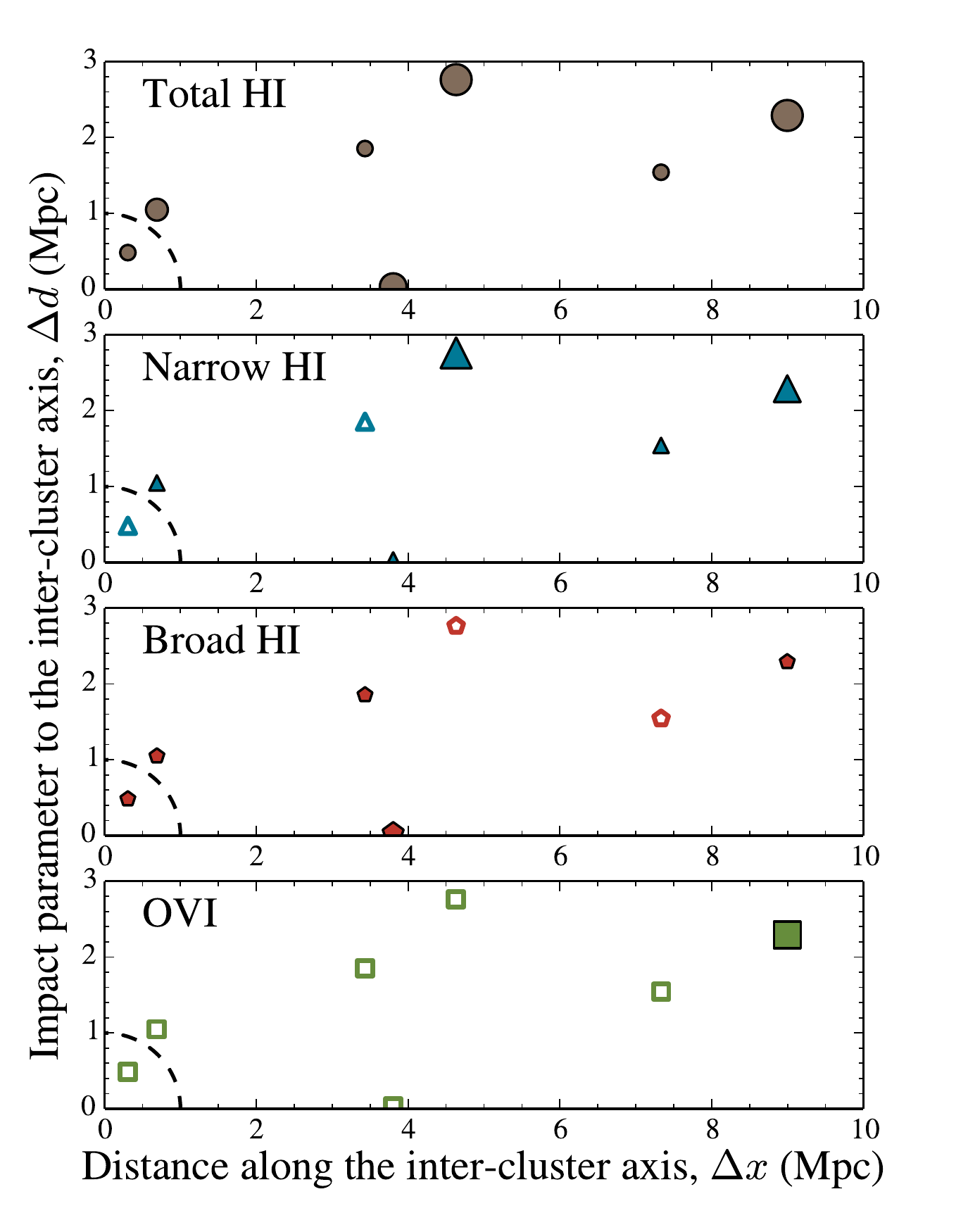}
    \end{minipage}
    
    \caption{Representation of the position of the Q1410 sightline
    with respect to different independent cluster-pairs in our
    sample. The $y$-axes correspond to the impact parameter to a given
    cluster-pair, $\Delta d$, while the $x$-axes correspond to the
    distance to the closest galaxy cluster in a given cluster-pair,
    along the cluster-pair axis, $\Delta x$, (see \Cref{fig:diagram}
    for a schematic diagram). The four panels correspond to different
    absorption line samples, from top to bottom: total \hi\, narrow
    \hi\ (NLAs; $b < 50$\kms), broad \hi\ (BLAs; $b \ge
    50$\kms), and total \ovi. Filled symbols correspond to portions of
    the sightline showing absorption lines within $\Delta v = 1000$\kms
    from the redshifts of a given cluster-pair, where the sizes of the
    symbols are proportional to the number of absorption lines. Empty
    symbols correspond to portions of the sightline showing no
    absorption. When multiple cluster-pairs lie at similar redshifts
    (i.e. from grouped ones), the absorption lines are associated to
    the one having the smallest impact parameter value and the rest of
    the cluster-pairs are obviated. We also show the typical virial
    radii of our sample of clusters with a quarter dashed circumference
    of radii $1$\mpc centred at the origin. See
    \Cref{sec:filament_geometry} for further details.}
    \label{fig:filament_geometry} \end{figure}

%In this section we quantify how the relative behaviors of $dN/dz$ for
%\hi, \ovi, NLA and BLA absorption line samples compare to that of their
%respective field expectations. 

\Cref{fig:excesses} shows the relative excesses of redshift number
densities of our absorption line samples compared to their respective
field expectations, defined as $\frac{dN}{dz}/\frac{dN}{dz}|_{\rm
  field}$, as a function of rest-frame velocity window ($\Delta v$;
left panels) and maximum impact parameter to the closest cluster-pair
axis ($\Delta d$; right panels), for fixed $\Delta d = 3$\mpc, $\Delta
v = 1000$\kms, respectively. The top panels show the results for our
total \hi (brown circles, solid line) and \ovi (green squares, dashed
line) samples, while the bottom panels show the results for our NLA
(blue triangles, solid line) and BLA (red pentagons, dashed line)
samples. Coloured light shaded areas represent the $\pm 1 \sigma$
statistical uncertainties.

Although subject to large statistical uncertainties, the relative
excess for BLAs tends to be the highest of all, reaching a value of
$\sim 6$ times its field expectation at $\Delta d = 3$\mpc and $\Delta
v = 1000$\kms. On the other hand, the excess of NLAs tends to be the
smallest of all, reaching a value of only $\sim 1.5$ times its field
expectation at the same scales. The relative excess of total \hi tends
to lie in between that of NLAs and BLAs, but is closer to that of NLAs
because these type of absorbers dominate the neutral hydrogen
sample. The sample of \ovi has the largest statistical uncertainties
(it is indeed the smallest sample), which makes its relative excess to
be consistent with all others even at the $1\sigma$ c.l. Strictly,
within $\sim 2\sigma$ c.l. {\it all} the reported excesses are
consistent with each other across different samples, and are also
consistent with their respective field expectations. Therefore, it is
important to test these trends with larger datasets.

\subsection{Equivalent widths distributions}

In \Cref{sec:Wr_dndz} we provide a comparison between equivalent widths
distributions for our different samples. We did not find statistically
significant differences between systems close to cluster-pairs and the
field expectation in terms of equivalent widths, at least from our
limited sample sizes.

\section{Discussion}\label{sec:discussion}

\subsection{Filamentary structure}\label{sec:filament_geometry}

Here we argue that our results are roughly consistent with a
filamentary structure for the absorbing gas close to
cluster-pairs. This is so because when we restrict the analysis to a
fixed $\Delta v =1000$\kms, the excess is maximized at impact
parameters of $\Delta d \sim 3$\mpc (or even $\Delta d \lesssim 3$\mpc
for broad \hi), while the typical separation between clusters in our
cluster-pairs are of the order of $\gtrsim 10-15$\mpc (see sixth and
seventh column of \Cref{tab:cpairs}). Moreover, we have also found that
when we consider scales far {\it outside} our fiducial values, we fully
recover the field expectation (see reasoning presented in
\Cref{sec:dndz_field}). From the cumulative results presented in
\Cref{sec:results}, we can directly calculate $dN/dz$ in independent
intervals instead by subtracting both the reported number of absorption
lines and redshift path in a given bin (i.e. $N_{\rm bin}$ and $\Delta
z_{\rm bin}$) of a smaller scale, to those at the scale of
interest. For instance, if we focus on the total \hi sample and
consider scales between $3$ to $100$\mpc as those of interest, we can
estimate $dN/dz$ as $\frac{dN}{dz} \sim \frac{(38-16)}{(0.25-0.06)}
\sim \frac{22}{0.19} \sim 116 \pm 25$ (see top panel of
\Cref{fig:dndz_hi_ovi} and \Cref{tab:hi_summary}). This number is fully
consistent with the field expectation \citep[e.g.][]{Danforth2008}, and
therefore we conclude the vast majority (if not all) of the observed
excesses come from scales within $\Delta v =1000$\kms and $\Delta d =
3$\mpc. However, because we did not impose a minimum distance between
the \ac{qso} sightline and the {\it closest} cluster of a cluster-pair,
there is also the possibility that our survey geometry does not
represent that of a filamentary structure. If those separations are all
$\lesssim 3$\mpc for instance, our survey could be probing a more
spherical (or disk) geometry instead.

\Cref{fig:filament_geometry} shows a geometrical representation of our
survey.
%The $y$-axes correspond to the impact parameter to a given cluster-pair
%inter-cluster axis (see the sixth column in \Cref{tab:cpairs}), while
%the $x$-axes correspond to the distance to the {\it closest} galaxy
%cluster in a given cluster-pair, {\it along} the inter-cluster axis
%(see the fourth column in \Cref{tab:cpairs}). The four panels
%correspond to our different absorption line samples, from top to
%bottom: total \hi, broad \hi~($b \ge 50$\kms), narrow \hi~($b <
%50$\kms), and total \ovi. Filled circles correspond to portions of the
%sightline showing absorption lines within $\Delta v = 1000$\kms from
%the redshifts of a given cluster-pair, where the sizes of the circles
%are proportional to the number of absorbers of each kind. When multiple
%cluster-pairs lie at similar redshifts (i.e. grouped cluster-pairs),
%the absorption lines are associated with the one having the {\it
%  smallest} impact parameter value, and the rest of the cluster-pairs
%are obviated (i.e. not shown). When the closest cluster-pair does not
%show absorption lines within $\Delta v = 1000$\kms from the redshifts
%of a given cluster-pair, we plot an open circle at its respective
%position and move to the next closest cluster-pair. The typical virial
%radii of our sample of clusters is shown by the quarter dashed
%circumference of radii $1$\mpc.
We observe that the distances probed by our survey cover scales between
$\Delta x \sim 0- 10$\mpc along the inter-cluster axes, roughly
uniformly. Therefore, we conclude that the geometry of our survey is
indeed consistent with that of a filamentary structure, but we also
stress that a larger sample must be analysed in order to better
constrain the geometry.

\subsection{Covering fractions of absorbing gas close to cluster-pairs}\label{sec:fcs}

Here we provide a first estimation of the covering fractions, $f_c^{\rm
  flmnt}$, of the absorbing gas close to cluster-pairs and compare them
with the random expectation, $f_c^{\rm rand}$. Our adopted fiducial
$\Delta v = \pm 1000$\kms corresponds to $\sim \pm 16$\mpc along the
line-of-sight (if cosmological). This is a larger scale compared to our
fiducial filament radius of $\sim 3$\mpc. Therefore, the excesses do
not necessarily come from single inter-cluster filaments. Although our
reported $dN/dz$ signals tend to keep increasing at smaller rest-frame
velocity differences, the samples also get smaller, which makes the
statistical uncertainties larger too (see Figures~\ref{fig:dndz_hi_ovi}
and \ref{fig:dndz_nla_bla}). By comparing the observed covering
fractions to random expectations, we can shed light into the origin of
the reported excesses in relation to the cluster-pairs themselves.

From \Cref{fig:filament_geometry} we observe that $7/7$ sightlines
close to cluster-pairs did show at least $1$ \hi\ absorber, which
implies a covering fraction of $f_c^{\rm flmt}({\rm HI}) \sim
1.00^{+0.00}_{-0.23}$.\footnote{The uncertainty is estimated assuming a
  binomial distribution for the number of `hits' given the $7$
  independent `trials', using the Bayesian formalism described by
  \citet{Cameron2011} with a flat prior.} Similarly, NLAs and BLAs were
both found in $5/7$ of the sightlines probing them (although different
subsamples; see second and third panel of
\Cref{fig:filament_geometry}), implying $f_c^{\rm flmt}({\rm NLA})
\approx f_c^{\rm flmt}({\rm BLA}) \approx 0.71^{+0.18}_{-0.26}$. In
contrast, \ovi absorbers were found in $1/7$ of the sightlines probing
them, implying a smaller covering fraction of $f_c^{\rm flmt}({\rm
  OVI}) \approx 0.14^{+0.26}_{-0.12}$. These results are summarized in
\Cref{tab:fcs} (upper half) and shown in \Cref{{fig:fcs}} (filled
symbols), using the same symbol/colour convention as in previous
figures.

    \begin{figure}
    \begin{minipage}{0.5\textwidth}
    \centering
    \includegraphics[width=1\textwidth]{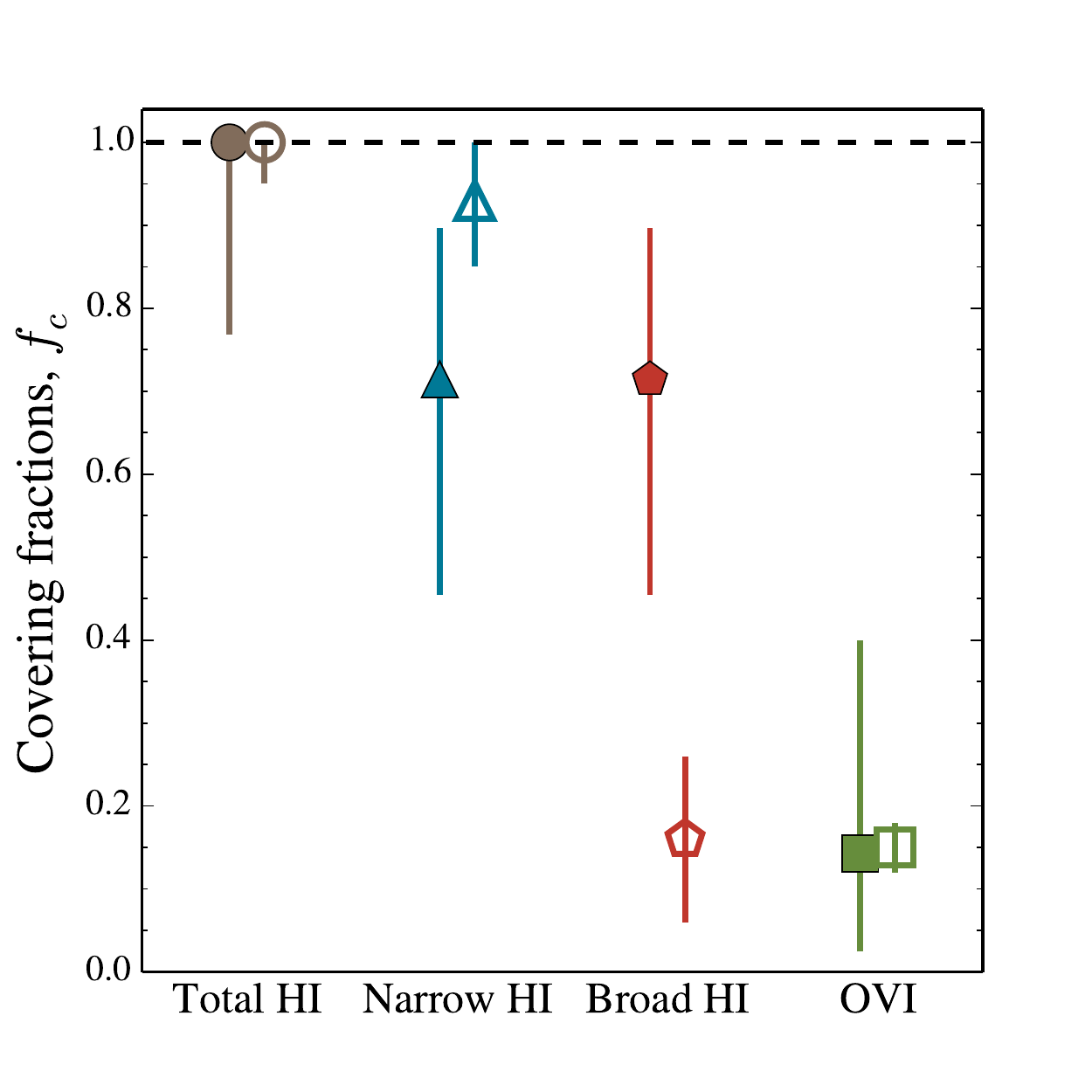}
    \end{minipage}
    
    \caption{Estimated covering fractions of gas close to cluster-pairs
    ($f_c^{\rm flmnt}$; solid symbols) and in the random
    expectation ($f_c^{\rm rand}$; open symbols), for our different
    samples of absorption lines using the same symbol/colour convention
    as in previous figures. We observe that broad \hi\ absorbers
    ($b\ge50$\kms) have about $\sim 4$ times a larger covering fraction
    in close to cluster-pairs than from the random expectation. See
    \Cref{sec:fcs} for further details.}  \label{fig:fcs}
    \end{figure}

As a control sample for a given absorber, we estimated $f_c^{\rm rand}$
as,

\begin{equation}
f_c^{\rm rand} =
\left\{
	\begin{array}{ll}
		\Delta z({\Delta v}) \ \frac{dN}{dz}|_{\rm field}  &  \mbox{if } \Delta z({\Delta v}) \frac{dN}{dz}|_{\rm field} < 1 \\
		1 &  \mbox{if } \Delta z({\Delta v}) \frac{dN}{dz}|_{\rm field} \ge 1 \\
	\end{array}
\right.
\label{eq:fc_field}
\end{equation}

\noindent where $\frac{dN}{dz}|_{\rm field}$ is the redshift number
density of lines in the field, and $\Delta z(\Delta v)$ is the
corresponding redshift path to a rest-frame velocity window $\Delta v$
evaluated at $z=0.35$ (the median redshift of our cluster-pair
sample). We note that we are neglecting the intrinsic clustering of
absorption lines with themselves, which is justified because our
$f_c^{\rm flmt}$ estimations are obtained from {\it independent}
structures.\footnote{For $f_c^{\rm flmt}$ obtained from non-independent   structures (e.g. a single well mapped filament), a
meaningful   estimation of $f_c^{\rm rand}$ must take clustering into
account. At   first order, there will be two opposing effects (when
clustering is   positive): clustering along the line-of-sight will
tend to decrease   $f_c^{\rm rand}$, while the clustering transverse
to the   line-of-sight will tend to increase it. Moreover, higher
order   correlations must also be considered to account for the joint
probability of having multiple `hits' in such large single   non-independent volume.}. The fifth column of \Cref{tab:fcs} (upper half)
summarizes the random expectations for our different samples using
$\Delta v = \pm 1000 = 2000$\kms. These results are also shown in
\Cref{fig:fcs} as open symbols. We observe that the covering fractions
for total \hi, NLAs and \ovi close to cluster-pairs are consistent
with their random expectations, while $f_c^{\rm flmt}({\rm BLA})$ is
about $\sim 4$ times larger than its random expectation (at the $\sim
2\sigma$ c.l.). Having consistency with the random expectations are
not surprising; as mentioned, $\Delta v = \pm 1000$\kms corresponds to
about $\sim \pm 16$\mpc along the line-of-sight (if cosmological)
around massive structures traced by galaxy cluster pairs. In this
scenario, we do expect that the $dN/dz|_{\rm field}$ values to be
dominated by absorption lines found in the overdense \ac{lss}. On the
other hand, having an excess in the covering fractions of BLAs with
respect to the random expectation implies that this type of gas is not
common over $\sim \pm 16$\mpc scales around cluster-pairs, and
therefore it has to come from smaller scales (i.e the cluster-pair
itself). To test this hypothesis, we have repeated the covering
fractions estimations using a smaller $\Delta v = \pm 500$\kms, and
the excess of BLAs remains large ($\sim 7$; see values in the bottom
half of \Cref{tab:fcs}). This behaviour favours the conclusion that most
of the BLAs in our sample are directly related to the cluster-pairs themselves.

\begin{table}
  \begin{minipage}{0.44\textwidth}
    \centering
    \caption{Estimated covering fractions of absorbing gas close to cluster-pairs.}\label{tab:fcs}
    \begin{tabular}{@{}lccccc@{}}
      \hline
      \multicolumn{1}{c}{Sample}   & $n_{\rm hits}$  & $n_{\rm trials}$  &$f_c^{\rm flmnt}$ & $f_c^{\rm rand}$ & Excess \\
      \multicolumn{1}{c}{(1)}      & (2)           & (3)             & (4)            & (5)            & (6)    \\
      \hline
      \multicolumn{6}{c}{$\Delta v = \pm 1000$\kms and $\Delta d = 3$\mpc} \\
      \hline \smallskip
      \hi         &   7  &  7    &   $1.00^{+0.00}_{-0.23}$& $1.00^{+0.00}_{-0.05}$  & $\sim 1$  \\ \smallskip
      NLA         &   5  &  7    &   $0.71^{+0.18}_{-0.26}$ & $0.93^{+0.07}_{-0.08}$ & $\sim 1$   \\ \smallskip
      BLA         &   5  &  7    &   $0.71^{+0.18}_{-0.26}$ & $0.16^{+0.10}_{-0.10}$ & $\sim 4$  \\ \smallskip
      \ovi        &   1  &  7    &   $0.14^{+0.26}_{-0.12}$ & $0.15^{+0.03}_{-0.03}$ & $\sim 1$  \\ 
      \hline
      \multicolumn{6}{c}{$\Delta v = \pm 500$\kms and $\Delta d = 3$\mpc} \\
      \hline \smallskip
      \hi         &   5  &  7    &   $0.71^{+0.18}_{-0.26}$ & $0.58^{+0.03}_{-0.03}$ & $\sim 1$  \\ \smallskip
      NLA         &   3  &  7    &   $0.43^{+0.25}_{-0.22}$ & $0.47^{+0.04}_{-0.04}$ & $\sim 1$  \\ \smallskip
      BLA         &   4  &  7    &   $0.57^{+0.22}_{-0.25}$ & $0.08^{+0.05}_{-0.05}$ & $\sim 7$  \\ \smallskip
      \ovi        &   1  &  7    &   $0.14^{+0.26}_{-0.12}$ & $0.08^{+0.02}_{-0.02}$ & $\sim 1$  \\
      \hline
     
    \end{tabular}
    \end{minipage}
    \begin{minipage}{0.44\textwidth}
    (1) Sample of absorbing gas.
    (2) Number of `hits' defined as sightlines showing absorption in a given sample, within $\Delta v =\{\pm1000, \pm500\}$\kms and within $\Delta d = 3$\mpc. 
    (3) Number of `trials' defined as the total number of sightlines to look for absorption. 
    (4) Covering fraction close to cluster-pairs estimated as $n_{\rm hits}/n_{\rm trials}$ (uncertainties correspond to those of a binomial $1\sigma$ c.l.).
    (5) Covering fraction in a random sightline for a given $\Delta v =\{\pm 1000,\pm 500\}$\kms.
    (6) Excess covering fraction defined as $f_{c}^{\rm flmnt}/f_{c}^{\rm rand}$. 
  \end{minipage}
\end{table}

\subsection{Could the observed excesses of gas be due to galaxy 
clusters/groups or individual galaxy halos?}\label{sec:halos}

%Another relevant question to shed light into the origin of the
%absorbing gas in these inter-cluster filaments, is whether galaxy
%clusters or individual galaxy halos could account for the observed
%excesses.

Regarding massive structures, we can see from
\Cref{fig:filament_geometry} that the vast majority of sightlines are
probing regions far away from the virial radii of {\it known} galaxy
clusters (see also the left panels of
\Cref{fig:cpairs_dist_spec_all}). However, because of the limited
completeness level of the \ac{redmapper} catalog, there could still
be unknown clusters or groups in such regions. In order to directly
address this question, one must survey the Q1410 field for individual
galaxies and \ac{lss} close to the Q1410 line-of-sight, which we leave
for future work (see \Cref{sec:future}). Still, in the following we
provide an assessment of the plausible incidence of gas associated to
individual halos from two indirect but independent arguments.

\subsubsection{Redshift number density of galaxies} Based on the
reasoning presented by \citet{Prochaska2011b}, one can estimate the
redshift number density of galaxies of luminosity $L \ge L_{\rm min}$,
$dN/dz|_{\rm gals}$, intersecting a given \ac{los} by assuming a given
cross-section for them. For galaxies with $L \ge 0.01-0.001 L^{*}$ and
assuming cross-sections given by the virial radius of galaxies with
unity covering fractions,\footnote{Which is a   conservative
assumption \citep[e.g. see][for counter
examples]{Wakker2009,Prochaska2011b,Johnson2014}.}
\citet{Prochaska2011b} find that $dN/dz|_{\rm gals} \approx 10-20$
(see their figure 8). This estimate is valid for the field and so we
need to correct for the fact that cluster-pairs generally probe
\ac{lss} overdensities. We use the excess of narrow \hi lines to
estimate the overall overdensity in our cluster-pair sample, as
$\sim 1-2$ times the mean density of the Universe (note that these
values are consistent with the expected overdensities traced by
cluster-pairs from cosmological simulations). Therefore, we estimate
$dN/dz|^{\rm   flmnt}_{\rm gals} \approx 10-40$. This number is lower
than the typical $dN/dz \sim 60-300$ observed for our samples of \hi
(either narrow or broad), but is comparable to the $dN/dz \sim 30-100$
observed for \ovi. Therefore, it seems unlikely that most of the
excess observed for narrow and broad \hi gas close to cluster-pairs is driven by galaxy halos of individual galaxies. On the other
hand, our reported excess of \ovi gas might well be produced (at least
partly) by individual galaxy halos (see also below).

\subsubsection{Metal absorption lines}
One can also infer the presence of galaxy halo material by means of
metal absorption lines, in particular from low-ionization species
\citep[e.g.][]{Werk2014}. There is only $1$ absorption system in our
cluster-pair sample showing metal absorption lines: the one at
$z\approx0.35$ from which all the $3$ observed \ovi components come
from (see the fourth panel in \Cref{fig:cpairs_dist_spec_all}). This
system has strong \hi absorption with column densities $N > 10^{14}$\cm
and shows a complex kinematic structure ($4$ components in total, $3$
narrow and $1$ broad). The second narrow component also shows the
presence of \cii, \siiii and \niii (see
\Cref{tab:abslines}). Therefore, although it seems very likely that an
important fraction of the absorbing gas in this system comes from an
individual galaxy halo or its immediate surroundings, this system only
accounts for $3/10 \sim 30$ per cent of NLAs and $1/6 \sim 16$ per cent
of BLAs in our sample. Again, from this independent reasoning we
reached the same conclusion as before, i.e. individual galaxy halos
could account for the observed excess in \ovi lines (although not a
requirement), but not for the majority of \hi gas.

\subsection{Statistical evidence of the \ac{whim}}\label{sec:whim}

Here we consider whether the observed trends could also be consistent
with the presence of a \acf{whim}. Ideally, one would require a full
characterization of the physical conditions of individual absorbers
using multiple species and comparing them with the expectations of
different models of ionization. However, because our sample is
dominated by absorption systems having no other species than \hi, this
approach is not feasible. Even when other species are present in
individual systems (e.g. \hi and \ovi) this approach requires the
uncertain assumption that the majority of the gas comes from a single
phase, which is controversial at the very least (e.g. Werk et al., in
prep.). In view of this intrinsic limitation for an individual
characterization, here we opted for a purely statistical approach.

The \ac{whim} is usually defined as gas at temperatures in the range of
$T \sim 10^{5}-10^{7}$\,K, implying a minimum Doppler parameter of $b =
40$\kms for \hi (i.e. assuming the broadening is purely
thermal). Therefore, \hi lines with $b < 40$\kms can not be caused by a
\ac{whim}. Non-thermal processes can also broaden absorption line
profiles, including turbulence, Hubble broadening and unresolved blends
\citep[e.g.][and references therein]{Garzilli2015}. In overdense
environments, we expect turbulence to be the dominant source of
non-thermal physical broadening (but see \Cref{sec:caveat} for a
discussion regarding unresolved blends). Assuming that the turbulence
contribution is higher for hotter gas, we should have $b_{\rm turb}
\approx \alpha b_{T}$, and hydrodynamical simulations suggest
$0 \le \alpha \le 1$ \citep[e.g.][]{Tepper-Garcia2012}. Assuming the most
extreme case of $\alpha = 1$, we have that the observed Doppler
parameter would be $b = \sqrt{2} b_{T}$, making a limit $b > 40
\sqrt{2}\approx 57$\kms extremely conservative in ensuring to trace \hi
gas at $T > 10^{5}$\,K. However, there could still be genuine \ac{whim}
\hi absorption lines in the $40 < b_{\rm lim} < 40 \sqrt{2}$\kms range
(i.e. when thermal broadening does dominate).

In this paper, we used a limiting Doppler parameter value of $b_{\rm
lim}=50$\kms instead, which is in between $40 < b_{\rm lim} < 40
\sqrt{2}$\kms. This limit was partly chosen for allowing a direct
comparison to previous published work \citep[e.g.][]{Danforth2010},
but also because it minimizes potential misidentification of lines
that are supposed to trace warm-hot gas but trace cool gas instead,
and viceversa (i.e. genuine warm-hot absorbers having $b< b_{\rm
lim}$, and genuine cool absorbers having $b \ge b_{\rm
lim}$).\footnote{This is   not the same as choosing the limit that
maximizes the difference   between observed incidences though (which
we did not try).} Indeed, using $b_{\rm lim}=50$\kms, there is only
$1$ ($1$) line in the range $40 < b < 50$\kms ($50 < b <
40\sqrt{2}$\kms) in our NLA (BLA) sample associated with cluster-pairs
(e.g. see points with white stars in \Cref{fig:abslines}). A limit of
$b_{\rm lim}=50$\kms also corresponds to a $\alpha = 0.75$, and
therefore is still quite conservative in ensuring that BLAs trace gas
at $T\ge 10^5$\,K, even with a substantial turbulence contribution.
Regardless of these considerations, we also note that about half of
the BLAs associated with cluster-pairs in our sample are actually {\it
very} broad, with Doppler parameters $b\sim 80-150$\kms (e.g. see
\Cref{fig:abslines}), which should make them more likely to
trace gas at \ac{whim} temperatures (but see \Cref{sec:caveat}).

% so they are most certainly coming from gas at \ac{whim} temperatures
% (but see \Cref{sec:caveat}).

One of our proposed diagnostics is to compare the excesses in the
incidence of narrow and broad \hi absorbers (and eventually \ovi when
larger samples are gathered) found close to cluster-pairs with respect
to their field expectations (see \Cref{fig:excesses}). Because inter-
cluster filaments correspond to overdense regions in the Universe, an
excess of gas is generally expected to occur, and indeed we have shown
that this is the case (see \Cref{sec:results}). Under the null-hypothesis that BLAs and NLAs probe gas in similar physical conditions
(i.e. similar physical entities), then we expect both these excesses
to behave in a consistent manner. On the other hand, if BLAs and NLAs
{\it are not} probing similar physical conditions, a different
behaviour for the excesses is expected instead. A \ac{whim} signature
associated with inter-cluster filaments may include the
relative excess of BLAs to be {\it higher} than that of NLAs, which is
exactly what we observed (although only at the $\sim 1\sigma$ c.l.;
see \Cref{fig:excesses}, bottom panels). By increasing the sample
sizes we may test for any statistically significant difference
between them.

Another proposed diagnostic is to constrain the overall geometry for
the excess of gas, in terms of both BLAs and NLAs (and \ovi when
larger samples are gathered). A \ac{whim} signature in this context
would require an increase in the covering fractions of BLAs towards
the inter-cluster axes compared to the random expectations, which is
also what we have observed (again, only at the $\sim2\sigma$ c.l.; see
\Cref{sec:fcs}). Moreover, one can also look for trends in the Doppler
parameters of \hi absorption lines with respect to impact parameter to
the inter-cluster axes, as a proxy of temperature.  Assuming simple
models for the ionization of the gas (e.g. purely collisional), one
can even use the inferred temperatures to estimate a total hydrogen
column density from the observed \hi one. A \ac{whim} signature should
produce, on average, higher hydrogen column densities for higher
temperatures.

Although all our tentative results ($\sim 1-2\sigma$ c.l.) may
be consistent with the presence of a \ac{whim} in inter-cluster
filaments, we emphasize that a larger sample must be analysed before
reaching a definite conclusion.

\subsection{Caveat}\label{sec:caveat}

Probably the most important source of concern in our experimental
design, is our limited ability to disentangle blends, which is key to
detect broad and shallow absorption features expected to arise in the
\ac{whim}. The importance of this systematic uncertainty depends on the
\ac{snr}, as the higher the \ac{snr} the easier it is to assess the
kinematic structure of the absorption feature
\citep[e.g.][]{Richter2006,Danforth2010}. We emphasize that this is an
intrinsic limitation of the absorption-line technique, meaning that all
these kind of observational samples are, to some extent, affected by
this issue. As described in \Cref{sec:abslines}, we attempted to avoid
this bias by fitting asymmetric lines with at least two
components. Although not impossible, we believe that the likelihood of
having misidentified multiple narrower blended components as a single
broader and symmetric one in a {\it large} fraction of our \hi sample
is low (see \Cref{fig:cpairs_dist_spec_all} for individual examples
close to cluster-pairs). Given that we are comparing the relative
incidences of lines between different samples drawn from spectra of
similar \ac{snr} (see \Cref{sec:results}), our statistical approach
seems adequate for minimizing this potential source of uncertainty (as
opposed to attempting a full physical characterization of individual
systems).

\subsection{Future prospects}\label{sec:future}

Despite our promising results, the existence of the \ac{whim} in
inter-cluster filaments still needs to be observationally confirmed;
our pilot survey was not design to draw statistically significant
results, but primarily to show that such a goal is currently possible
with existing instrumentation. In this section, we enumerate remaining
work for providing a firm detection of the elusive \ac{whim} in the
context of our methodology.

\subsubsection{Increase the sample sizes}
Increasing the sample sizes is a key requirement. In this respect, we
are actively working on two fronts: (i) pursuing new \ac{hst}/\ac{cos}
observations of targeted \ac{qso} sightlines intersecting multiple
cluster-pairs; and (ii) searching in the \ac{hst} archive for already
observed \acp{qso} whose sightline intersects single or multiple
cluster-pairs. We estimate that the \ac{hst}/\ac{cos} archive will
allow us to extend the present work to tens of sightlines, but approach
(i) will still be necessary for efficient follow up observations
(e.g. galaxy surveys; see \Cref{sec:gals}).

\subsubsection{Survey for galaxies around the \ac{qso}
sightline}\label{sec:gals} As discussed in \Cref{sec:halos}, we need
to survey galaxies around our \ac{qso} sightline in order to {\it
directly} rule out the potential association of BLAs with the halos of
individual galaxies. To this end, spectroscopic redshifts are needed
(current galaxy photometric redshift uncertainties are too large for
meaningful associations with absorption lines). We are currently
pursuing \ac{mos} and \ac{ifu} observations around the Q1410 field. We
will use \ac{mos} surveys to assess the distribution of galaxies over
$~0.3-10$\mpc scales. This will be important not only for determining
whether galaxy groups or clusters are responsible of our observed
excesses of absorbing gas, but also to determine the actual geometry
of the \acp{lss} intersected by the Q1410 sightline, including: (i)
assess whether these putative inter-cluster filaments are
straight or bent; and (ii) refine the cosmological redshift of the
structures at the position of the Q1410 sightline. The \ac{ifu}
observations will primarily focus on mapping galaxies on scales within
$\lesssim 100-300$\kpc to the \ac{qso} sightline (the typical \ac{cgm}
scales) at a very high completeness level, including faint star-forming galaxies with no detectable continuum but having bright enough
emission lines.

\subsubsection{Comparison to hydrodynamical simulations}
Another key aspect of this project, is the comparison of our
observational results to those obtained from cosmological
simulations. Our experimental design offers a unique opportunity to
test the prediction of the $\Lambda$CDM paradigm in the largest and
densest filaments of the cosmic web, while also constraining current
galaxy evolution models. As we have shown, \hi dominates the gaseous
content found in inter-cluster filaments (see \Cref{sec:results}), and
is very likely that they originate far away from individual galaxy
halos (see \Cref{sec:halos}). This makes a direct comparison to
simulations straightforward because we expect this gas to be unaffected
by the uncertain baryonic processes occurring in galaxies (e.g. SNe/AGN
feedback). On the other hand, the subsample of absorption systems
showing metal absorption (e.g. those with \ovi) can put constraints on
these uncertain `sub-grid physic' models for galaxy formation. However,
the need for a full treatment of shocks in the gas limits the approach
to being hydrodynamical. If the predictions of \ac{whim} signatures in
inter-cluster filaments match our observational results, they will
provide yet another piece of evidence supporting the existence of this
elusive medium.

\section{Summary}\label{sec:summary}

The \acf{whim} has been predicted to account for a significant
fraction of the baryons at low-$z$, but its actual existence has
eluded a firm observational confirmation. In this paper, we have
presented a novel approach for detecting the \ac{whim}, by targeting
regions of the cosmic web where its presence is predicted to be
ubiquitous: the putative filaments connecting galaxy clusters. As a
proof of concept, we selected a single bright \ac{qso} (namely Q1410),
whose exceptional sightline passes within $\Delta d = 3$\mpc from the
$7$ inter-cluster axes connecting $7$ independent cluster-pairs at
redshifts $0.1 \le z \le 0.5$, and observed it with \ac{hst}/\ac{cos}.
We performed a full characterization of absorption features in the
\ac{fuv} spectrum of Q1410, independently of the presence of known
\ac{lss} traced by the galaxy cluster-pairs.  From this dataset, we
conducted a survey of diffuse gas along the \ac{qso} sightline with
special focus on \hi and \ovi absorption lines. This survey allowed us
to provide, for the first time, a systematic and statistical
measurement of the incidence, $dN/dz$, of intervening \hi and \ovi
absorption lines close to cluster-pairs. We split the sample of
\hi~\lya into broad (BLA) and narrow (NLA) using a Doppler parameter
limit of $b_{\rm lim} = 50$\kms, which ensures BLAs to be mostly from
gas at temperatures $T \ge 10^{5}$\,K, even when accounting for
turbulence. We quantified the incidence of \hi, NLAs, BLAs and \ovi
absorption lines close to cluster-pairs by varying the minimum rest-frame velocity window, $\Delta v$, and the minimum impact parameter to
the inter-cluster axes, $\Delta d$, and found that the incidence of
diffuse gas is maximized at $\Delta v \lesssim 1000$\kms and $\Delta d
\lesssim 3$\mpc. At these scales we report:

\begin{itemize}
\item $dN/dz({\rm HI})|_{\rm flmnt} = 287^{+91}_{-71}$, which
  corresponds to $2.3^{+0.8}_{-0.6}$ times the field expectation;

\item $dN/dz({\rm NLA})|_{\rm flmnt} = 179^{+77}_{-56}$ which corresponds to
  $1.7^{+0.7}_{-0.6}$ times the field expectation;

\item $dN/dz({\rm BLA})|_{\rm flmnt} = 108^{+65}_{-43}$ which
  corresponds to $6.1^{+4.2}_{-3.2}$ times the field expectation; and,

\item $dN/dz({\rm OVI})|_{\rm flmnt} = 58^{+57}_{-32}$ which
  corresponds to $4.1^{+4.3}_{-2.7}$ times the field expectation.
\end{itemize}

Although individually these excesses are only at the $\sim 1-2\sigma$
c.l., in concert they suggest a physical overdensity close to cluster-pairs. Our results are also roughly consistent with a filamentary
geometry for the absorbing gas connecting cluster-pairs, with covering
fractions of: $f_c({\rm HI})|_{\rm flmnt} = 1.0^{+0.0}_{-0.2}$;
$f_c({\rm NLA})|_{\rm flmnt} = 0.7^{+0.2}_{-0.3}$; $f_c({\rm
BLA})|_{\rm flmnt} = 0.7^{+0.2}_{-0.3}$; and $f_c({\rm   OVI})|_{\rm
flmnt} = 0.14^{+0.3}_{-0.1}$. Our resported covering fractions of
total \hi, NLAs and \ovi are all consistent with their random
expectations. In contrast, the reported covering fraction of BLAs is a
factor of $\sim 4$ larger than the random expectation. Because a rest-frame velocity window of $\Delta v \approx \pm 1000$\kms corresponds
to a rather large co-moving distance along the \ac{los} (i.e $\sim \pm
16$\mpc), having consistency with the random expectation is not
surprising, and suggests that most of the excesses of NLAs and \ovi
absorption come from the overall \ac{lss} overdensities around massive
structures traced by galaxy cluster pairs. On the other hand, a higher
covering fraction of BLAs close to cluster-pairs compared to the
random expectation, suggests that the excess of BLAs is physically
associated to the cluster-pairs themselves. Indeed, we also
reported covering fractions using a $\Delta v \approx \pm 500$\kms and
reached the same conclusion.

Based on statistical arguments, we also concluded that most of the
reported excesses of NLAs and BLAs cannot be attributed to individual
galaxy halos but rather to truly intergalactic material. In contrast,
the reported excess of \ovi\ most likely comes from gas associated to
individual galaxy halos or their immediate surroundings.

We argued that a behaviour in which BLAs show larger relative excesses
compared to that of NLAs (as tentatively reported here), may be a
direct signature of the \ac{whim}, especially if identified in
the regions close to cluster-pairs. With a larger sample of QSOs and a
careful accounting of systematic effects, the technique we have
presented here should therefore enable a firm detection of the
\ac{whim} in inter-cluster filaments.

\section*{Acknowledgements}

{We thank the anonymous referee for their constructive criticism
that improved the paper. We thank E. Rozo and E. Rykoff for providing
us with the extended version of their \ac{redmapper} catalog used in
this work.} We thank Ryan Cooke, Kathy Cooksey, Joe Hennawi, Akio
Inoue, John O'Meara, Philipp Richter, John Stocke and Todd Tripp for
helpful discussions. N.T. acknowledges support from the IMPS
Fellowship\footnote{\url{http://imps.ucolick.org}} that allowed him to
conduct independent research in the areas of intergalactic or
interstellar media at University of California, Santa Cruz. N.T.
is partially funded by NASA grant HST-GO-134491.008-A and JXP is
partially funded by NSF AST-1412981.

We thank contributors to SciPy\footnote{\url{http://www.scipy.org}},
Matplotlib\footnote{\url{http://www.matplotlib.sourceforge.net}},
Astropy
\footnote{\url{http://www.astropy.org}
  \citep{AstropyCollaboration2013}}, the Python programming
language\footnote{\url{http://www.python.org}}, and the free and
open-source community.

This work was mainly based on observations made with the NASA/ESA
Hubble Space Telescope under program GO 12958, obtained at the Space
Telescope Science Institute, which is operated by the Association of
Universities for Research in Astronomy, Inc., under NASA contract NAS
5-26555.

This research has made use of: the NASA/IPAC Extragalactic Database
(NED)\footnote{\url{http://ned.ipac.caltech.edu}}; the NASA's
Astrophysics Data System
(ADS)\footnote{\url{http://ads.harvard.edu}}; and data products
from the SDSS, SDSS-II\footnote{\url{http://www.sdss.org/}} and
SDSS-III\footnote{\url{http://www.sdss3.org}}.

\bibliographystyle{mn2e_trunc8.bst}
\bibliography{/home/ntejos/lit/bib/IGM}

\appendix

\section{Properties of clusters close to Q1410 compared to a control sample}\label{sec:cluster_control}

\begin{figure*}
    
 \begin{minipage}{0.45\textwidth}
    \centering
    \includegraphics[width=0.95\textwidth]{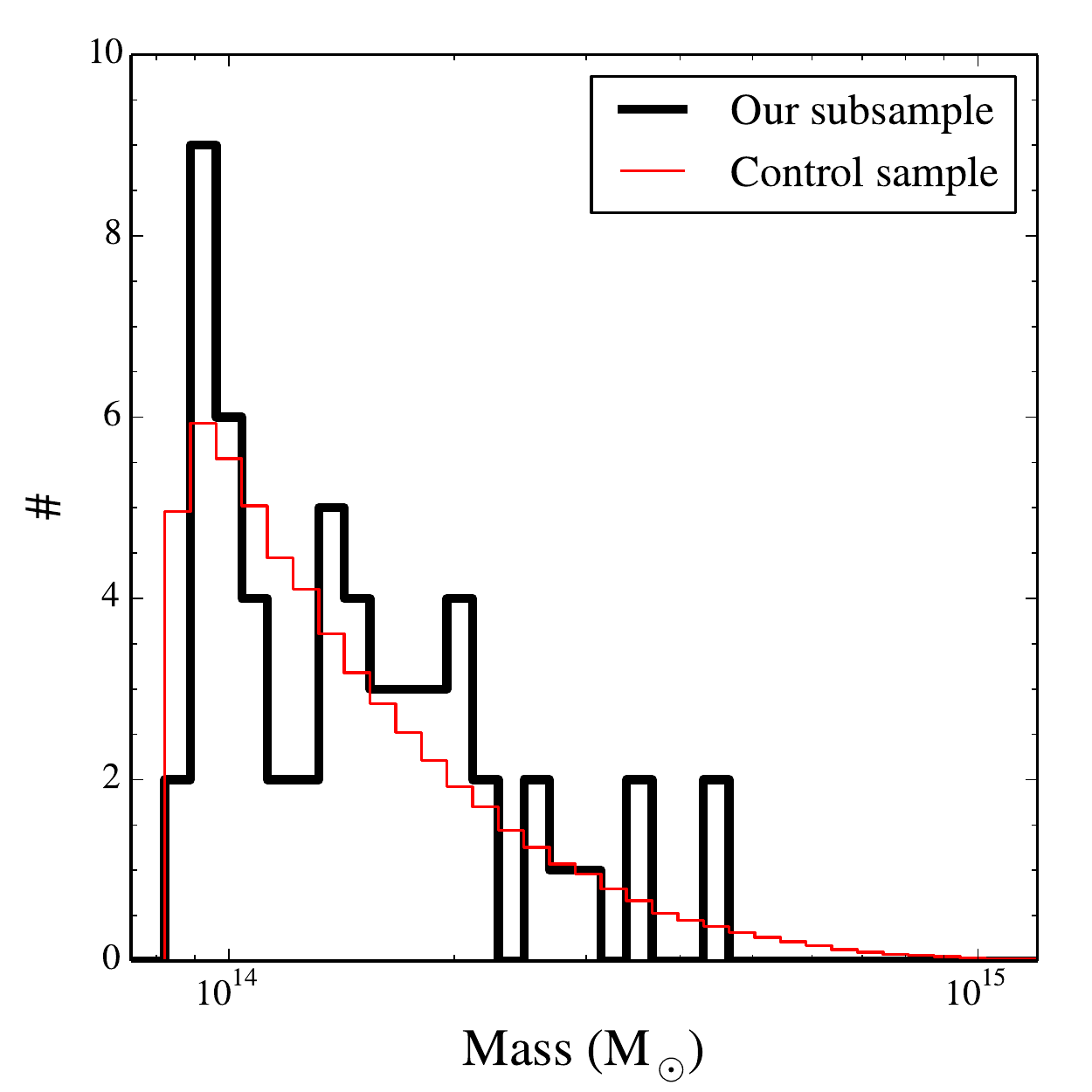}
\end{minipage}
\begin{minipage}{0.45\textwidth}
    \centering
    \includegraphics[width=0.95\textwidth]{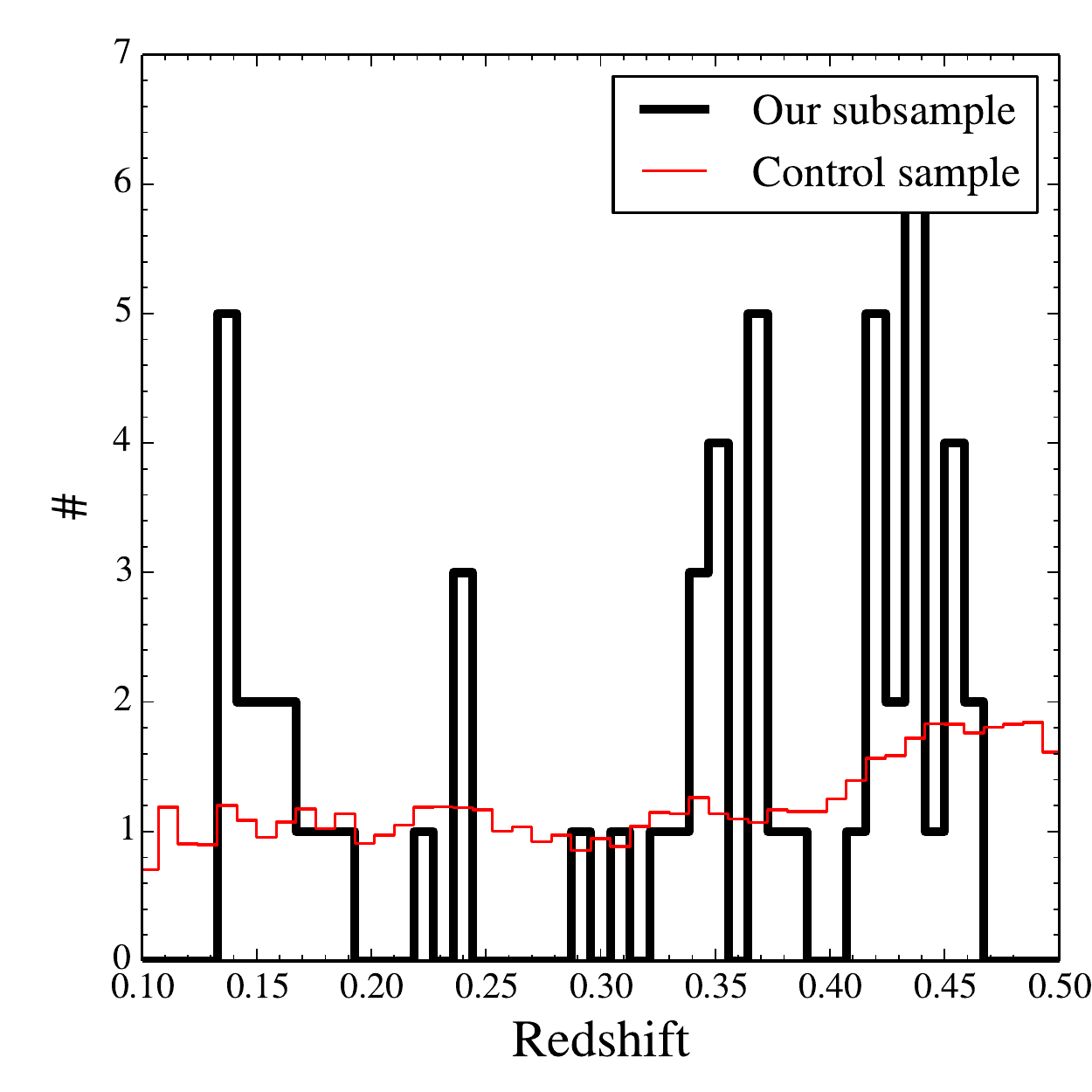}
\end{minipage}

\caption{Comparison between mass (left panel) and redshift (right
  panel) distributions between our subsample of $57$ \ac{redmapper}
  clusters within $20$\mpc from the Q1410 sightline, having richness
  values $\ge 10$ and lying between $0.1 \le z \le 0.5$ (thick black
  line), and normalized control samples satisfying only the richness
  and redshift range criteria (thin red line; see
  \Cref{sec:cluster_control} for details). A Kolmogorov-Smirnov test to
  the unbinned distributions give no statistically significant
  differences between the samples.}\label{fig:cluster_control}

\end{figure*}

In this section we investigate whether our subsample of clusters close
to the Q1410 sightline is a fair representation of the cluster
population in the whole \ac{redmapper} catalog. To do so, we will
compare the mass (estimated from \Cref{eq:mass}) and redshift
distributions of our sample to appropriate control samples drawn from
the \ac{redmapper} catalog.

The left panel of \Cref{fig:cluster_control} shows the mass
distribution for our subsample of $57$ clusters within $20$\mpc of
Q1410, having richness values $>10$ and lying between $0.1 \le z \le
0.5$ (thick black line), and the normalized mass distribution from a
control sample satisfying only the richness and redshift range criteria
($162\,144$ clusters in total; thin red line). We see no apparent
difference between these two distributions, and no statistically
significant differences are detected either: the Kolmogorov-Smirnov
test over the full unbinned samples gives a $\approx 60\%$ probability
of both being drawn from the same parent distribution.

In the case of the redshift distribution, we can not directly compare
that of the $162\,144$ clusters satisfying only the richness and
redshift range criteria, to our subsample. This is so because our
subsample of clusters are defined by a {\it cylinder} (i.e. constant
volume as a function of redshift), rather than fixed solid angle
(increasing volume as a function of redshift). In order to provide an
appropriate control sample, we looked at the redshift distribution of
\ac{redmapper} clusters having richness values $>10$ and lying between
$0.1 \le z \le 0.5$, in $5000$ cylinders of radius $20$\mpc, selected
at random positions between R.A. $\in [140,222]$ degrees and Dec. $\in
[4,56]$ degrees (i.e. well within the \ac{sdss} footprint).\footnote{We
  note that the expected number of clusters per random sightline is
  $\sim 32$ (see \Cref{sec:how_unusual}), and so $162\,144/32 \approx
  5000$ should cover a significant fraction of the cluster catalog in
  our chosen \ac{redmapper} subvolumes.} The right panel of
\Cref{fig:cluster_control} shows the redshift distribution for our
subsample of $57$ clusters (thick black line), and the normalized
redshift distribution from our aforementioned control sample (thin red
line). Again, no statistically significant difference is detected: the
Kolmogorov-Smirnov test over the full unbinned samples gives a $\approx
20\%$ probability of both being drawn from the same parent
distribution.
%We do note that the redshift distribution for the \ac{redmapper}
%cluster catalog seem to have decreased from $z\sim 0.43$ to $z\sim
%0.4$, but a proper interpretation of such an effect is beyond the
%scope of this paper. \comment{check if this is due to a change in
%filter bands to estimate the red sequence. i.e. artifact.}

From these comparisons we conclude that no strong bias is present in
our subsample of clusters close to the Q1410 sightline.

\section{Comparison between absorption line significance estimation methods}\label{sec:significance}
 
\begin{figure}
 \begin{minipage}{0.45\textwidth}
    \centering
    \includegraphics[width=1\textwidth]{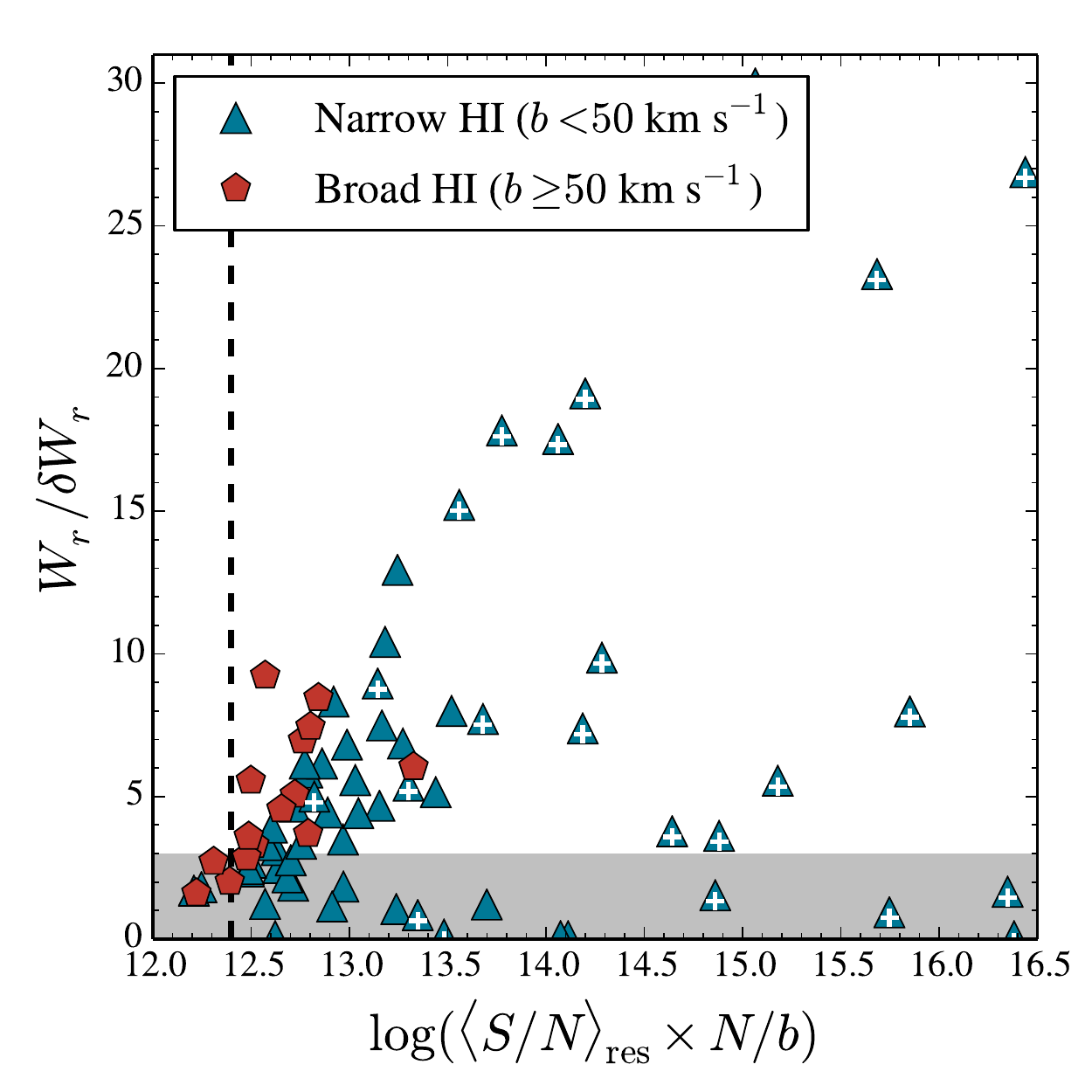}
  \end{minipage}

\caption{Comparison between two different methods for estimating the
    significance level of absorption line features. The $x$-axis
    corresponds to the $\log (\langle S/N \rangle_{\rm res}
    \times \ N / b)$ values proposed by \citet{Richter2006} (at
    least for broad \hi\ lines) where $N$ is in units of cm$^{-2}$
    and $b$ in \kms, while the $y$-axis corresponds to our $W_r/\delta
    W_r$ (see \Cref{sec:ew}). Blue triangles correspond to narrow
    \hi\ absorption components ($b<50$\,\kms), while red pentagons
    correspond to broad \hi\ absorption components ($b \ge
    50$\,\kms), both found across the full Q1410 spectrum. White
    crosses mark components having multiple transitions, hence reliable
    (label `a'; see \Cref{sec:reliability}). The grey shaded area
    correspond to values $W_r/\delta W_r < 3$, while the vertical
    dashed line correspond to a value of $\langle S/N \rangle_{\rm
    res} \times \ N / b = 2.5 \times 10^{12}$ cm$^{-2}$ $($km
    s$^{-1})^{-1}$. See \Cref{sec:significance} for further
    details.} \label{fig:significances}

\end{figure}

Here we compare two methods for estimating the significance level of
absorption features, namely our $W_{\rm r}/\delta W_{\rm r}$ criterion
(see \Cref{sec:ew}) and $\langle S/R \rangle_{\rm res} \times N / b$
proposed by \citet{Richter2006} for broad \hi\ absorption systems. The
motivation for the later being that what matters to confidently detect
a broad absorption line is both \ac{snr} {\it and} the optical depth at
the line centres, $\tau_0 \propto N / b$, and so it will not be
appropriate to use the commonly adopted formalism based on a minimum
equivalent width threshold for unresolved
lines. 

\Cref{fig:significances} shows a comparison between these two
quantities $\langle S/R \rangle_{\rm res} \times N / b$ in the
$x$-axis and $W_{\rm r}/\delta W_{\rm r}$ in the $y$-axis), for our
sample of \hi~absorption lines. Blue triangles correspond to \hi\
absorption components with $b<50$\,\kms (narrow), while red 
pentagons correspond to \hi\ absorption components with $b \ge
50$\,\kms (broad) over the full Q1410 spectrum. White crosses mark
components having multiple transitions, which we always account as
reliable (label `a'; see \Cref{sec:reliability}). The grey shaded area
corresponds to values $W_{\rm r}/\delta W_{\rm r} < 3$.

Restricting ourselves to broad \hi\ lines (red pentagons in
\Cref{fig:significances}), we see a clear correlation between these
two criteria. The dashed vertical line in \Cref{fig:significances}
corresponds to a value of $\langle S/R \rangle_{\rm res} \times N / b
\ge 2.5 \times 10^{12}$\, cm$^{-2}$ $($km s$^{-1})^{-1}$ for our
\ac{cos} data, which is needed to make both approaches roughly
consistent with each other for broad \hi\ lines. Such a value is
similar to the one reported by \citet{Richter2006}, $\langle S/R
\rangle_{\rm res} \times N / b \ge 3 \times 10^{12}$\, cm$^{-2}$ $($km
s$^{-1})^{-1}$ in their STIS data.

For narrow \hi lines (blue triangles in \Cref{fig:significances})
however, no clear correlation is present between these two methods,
where several components having $W_{\rm r}/\delta W_{\rm r} < 3$, can
still have large values of $\langle S/R \rangle_{\rm res} \times N /
b$. We note that many of the large $\langle S/R \rangle_{\rm res}
\times N / b$ components are reliable (white crosses in
\Cref{fig:significances}), but many others at the lower end are not. In
terms of $W_{\rm r}/\delta W_{\rm r}$, we see that some reliable lines
fall below the limit of $W_{\rm r}/\delta W_{\rm r}<3$. This is due to
the fact that our adopted conservative method for estimating $W_{\rm
  r}$ (see \Cref{sec:ew}) will give unrealistically large
uncertainties, $\delta W_{\rm r}$, for unconstrained or saturated
components. Indeed the vast majority of reliable components with
$W_{\rm r}/\delta W_{\rm r}<3$ are saturated lines for which the column
densities are not well determined.

Our proposed significance estimation based on $W_{\rm r}/\delta W_{\rm
  r}$ (and the presence of multiple transitions in a given component;
see \Cref{sec:reliability}) can be straightforwardly applicable to any
absorption line irrespective of its Doppler parameter and ionic
transition. Therefore it has the advantage of allowing an homogeneous
analysis.

\section{Estimation of redshift paths and absorption distances}\label{sec:dz}

We estimate $\Delta z(\Delta d,\Delta v, W_{\rm r}^{\rm min})$ and
$\Delta X(\Delta d,\Delta v, W_{\rm r}^{\rm min})$ as follows. First,
for a given transition we just considered regions in the Q1410 spectrum
probing rest-frame velocity differences within $\Delta v$ from {\it
  any} cluster-pair within $\Delta d$ from the Q1410 sightline, and
within our redshift range of $0.1 \le z \le 0.5$. We also masked out
regions with fluxes below $50\%$ the value of the continuum fit, to
account for the fact that we are usually biased against finding
absorption systems on top of strong absorption lines. We also masked
out regions within $\pm 200$\kms from strong Galactic absorption could
exist, namely \cii~$\lambda 1334.53$, \nv~$\lambda\lambda
1238.82,1242.80$, \oi~$\lambda 1302.17$, O~{\sc i}*~$\lambda 1304.86$,
O~{\sc i}**~$\lambda 1306.03$, \siii~$\lambda\lambda 1260.42,1304.37$,
P~{\sc iii}~$\lambda 1334.81$, S~{\sc ii}~$\lambda\lambda
1253.81,1259.52$ and \feii~$\lambda 1260.53$. We then calculated the
minimum rest-frame equivalent width to observe a transition at
rest-frame wavelenght, $\lambda_0$, along the spectrum as,

\begin{equation}
W_{\rm r,\lambda_0}^{\rm min} = 3 \frac{\lambda_0/R}{\langle S/R
  \rangle}
\label{eq:wmin}
\end{equation}

\noindent where $R$ is the resolution of the spectrograph (taken to be
$R= 20\,000$), and $\langle S/R \rangle$ is the signal-to-noise ratio
of the spectrum smoothed over $2$ pixels using a box-car filter
(i.e. $\sim 1$ resolution element). We then identified chunks of the
spectrum satisfying the criteria of $W_{\rm r,\lambda_0}^{\rm min} \ge
W_{\rm r}^{\rm min}$, and recorded each corresponding minimum and
maximum redshift, $z^{\rm min}_i(\Delta d, \Delta v, W_{\rm r}^{\rm
  min})$ and $z^{\rm max}_i(\Delta d, \Delta v, W_{\rm r}^{\rm min})$,
in the $i$-th spectral chunk. We then computed the redshift path as,

\begin{equation}
\Delta z(\Delta d, \Delta v, W_{\rm r}^{\rm min}) = \sum_{i} (z^{\rm
  max}_i - z^{\rm min}_i) \ \rm{,}
\label{eq:dzi}
\end{equation}

\noindent and the absorption distance as, 
\begin{equation}
\Delta X(\Delta d, \Delta v, W_{\rm r}^{\rm min}) = \sum_{i}
\int_{z^{\rm min}_i}^{z^{\rm max}_i} \frac{(1+z)^2}{ \sqrt{\Omega_{\rm
      m}(1+z)^3 + \Omega_{\rm \Lambda} }} {\rm d}z \ \rm{,}
\label{eq:dXi}
\end{equation}

\noindent conforming our adopted cosmology.

\section{Equivalent widths distributions}\label{sec:Wr_dndz}
 
    \begin{figure*}
    
    \begin{minipage}{0.4\textwidth}
    \centering
    \includegraphics[width=1\textwidth]{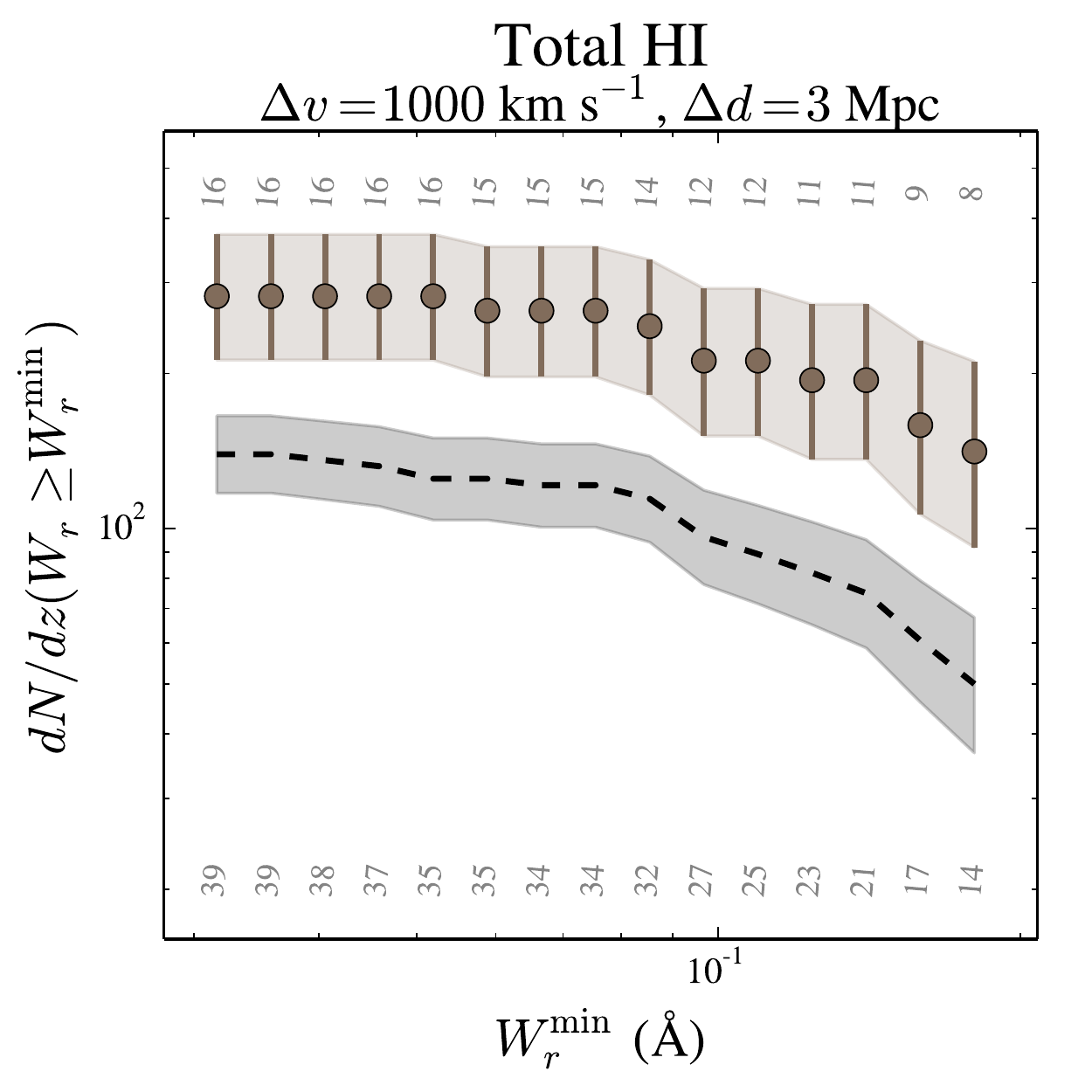}
    \end{minipage}
    \begin{minipage}{0.4\textwidth}
    \centering
    \includegraphics[width=1\textwidth]{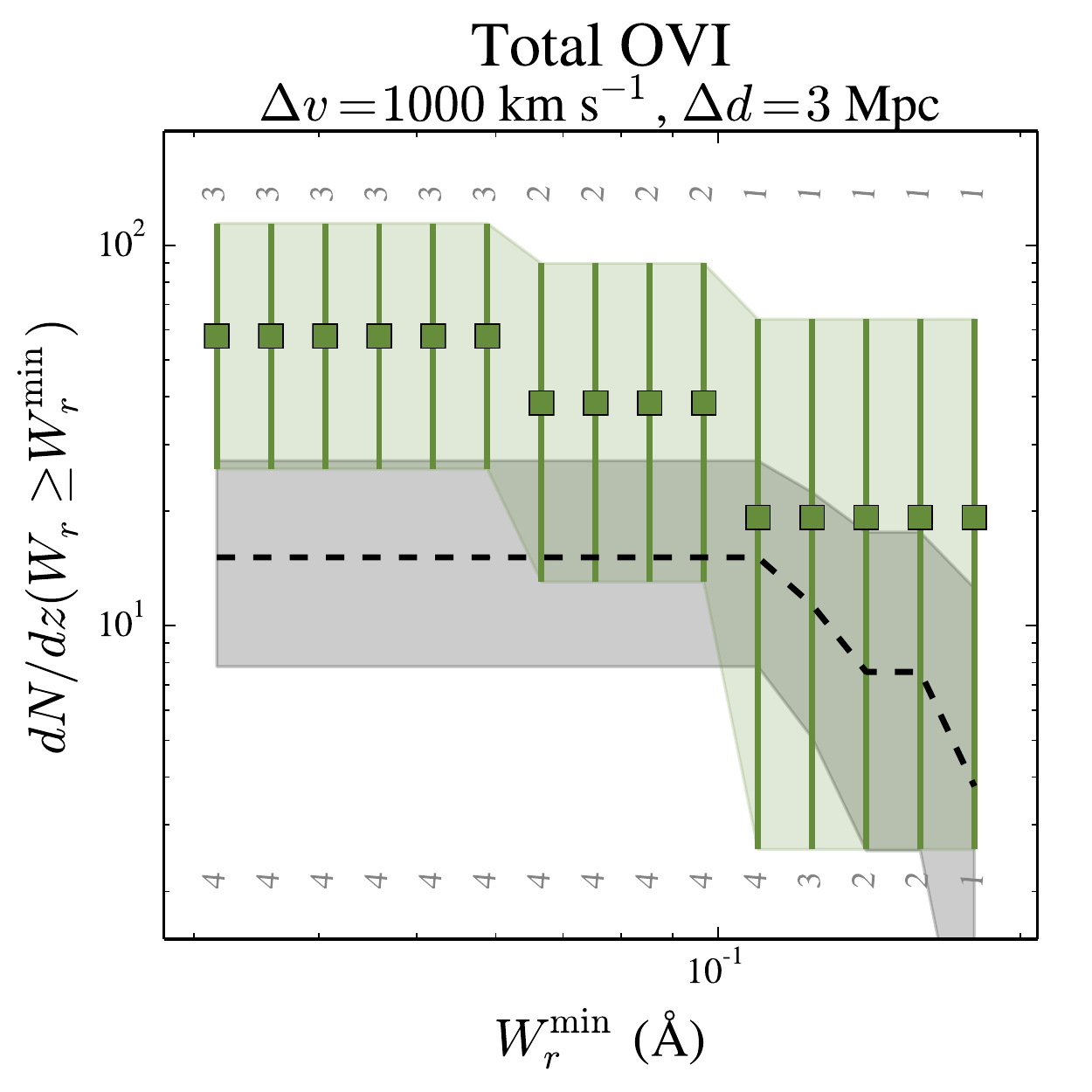}
    \vspace{2ex}
    \end{minipage}

    \begin{minipage}{0.4\textwidth}
    \centering
    \includegraphics[width=1\textwidth]{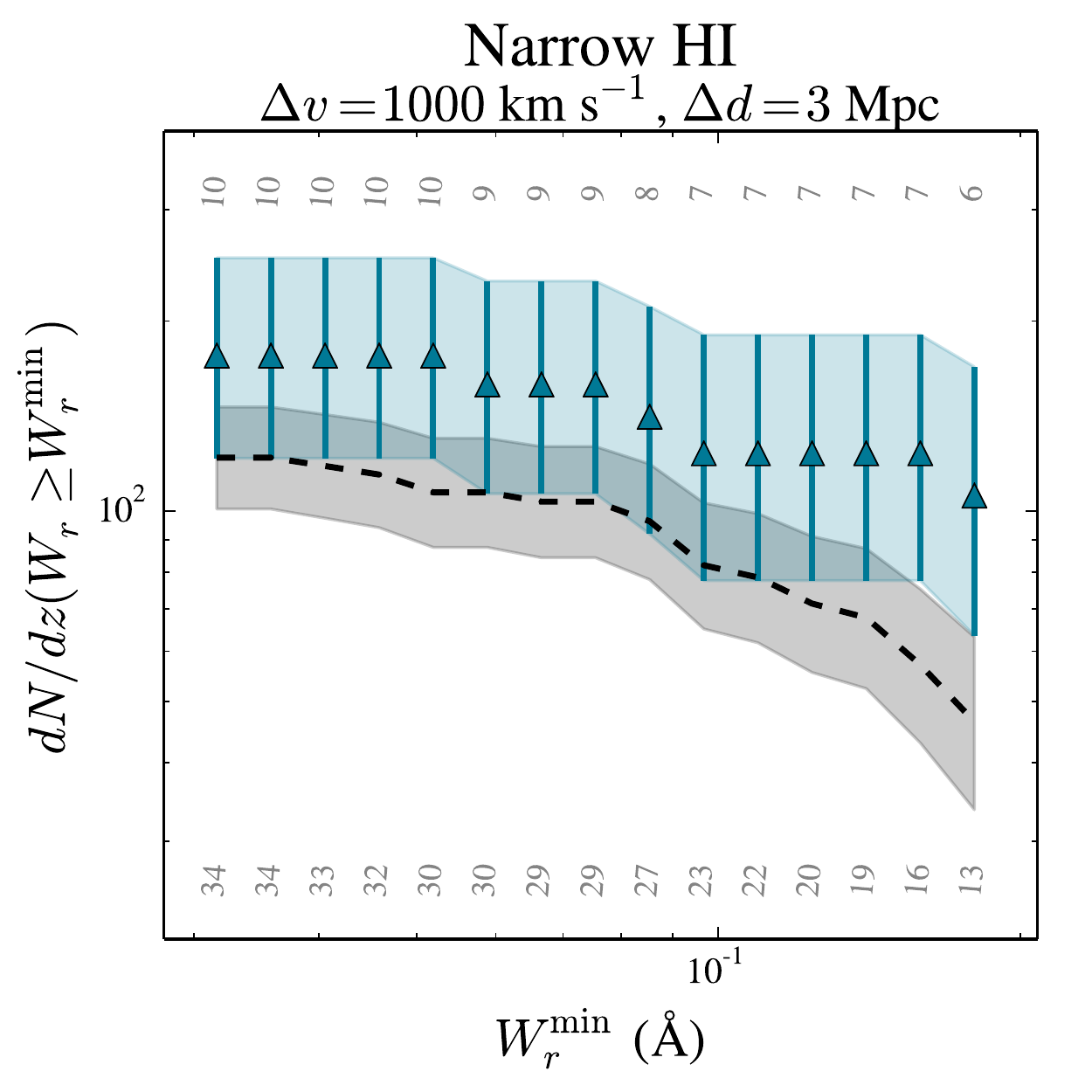}
    \end{minipage}
    \begin{minipage}{0.4\textwidth}
    \centering
    \includegraphics[width=1\textwidth]{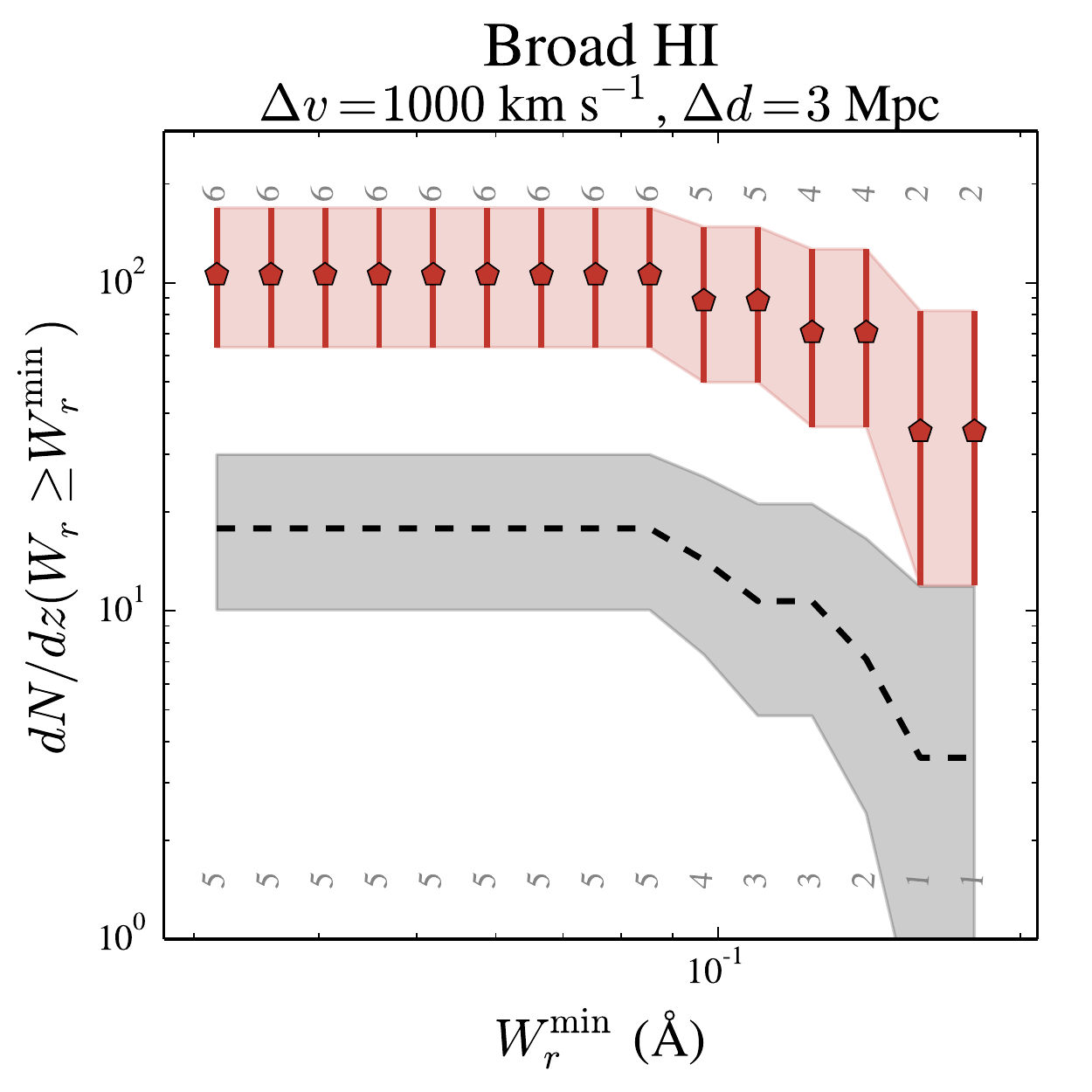}
    \end{minipage}
    
    \caption{Redshift number density of our different samples of
    absorption lines associated to our cluster-pairs (i.e. using our
    fiducial values of $\Delta d =3$\mpc and $\Delta v =1000$\kms),
    as a function of minimum rest-frame equivalent width $W_r^{\rm
    min}$. The top panels show the results for our total \hi (left
    panel) and \ovi (right panel), while the bottom panels show the
    results for narrow (NLA; $b < 50$\kms) and broad (BLA; $b \geq
    50$\kms) \hi, using the same colour/symbol convention as in
    Figure~\ref{fig:dndz_hi_ovi} to \ref{fig:fcs}. Note that bins
    are not independent from each other, as emphasized by the light
    coloured areas. The field distributions estimated from our Q1410
    sightline are represented by the dashed line with its $\pm 1\sigma$
    uncertainty represented by the darker grey region. The total number
    of lines per bin are given in grey numbers on top and bottom of
    each panel, for our cluster-pairs and field samples
    respectively. See \Cref{sec:Wr_dndz} for further details.}
    \label{fig:Wr_dndz} \end{figure*}

In this section we explore how do the equivalent widths distributions
from our sample of lines associated with cluster-pairs (i.e. adopting
our fiducial values of $\Delta d = 3$\mpc and $\Delta v = 1000$\kms),
compare to those of the field expectation.

\Cref{fig:Wr_dndz} shows the redshift number density, $dN/dz$, of our
different samples of absorption lines as a function of minimum
rest-frame equivalent width, $W_{\rm r}^{\rm min}$, associated with our
cluster-pairs: total \hi (top left panel; brown circles), \ovi (top
right panel; green squares), narrow \hi\ (bottom left panel; blue
triangles) and broad \hi\ (bottom right panel; red pentagons). The
field distributions estimated from our Q1410 sightline are represented
by the dashed line with its $\pm 1\sigma$ uncertainty represented by
the darker grey region. Such field estimation comes from lines not
being in the sample associated with cluster pairs, and its redshift
path is estimated by excluding that of cluster-pairs too.

We observe that the equivalent widths distributions are not remarkably
different between lines associated with cluster-pairs and the field
expectation, and hence the relative excess in $dN/dz$ seems to remain
somewhat constant as a function of $W_{\rm r}^{\rm min}$. The
Kolmogorov-Smirnov test (K-S) for the unbinned $W_{\rm r}$
distributions between our cluster-pair and field samples give no
statistically significant differences between them, but we note any
possible real difference would be hard to detect because the samples
are quite small ($\lesssim 10$).

%Similarly than in \Cref{sec:excesses}, in \Cref{fig:excesses_Wr} we
%show the $dN/dz$ excess in our cluster-pair sample with respect to the
%field, as a function of minimum equivalent width. We observe that the
%excess distributions remain almost flat, although a mild trend might be
%present for narrow \hi\ lines.
We conclude that, in terms of equivalent widths, our cluster-pair
absorption line samples and the field expectations are not
significantly different, at least from what can be obtained in our
limited sample.

\section{Q1410 HST/COS spectrum}\label{sec:spectrum}
\Cref{fig:spectrum} shows the reduced Q1410 spectrum (black line), its
corresponding uncertainty (green lines) and our adopted
pseudo-continuum fit (blue dotted line). The figure also shows our
Voigt profile fit solutions and residuals (red lines and grey dots
respectively; see \Cref{sec:voigt_fit}), and their corresponding IDs
and reliability labels (see \Cref{sec:reliability} and
\Cref{tab:abslines}).

\begin{figure*}
 \begin{minipage}{1\textwidth}
    \centering
    \includegraphics[height=21cm]{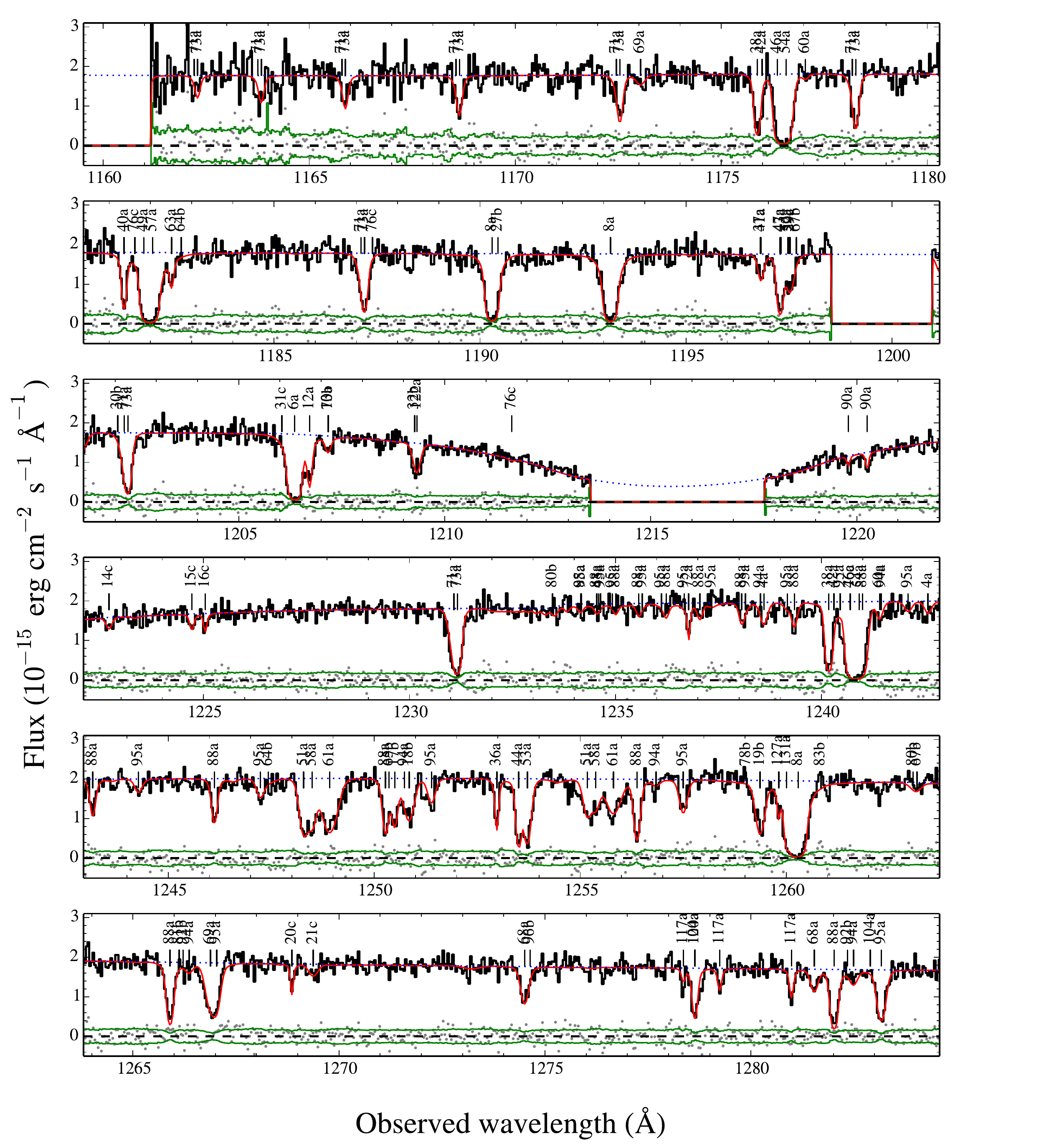}
  \end{minipage}

\caption{Observed \ac{hst}/\ac{cos} \ac{fuv} spectrum of
\ac{qso} Q1410 (black line), and its respective uncertainty (green
lines). The red line correspond to our fit of the spectrum (see
\Cref{sec:voigt_fit}), while the blue dotted line corresponds to the
assumed unabsorbed pseudo-continuum (i.e. including broad emission
lines and the Milky Way's DLA; see \Cref{data:obs}). Vertical tick
lines indicate individual absorption lines, where the numbers
correspond to the IDs given in \Cref{tab:abslines} and the letters
indicate their reliability (see \Cref{sec:reliability}). Grey points
show the difference between our model and the observed data. We see
that distribution of these residuals are consistent with the
uncertainty of the data.}  \label{fig:spectrum}

\end{figure*}

\begin{figure*}
 \begin{minipage}{1\textwidth}
    \centering
    \includegraphics[height=21cm]{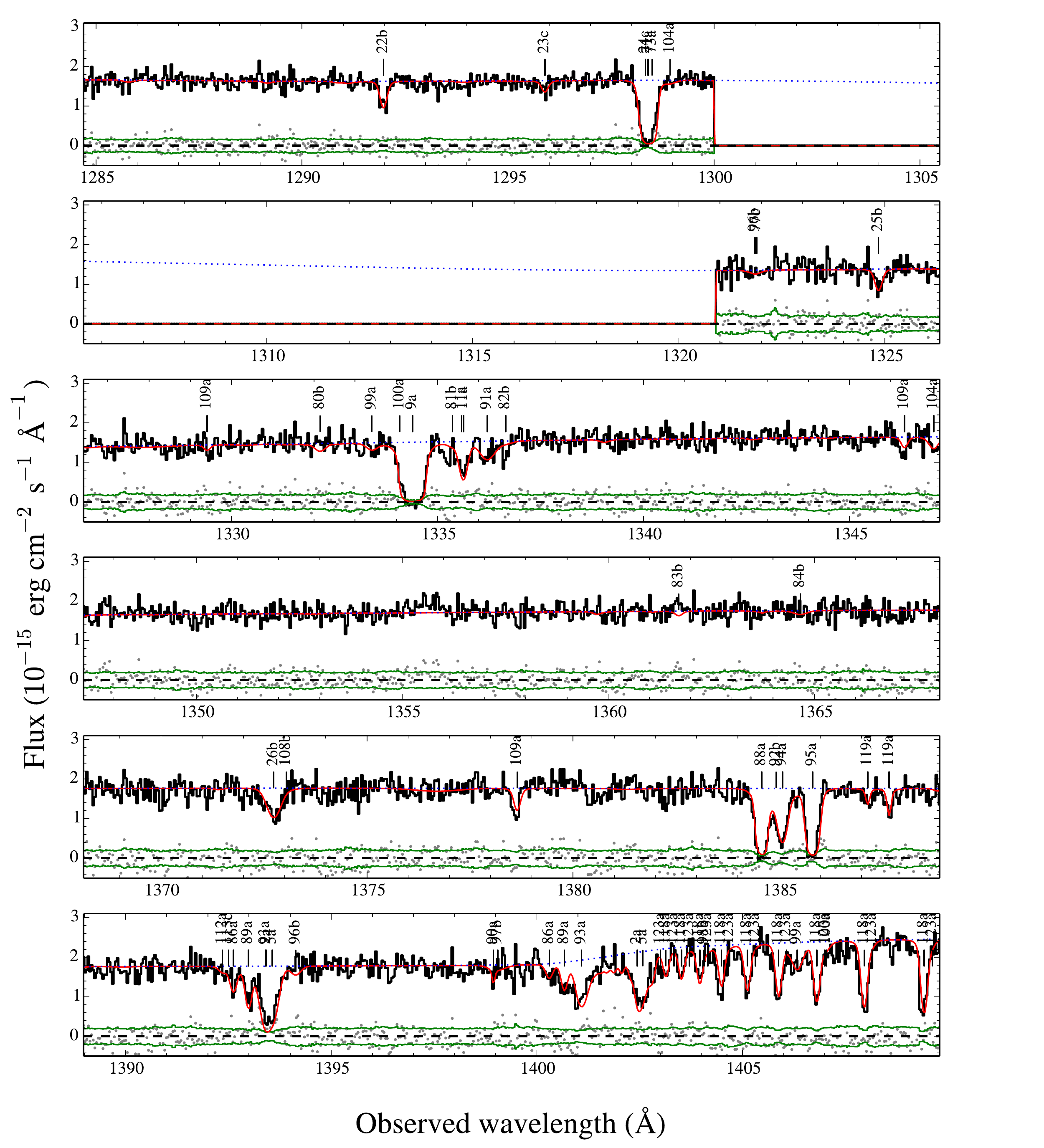}
  \end{minipage}
\contcaption{}
\end{figure*}

\begin{figure*}
 \begin{minipage}{1\textwidth}
    \centering
    \includegraphics[height=21cm]{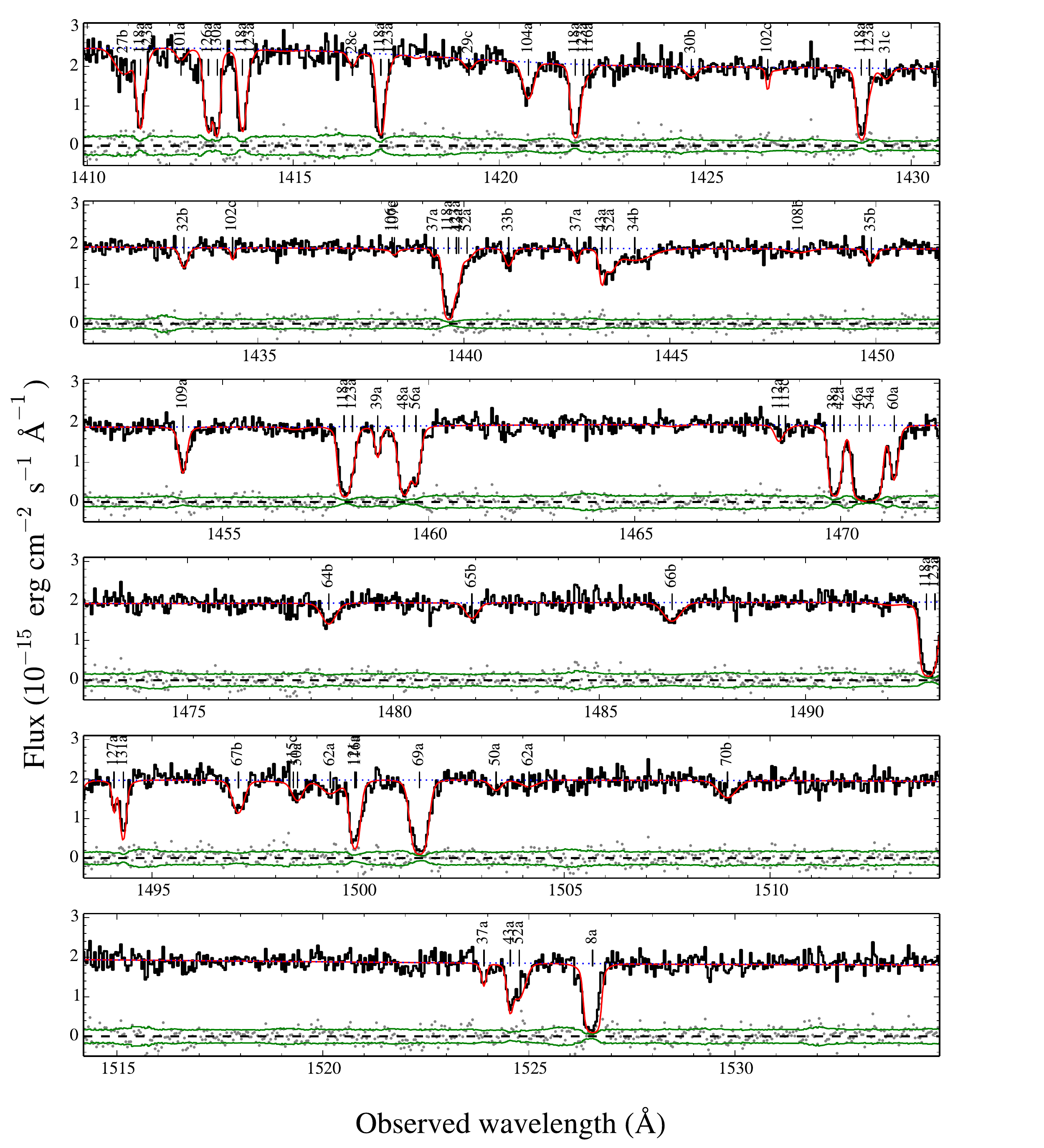}
  \end{minipage}
\contcaption{}
\end{figure*}

 \begin{figure*}
 \begin{minipage}{1\textwidth}
    \centering
    \includegraphics[height=21cm]{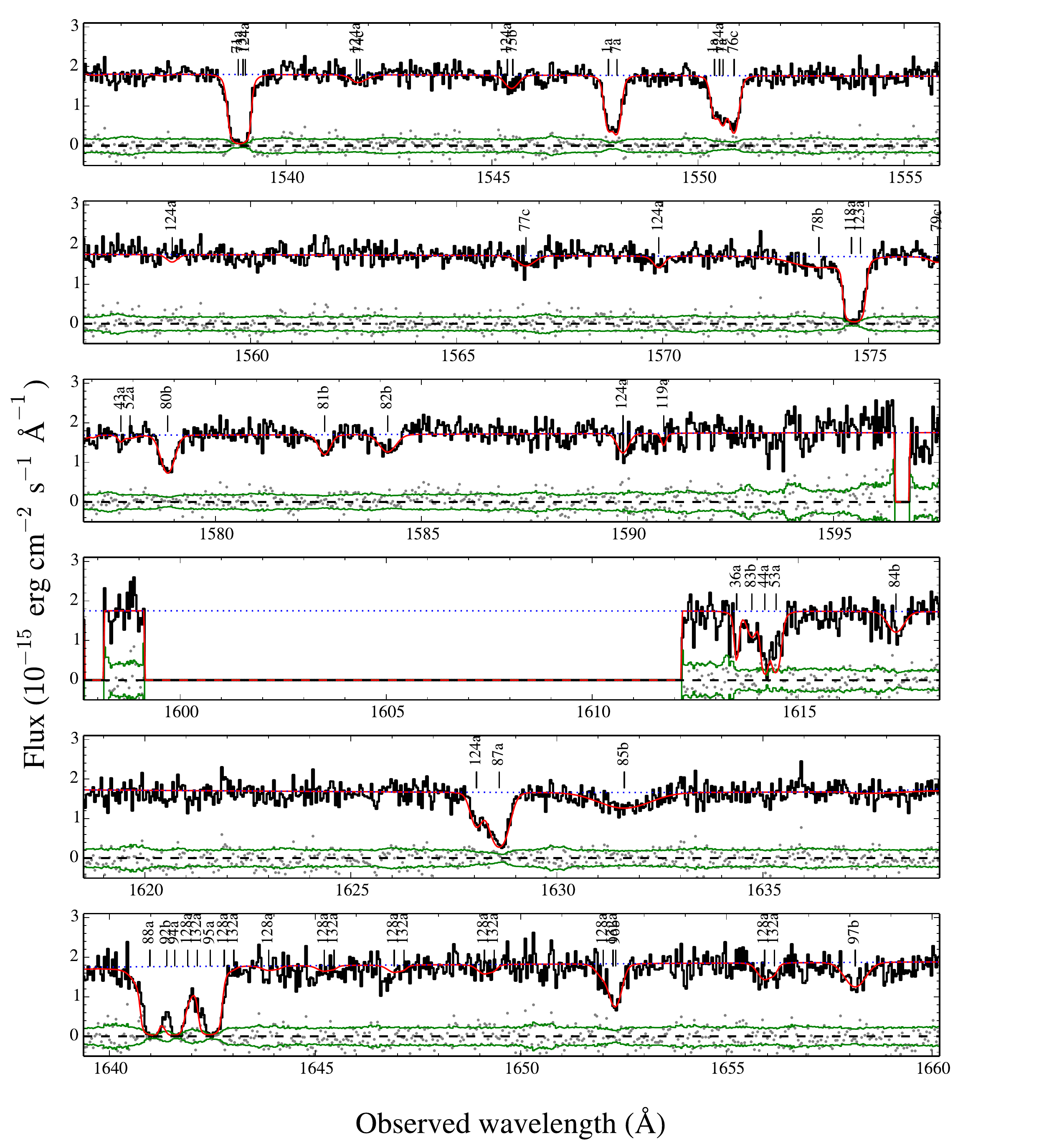}
  \end{minipage}
\contcaption{}
\end{figure*}

\begin{figure*}
 \begin{minipage}{1\textwidth}
    \centering
    \includegraphics[height=21cm]{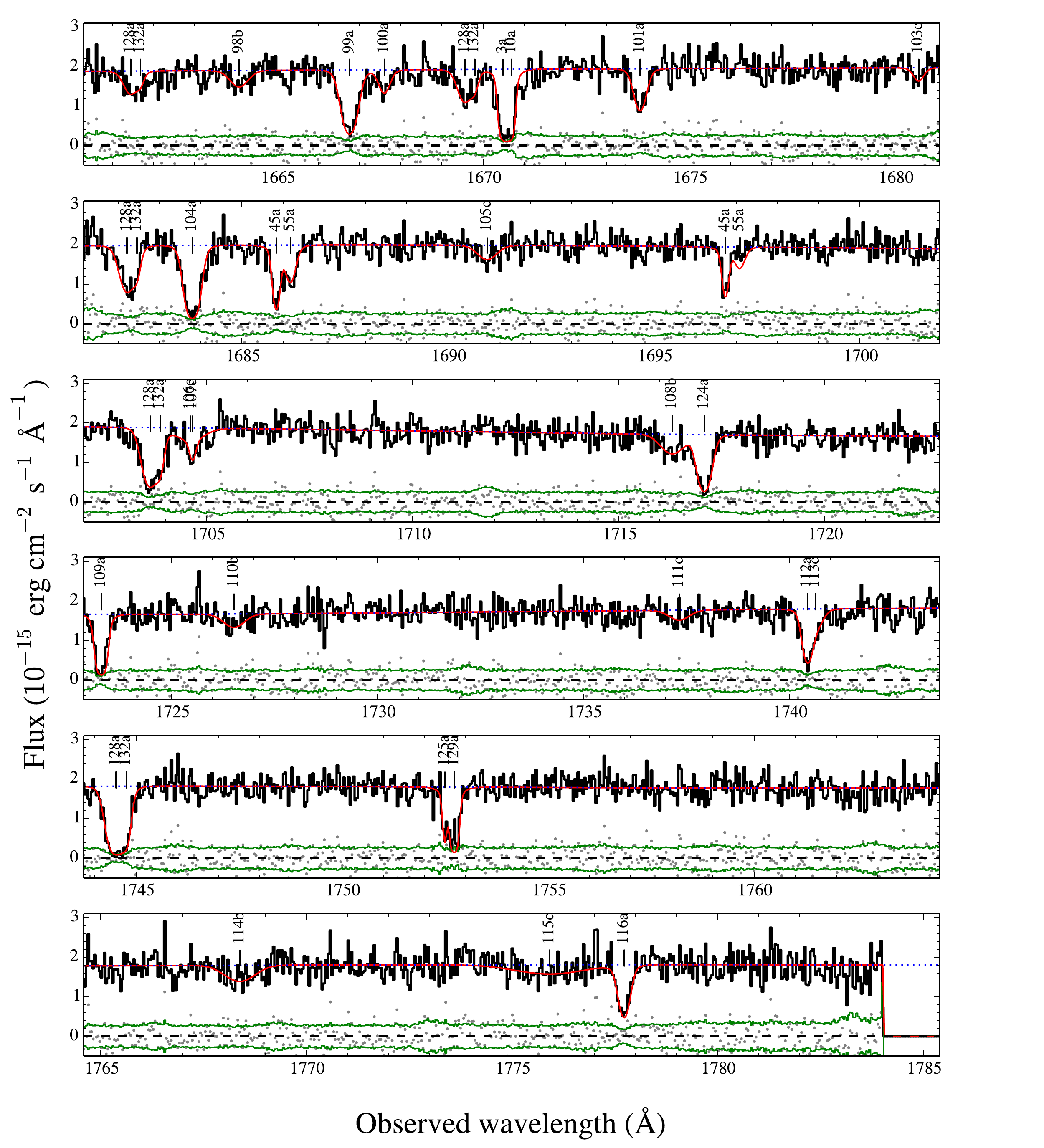}
  \end{minipage}
\contcaption{}
\end{figure*}

\section{Visual association of absorption line systems with cluster-pairs}\label{sec:visual}

\Cref{fig:cpairs_dist_spec_all} shows an schematic view of the
association between \hi\ and \ovi\ absorption line systems and our
known cluster-pairs obtained from the \ac{redmapper} catalog (see
\Cref{sec:field}). These plots are for illustrative purposes only; in
cases with multiple grouped cluster-pairs (see \Cref{sec:grouped}) the
chosen redshifts for defining rest-frames are arbitrary, but such
choice is not used in our scientific analyses.
\begin{figure*}
    \begin{minipage}{1\textwidth}
    \centering
    \includegraphics[width=0.9\textwidth]{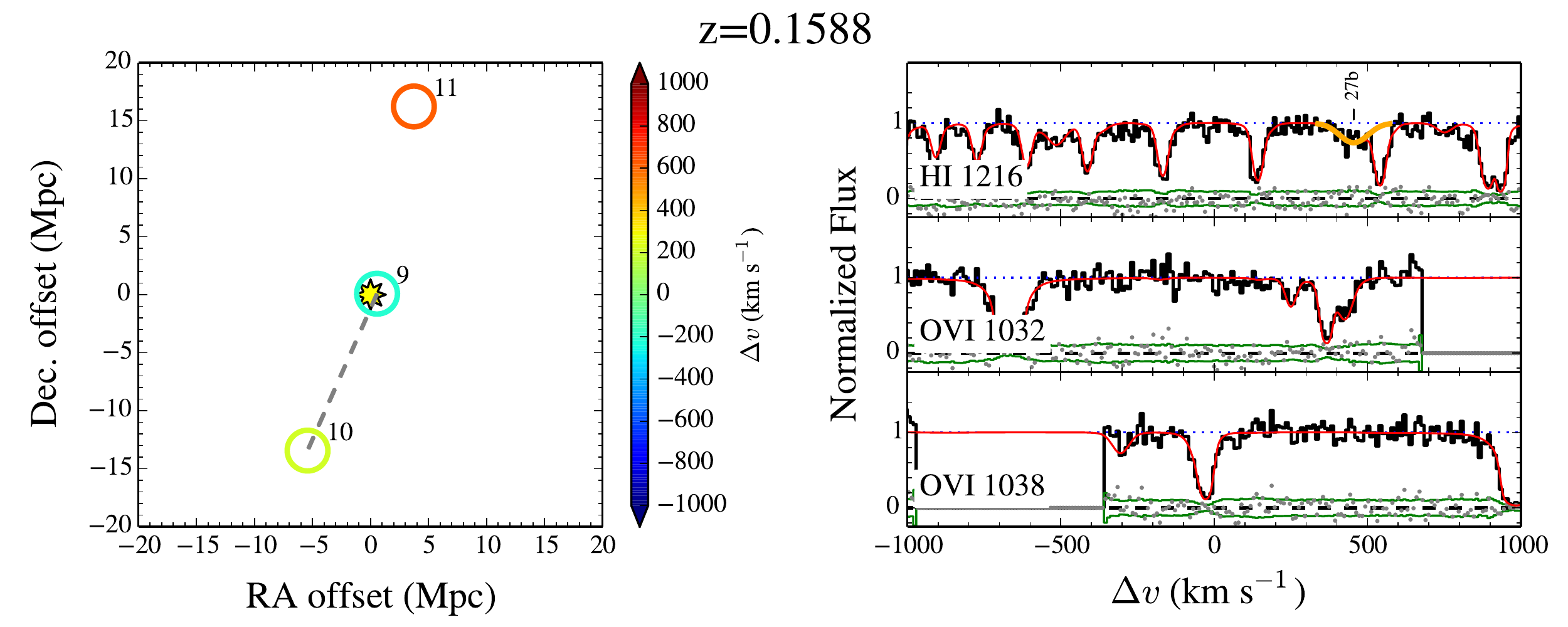}
    \end{minipage}

    \begin{minipage}{1\textwidth}
    \centering
    \includegraphics[width=0.9\textwidth]{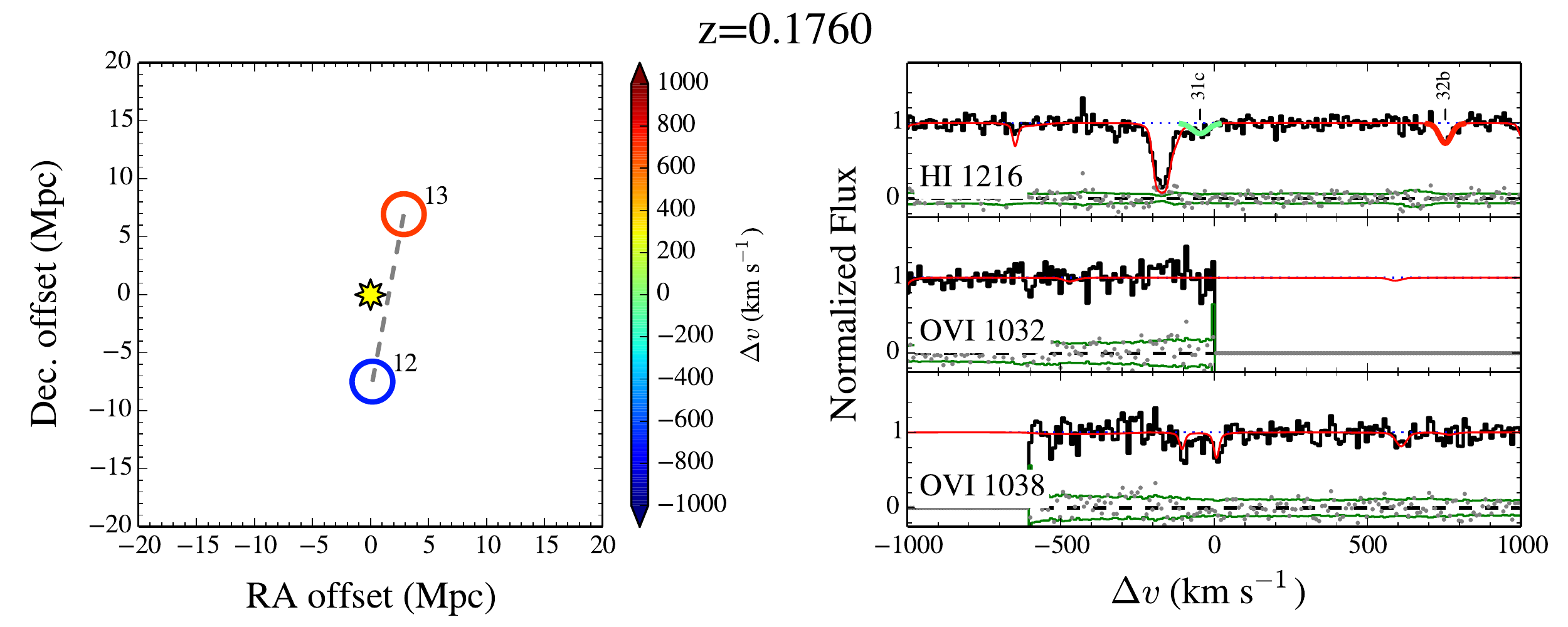}
    \end{minipage}
     
    \begin{minipage}{1\textwidth}
    \centering
    \includegraphics[width=0.9\textwidth]{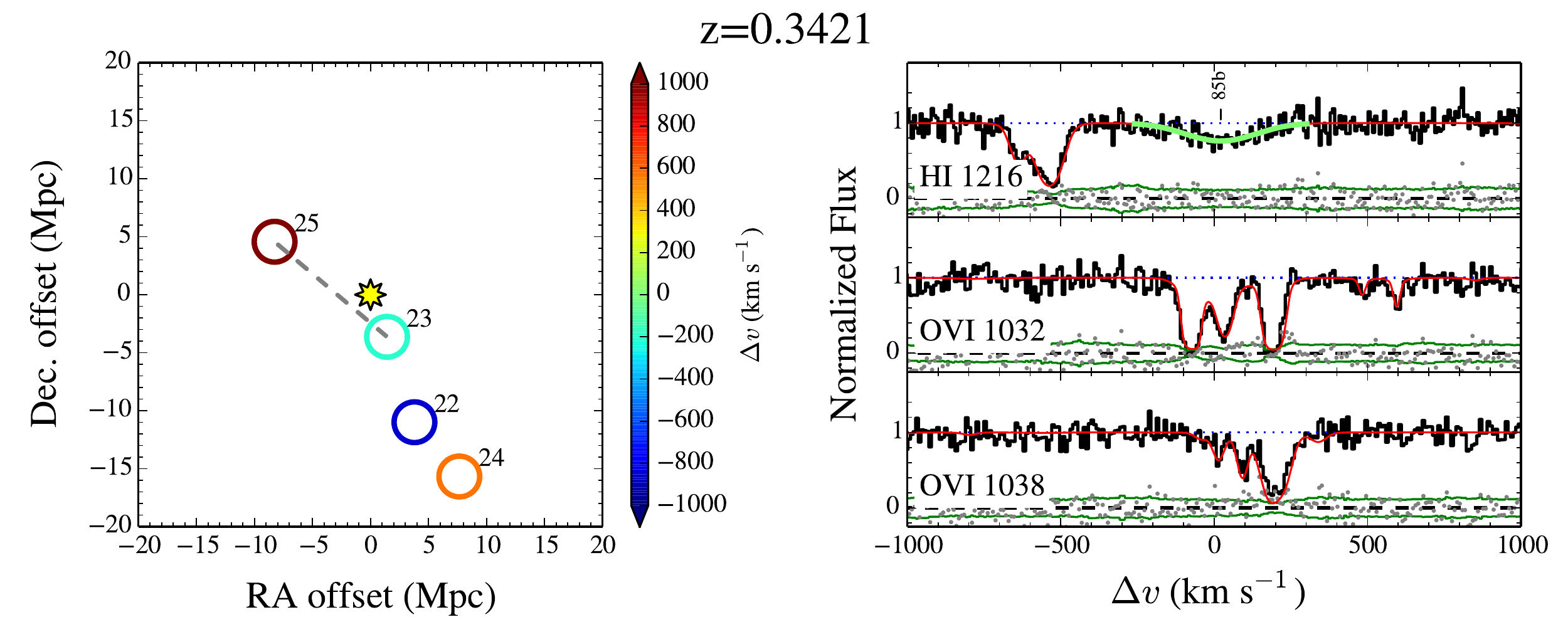}
    \end{minipage}
     
    \caption{Schematic view of the clusters and cluster-pairs around
    the Q1410 sightline at different given redshifts (see top label in
    subpanels). The left panels correspond to the projected in the sky
    distribution of clusters (circles) and cluster-pairs (dashed lines)
    around the Q1410 sightline (yellow star) in co-moving \mpc. Numbers
    close to the circles correspond to the cluster IDs as given in
    \Cref{tab:clusters}. The colour of the circles correspond to the
    rest-frame velocity difference of each cluster with respect to the
    given redshift, according to the colour bar scale on the right of
    the panel. The right panels correspond to portions of the
    normalized Q1410 HST/COS spectrum within a rest-frame velocity
    window of $\pm 1000$\kms from the given redshift, for the \hi\
    \lya\ transition (top) and the \ovi\ $\lambda\lambda
    1032,1038$~\AA transitions (middle and bottom, respectively). The
    spectrum itself is represented by the black lines and the
    uncertainty is represented by the green line. Our fits to
    associated absorption line systems are represented by the thick
    coloured lines according to the colour bar scale on the left of the
    panel, while the fits for unassociated absorption line systems
    (i.e. at different redshifts) are represented by the red
    lines. Absorption line ID and reliability flags are given for
    associated absorption line systems.}\label{fig:cpairs_dist_spec_all}

\end{figure*}

\begin{figure*}
    \begin{minipage}{1\textwidth}
    \centering
    \includegraphics[width=0.85\textwidth]{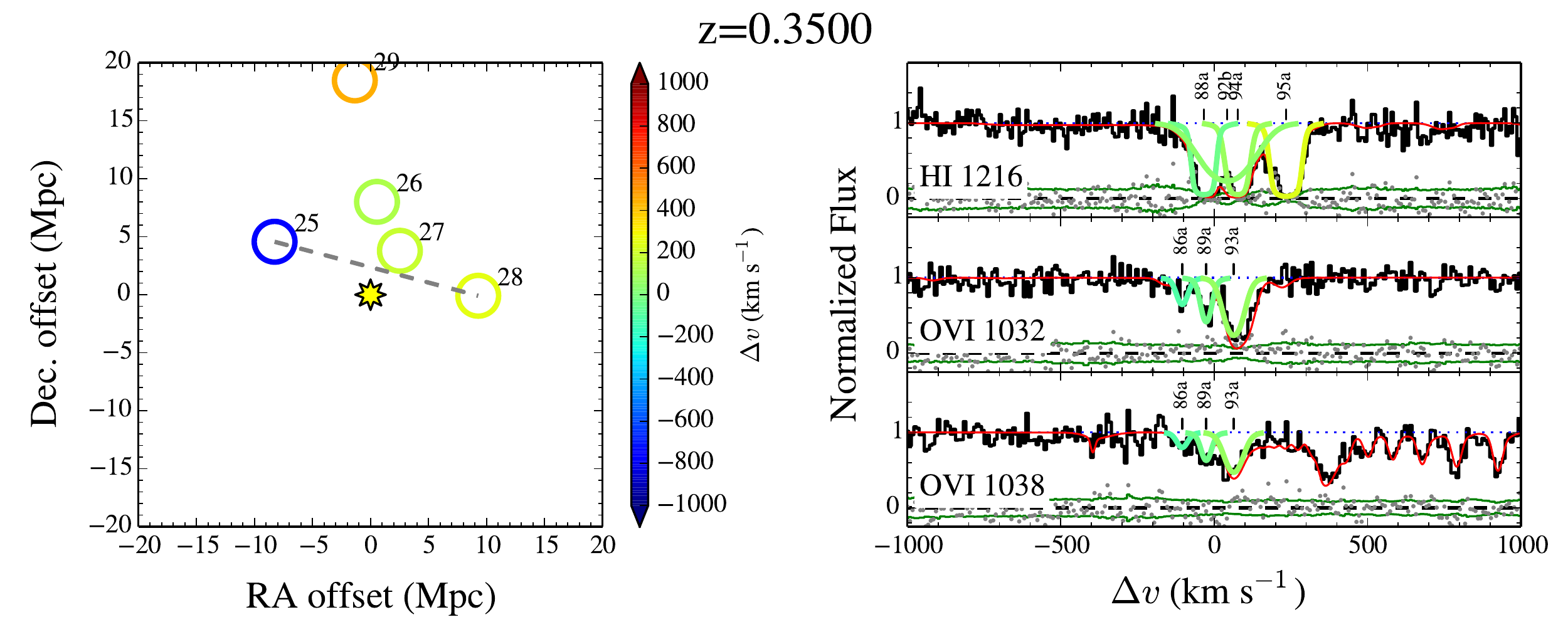}
    \end{minipage}
    \begin{minipage}{1\textwidth}
    \centering
    \includegraphics[width=0.85\textwidth]{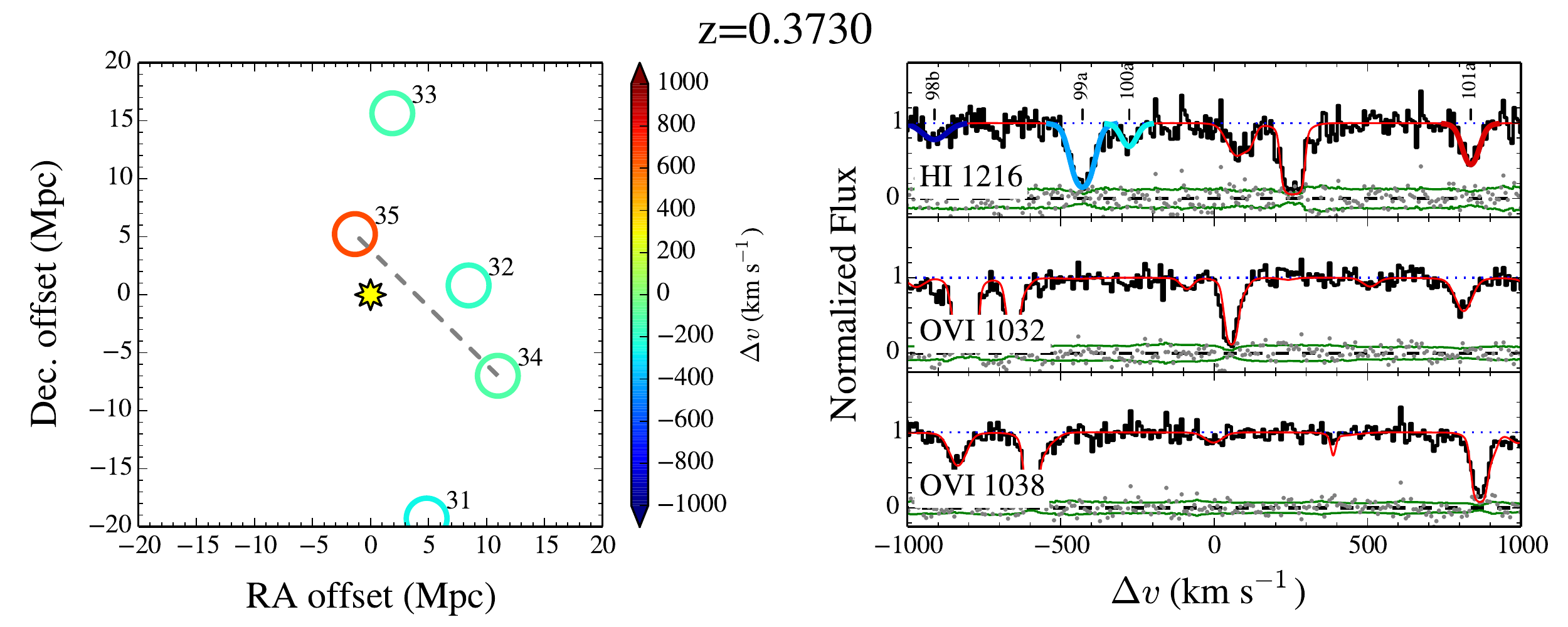}
    \end{minipage}
    \begin{minipage}{1\textwidth}
    \centering
    \includegraphics[width=0.85\textwidth]{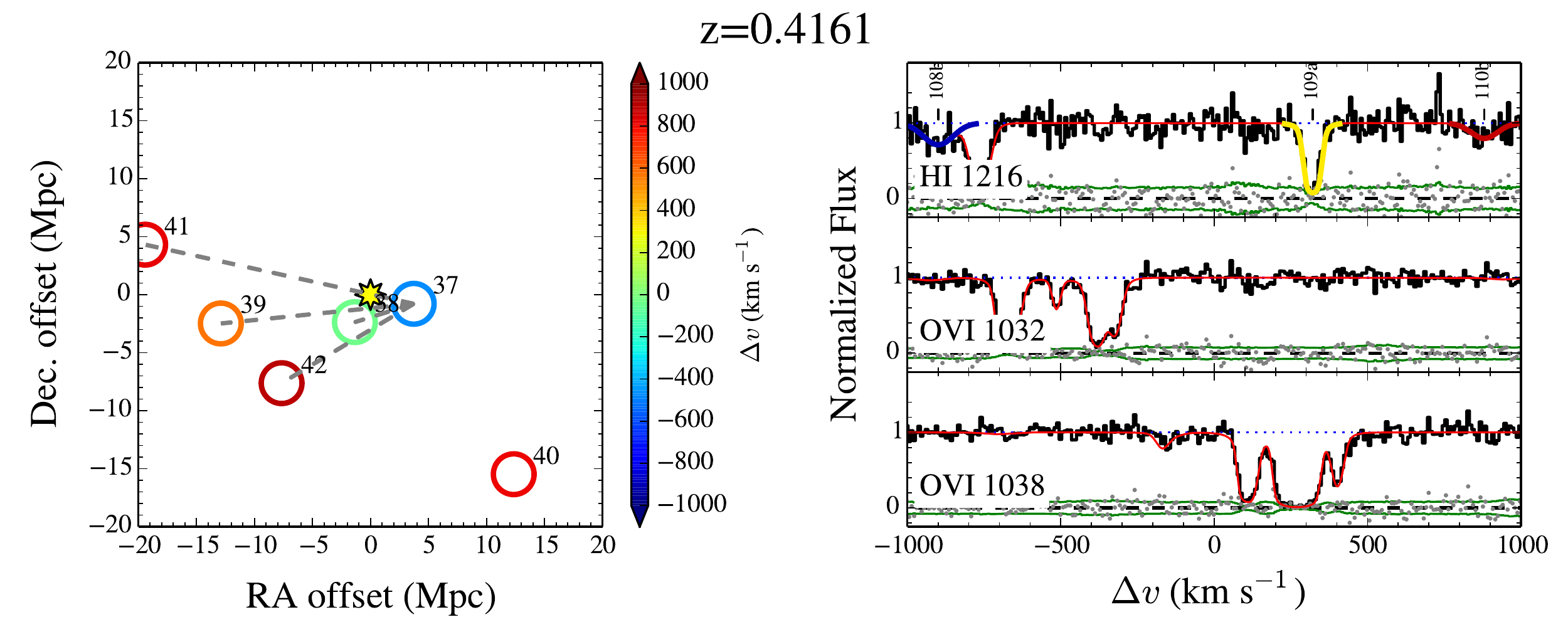}
    \end{minipage}
    %\contcaption{}
%\end{figure*}

%\begin{figure*}
\begin{minipage}{1\textwidth}
    \centering
    \includegraphics[width=0.85\textwidth]{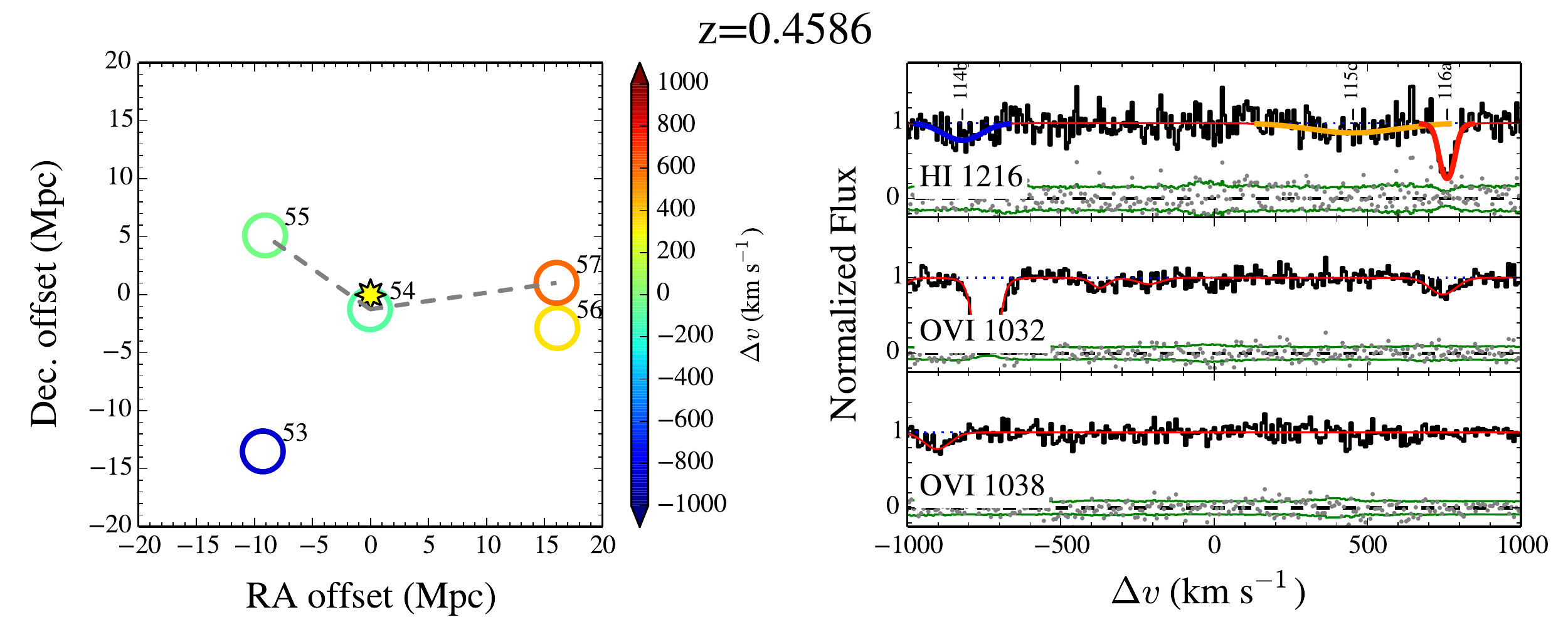}
    \end{minipage}
\contcaption{}
\end{figure*}

\section{Summary tables for our incidence measurements}\label{sec:summary_tables}

Tables~\ref{tab:hi_summary} to \ref{tab:bla_summary} show a summary of
our main $dN/dz$ and $dN/dX$ calculations for our different samples of
absorption lines.

    \begin{table*}
    \centering
    \scriptsize
    \begin{minipage}{0.84\textwidth}
    \centering
    \caption{Summary of relevant quantities for our sample of total \hi\ absorption lines.}\label{tab:hi_summary}
    \begin{tabular}{@{}cccccccccccc@{}}

\hline                            
$\Delta d$ & $\Delta v$ & $W_r^{\rm min}$ &$\Delta z$ & $\Delta X$&  $N$  & $\frac{dN}{dz}$ & $\frac{dN}{dX}$ & $\frac{dN}{dz}|_{\rm field}$ & $\frac{dN}{dX}|_{\rm field}$ &$\frac{dN}{dz}/\frac{dN}{dz}|_{\rm field}$ & $\frac{dN}{dX}/\frac{dN}{dX}|_{\rm field}$ \\ 

(\mpc)     &  (\kms)    &      (\AA)        &            &              &       &                    &                      &                                      &                                      &                                                        &                                                       \\ 

(1)         &  (2)        &      (3)           &      (4)   & (5)          &  (6)  &    (7)             & (8)                  &     (9)                              & (10)                                  &   (11)                                                  &   (12)                                                  \\ 

\hline 

3.0&500&0.04&0.03&0.05&11&$358^{+144}_{-107}$&$238^{+96}_{-71}$&$123^{+26}_{-23}$&$87^{+18}_{-16}$&$2.9^{+1.2}_{-0.9}$&$2.7^{+1.1}_{-0.9}$ \\ 
3.0&630&0.04&0.04&0.06&11&$293^{+118}_{-87}$&$195^{+78}_{-58}$&$123^{+26}_{-23}$&$87^{+18}_{-16}$&$2.4^{+1.0}_{-0.7}$&$2.2^{+0.9}_{-0.7}$ \\ 
3.0&794&0.04&0.05&0.07&14&$305^{+105}_{-81}$&$203^{+70}_{-54}$&$123^{+26}_{-23}$&$87^{+18}_{-16}$&$2.5^{+0.9}_{-0.7}$&$2.3^{+0.8}_{-0.7}$ \\ 
3.0&1000&0.04&0.06&0.08&16&$287^{+91}_{-71}$&$191^{+61}_{-47}$&$123^{+26}_{-23}$&$87^{+18}_{-16}$&$2.3^{+0.8}_{-0.6}$&$2.2^{+0.7}_{-0.6}$ \\ 
3.0&1313&0.04&0.07&0.11&17&$237^{+73}_{-57}$&$159^{+49}_{-38}$&$123^{+26}_{-23}$&$87^{+18}_{-16}$&$1.9^{+0.6}_{-0.5}$&$1.8^{+0.6}_{-0.5}$ \\ 
3.0&1724&0.04&0.09&0.13&17&$188^{+58}_{-45}$&$126^{+39}_{-30}$&$123^{+26}_{-23}$&$87^{+18}_{-16}$&$1.5^{+0.5}_{-0.4}$&$1.5^{+0.5}_{-0.4}$ \\ 
3.0&2264&0.04&0.12&0.17&18&$156^{+46}_{-37}$&$105^{+31}_{-25}$&$123^{+26}_{-23}$&$87^{+18}_{-16}$&$1.3^{+0.4}_{-0.3}$&$1.2^{+0.4}_{-0.3}$ \\ 
3.0&2972&0.04&0.14&0.21&21&$151^{+41}_{-33}$&$101^{+27}_{-22}$&$123^{+26}_{-23}$&$87^{+18}_{-16}$&$1.2^{+0.4}_{-0.3}$&$1.2^{+0.3}_{-0.3}$ \\ 
3.0&3903&0.04&0.16&0.24&25&$152^{+37}_{-30}$&$102^{+25}_{-20}$&$123^{+26}_{-23}$&$87^{+18}_{-16}$&$1.2^{+0.3}_{-0.3}$&$1.2^{+0.3}_{-0.3}$ \\ 
3.0&5125&0.04&0.18&0.27&26&$143^{+34}_{-28}$&$97^{+23}_{-19}$&$123^{+26}_{-23}$&$87^{+18}_{-16}$&$1.2^{+0.3}_{-0.3}$&$1.1^{+0.3}_{-0.2}$ \\ 
3.0&6729&0.04&0.19&0.28&26&$134^{+32}_{-26}$&$91^{+22}_{-18}$&$123^{+26}_{-23}$&$87^{+18}_{-16}$&$1.1^{+0.3}_{-0.2}$&$1.1^{+0.3}_{-0.2}$ \\ 
3.0&8835&0.04&0.22&0.32&32&$147^{+31}_{-26}$&$101^{+21}_{-18}$&$123^{+26}_{-23}$&$87^{+18}_{-16}$&$1.2^{+0.3}_{-0.2}$&$1.2^{+0.3}_{-0.2}$ \\ 
3.0&11601&0.04&0.25&0.36&38&$151^{+29}_{-24}$&$105^{+20}_{-17}$&$123^{+26}_{-23}$&$87^{+18}_{-16}$&$1.2^{+0.3}_{-0.2}$&$1.2^{+0.3}_{-0.2}$ \\ 
3.0&15232&0.04&0.30&0.42&43&$146^{+26}_{-22}$&$103^{+18}_{-16}$&$123^{+26}_{-23}$&$87^{+18}_{-16}$&$1.2^{+0.2}_{-0.2}$&$1.2^{+0.2}_{-0.2}$ \\ 
3.0&20000&0.04&0.33&0.47&48&$145^{+24}_{-21}$&$103^{+17}_{-15}$&$123^{+26}_{-23}$&$87^{+18}_{-16}$&$1.2^{+0.2}_{-0.2}$&$1.2^{+0.2}_{-0.2}$ \\ 
\hline1.0&1000&0.04&0.01&0.02&3&$203^{+199}_{-112}$&$140^{+137}_{-77}$&$123^{+26}_{-23}$&$87^{+18}_{-16}$&$1.6^{+1.6}_{-0.9}$&$1.6^{+1.6}_{-0.9}$ \\ 
1.4&1000&0.04&0.03&0.04&6&$231^{+139}_{-92}$&$150^{+90}_{-60}$&$123^{+26}_{-23}$&$87^{+18}_{-16}$&$1.9^{+1.1}_{-0.8}$&$1.7^{+1.1}_{-0.7}$ \\ 
2.1&1000&0.04&0.04&0.06&8&$184^{+91}_{-64}$&$124^{+61}_{-43}$&$123^{+26}_{-23}$&$87^{+18}_{-16}$&$1.5^{+0.8}_{-0.5}$&$1.4^{+0.7}_{-0.5}$ \\ 
3.0&1000&0.04&0.06&0.08&16&$287^{+91}_{-71}$&$191^{+61}_{-47}$&$123^{+26}_{-23}$&$87^{+18}_{-16}$&$2.3^{+0.8}_{-0.6}$&$2.2^{+0.7}_{-0.6}$ \\ 
4.1&1000&0.04&0.06&0.09&16&$280^{+89}_{-69}$&$187^{+59}_{-46}$&$123^{+26}_{-23}$&$87^{+18}_{-16}$&$2.3^{+0.8}_{-0.6}$&$2.2^{+0.7}_{-0.6}$ \\ 
5.7&1000&0.04&0.06&0.09&16&$279^{+89}_{-69}$&$186^{+59}_{-46}$&$123^{+26}_{-23}$&$87^{+18}_{-16}$&$2.3^{+0.8}_{-0.6}$&$2.1^{+0.7}_{-0.6}$ \\ 
7.8&1000&0.04&0.07&0.10&16&$242^{+77}_{-60}$&$164^{+52}_{-41}$&$123^{+26}_{-23}$&$87^{+18}_{-16}$&$2.0^{+0.7}_{-0.5}$&$1.9^{+0.6}_{-0.5}$ \\ 
10.7&1000&0.04&0.08&0.12&17&$218^{+67}_{-53}$&$147^{+45}_{-35}$&$123^{+26}_{-23}$&$87^{+18}_{-16}$&$1.8^{+0.6}_{-0.5}$&$1.7^{+0.5}_{-0.4}$ \\ 
14.8&1000&0.04&0.09&0.14&19&$205^{+59}_{-47}$&$140^{+40}_{-32}$&$123^{+26}_{-23}$&$87^{+18}_{-16}$&$1.7^{+0.5}_{-0.4}$&$1.6^{+0.5}_{-0.4}$ \\ 
20.3&1000&0.04&0.11&0.16&21&$193^{+52}_{-42}$&$132^{+36}_{-29}$&$123^{+26}_{-23}$&$87^{+18}_{-16}$&$1.6^{+0.5}_{-0.4}$&$1.5^{+0.4}_{-0.4}$ \\ 
27.9&1000&0.04&0.12&0.17&23&$195^{+50}_{-40}$&$133^{+34}_{-28}$&$123^{+26}_{-23}$&$87^{+18}_{-16}$&$1.6^{+0.4}_{-0.4}$&$1.5^{+0.4}_{-0.4}$ \\ 
38.4&1000&0.04&0.15&0.22&27&$180^{+42}_{-35}$&$124^{+29}_{-24}$&$123^{+26}_{-23}$&$87^{+18}_{-16}$&$1.5^{+0.4}_{-0.3}$&$1.4^{+0.4}_{-0.3}$ \\ 
52.9&1000&0.04&0.17&0.24&27&$160^{+37}_{-31}$&$111^{+26}_{-21}$&$123^{+26}_{-23}$&$87^{+18}_{-16}$&$1.3^{+0.3}_{-0.3}$&$1.3^{+0.3}_{-0.3}$ \\ 
72.7&1000&0.04&0.19&0.28&29&$150^{+33}_{-28}$&$105^{+23}_{-19}$&$123^{+26}_{-23}$&$87^{+18}_{-16}$&$1.2^{+0.3}_{-0.3}$&$1.2^{+0.3}_{-0.3}$ \\ 
100.0&1000&0.04&0.24&0.35&39&$160^{+30}_{-25}$&$113^{+21}_{-18}$&$123^{+26}_{-23}$&$87^{+18}_{-16}$&$1.3^{+0.3}_{-0.2}$&$1.3^{+0.3}_{-0.2}$ \\ 
\hline3.0&1000&0.03&0.06&0.09&16&$282^{+90}_{-70}$&$188^{+60}_{-47}$&$139^{+26}_{-22}$&$100^{+19}_{-16}$&$2.0^{+0.7}_{-0.6}$&$1.9^{+0.7}_{-0.6}$ \\ 
3.0&1000&0.04&0.06&0.09&16&$282^{+90}_{-70}$&$188^{+60}_{-47}$&$139^{+26}_{-22}$&$100^{+19}_{-16}$&$2.0^{+0.7}_{-0.6}$&$1.9^{+0.7}_{-0.6}$ \\ 
3.0&1000&0.04&0.06&0.09&16&$282^{+90}_{-70}$&$188^{+60}_{-47}$&$136^{+26}_{-22}$&$97^{+18}_{-16}$&$2.1^{+0.8}_{-0.6}$&$1.9^{+0.7}_{-0.6}$ \\ 
3.0&1000&0.05&0.06&0.09&16&$282^{+90}_{-70}$&$188^{+60}_{-47}$&$132^{+26}_{-22}$&$94^{+18}_{-15}$&$2.1^{+0.8}_{-0.7}$&$2.0^{+0.7}_{-0.6}$ \\ 
3.0&1000&0.05&0.06&0.09&16&$282^{+90}_{-70}$&$188^{+60}_{-47}$&$125^{+25}_{-21}$&$89^{+18}_{-15}$&$2.3^{+0.8}_{-0.7}$&$2.1^{+0.8}_{-0.7}$ \\ 
3.0&1000&0.06&0.06&0.09&15&$265^{+88}_{-68}$&$176^{+58}_{-45}$&$125^{+25}_{-21}$&$89^{+18}_{-15}$&$2.1^{+0.8}_{-0.7}$&$2.0^{+0.7}_{-0.6}$ \\ 
3.0&1000&0.07&0.06&0.09&15&$265^{+88}_{-68}$&$176^{+58}_{-45}$&$121^{+25}_{-21}$&$87^{+18}_{-15}$&$2.2^{+0.8}_{-0.7}$&$2.0^{+0.8}_{-0.6}$ \\ 
3.0&1000&0.08&0.06&0.09&15&$265^{+88}_{-68}$&$176^{+58}_{-45}$&$121^{+25}_{-21}$&$87^{+18}_{-15}$&$2.2^{+0.8}_{-0.7}$&$2.0^{+0.8}_{-0.6}$ \\ 
3.0&1000&0.09&0.06&0.09&14&$247^{+85}_{-65}$&$165^{+57}_{-44}$&$114^{+24}_{-20}$&$82^{+17}_{-14}$&$2.2^{+0.9}_{-0.7}$&$2.0^{+0.8}_{-0.7}$ \\ 
3.0&1000&0.10&0.06&0.09&12&$212^{+81}_{-60}$&$141^{+54}_{-40}$&$96^{+22}_{-18}$&$69^{+16}_{-13}$&$2.2^{+1.0}_{-0.8}$&$2.0^{+0.9}_{-0.7}$ \\ 
3.0&1000&0.11&0.06&0.09&12&$212^{+81}_{-60}$&$141^{+54}_{-40}$&$89^{+22}_{-18}$&$64^{+15}_{-13}$&$2.4^{+1.0}_{-0.9}$&$2.2^{+1.0}_{-0.8}$ \\ 
3.0&1000&0.12&0.06&0.09&11&$194^{+78}_{-58}$&$129^{+52}_{-39}$&$82^{+21}_{-17}$&$59^{+15}_{-12}$&$2.4^{+1.1}_{-0.9}$&$2.2^{+1.0}_{-0.8}$ \\ 
3.0&1000&0.14&0.06&0.09&11&$194^{+78}_{-58}$&$129^{+52}_{-39}$&$75^{+20}_{-16}$&$54^{+14}_{-12}$&$2.6^{+1.2}_{-1.0}$&$2.4^{+1.1}_{-0.9}$ \\ 
3.0&1000&0.16&0.06&0.09&9&$159^{+73}_{-52}$&$106^{+48}_{-35}$&$61^{+19}_{-15}$&$43^{+13}_{-10}$&$2.6^{+1.4}_{-1.1}$&$2.4^{+1.3}_{-1.0}$ \\ 
3.0&1000&0.18&0.06&0.09&8&$141^{+70}_{-49}$&$94^{+47}_{-33}$&$50^{+17}_{-13}$&$36^{+12}_{-9}$&$2.8^{+1.6}_{-1.3}$&$2.6^{+1.5}_{-1.2}$ \\ 

\hline
\end{tabular}
    
  \end{minipage}
\begin{minipage}{0.84\textwidth}
(1) Maximum transverse separation between cluster-pair axes and the Q1410 sightline (see \Cref{sec:dndz_method}).
(2) Maximum velocity difference to any cluster-pair within $\Delta d$ from the Q1410 sightline (see \Cref{sec:dndz_method}).
(3) Minimum rest-frame equivalent width (see \Cref{sec:dndz_method}).
(4) Redshift path (see \Cref{sec:dndz_method} and \Cref{sec:dz}).
(5) Absorption distance (see \Cref{sec:dz}).
(6) Total number of absorption components having $W_{\rm r} \ge W_{\rm r}^{\rm min}$ within $\Delta d$ and $\Delta v$ from cluster-pairs (see \Cref{sec:dndz_method}).
(7) Redshift number density associated to absorption components having $W_{\rm r} \ge W_{\rm r}^{\rm min}$ within $\Delta d$ and $\Delta v$ from cluster-pairs (see \Cref{sec:dndz_method}).
(8) Absorption distance number density to absorption components having $W_{\rm r} \ge W_{\rm r}^{\rm min}$ within $\Delta d$ and $\Delta v$ from cluster-pairs (analogous to $dN/dz$ but see also \Cref{sec:dz})
(9) Field expectation for the redshift number density (see \Cref{sec:dndz_field}).
(10) Field expectation for the absorption distance number density (see \Cref{sec:dndz_field}).
(11) Excess of the measured redshift number density compared to the field expectation (see \Cref{sec:excesses}).
(12) Excess of the measured absorption distance number density compared to the field expectation (see \Cref{sec:excesses}).

\end{minipage}
\end{table*}

    \begin{table*}
    \centering
    \scriptsize
    \begin{minipage}{0.84\textwidth}
    \centering
    \caption{Summary of relevant quantities for our sample of total \ovi\ absorption lines.}\label{tab:ovi_summary}
    \begin{tabular}{@{}cccccccccccc@{}}

\hline                            
$\Delta d$ & $\Delta v$ & $W_r^{\rm min}$ &$\Delta z$ & $\Delta X$&  $N$  & $\frac{dN}{dz}$ & $\frac{dN}{dX}$ & $\frac{dN}{dz}|_{\rm field}$ & $\frac{dN}{dX}|_{\rm field}$ &$\frac{dN}{dz}/\frac{dN}{dz}|_{\rm field}$ & $\frac{dN}{dX}/\frac{dN}{dX}|_{\rm field}$ \\ 

(\mpc)     &  (\kms)    &      (\AA)        &            &              &       &                    &                      &                                      &                                      &                                                        &                                                       \\ 

(1)         &  (2)        &      (3)           &      (4)   & (5)          &  (6)  &    (7)             & (8)                  &     (9)                              & (10)                                  &   (11)                                                  &   (12)                                                  \\ 

\hline 

3.0&500&0.06&0.03&0.05&3&$100^{+98}_{-55}$&$66^{+64}_{-36}$&$14^{+12}_{-8}$&$10^{+8}_{-6}$&$7.1^{+7.4}_{-4.7}$&$6.9^{+7.2}_{-4.5}$ \\ 
3.0&630&0.06&0.04&0.06&3&$82^{+80}_{-45}$&$54^{+53}_{-30}$&$14^{+12}_{-8}$&$10^{+8}_{-6}$&$5.8^{+6.1}_{-3.8}$&$5.6^{+5.9}_{-3.7}$ \\ 
3.0&794&0.06&0.04&0.07&3&$70^{+69}_{-39}$&$46^{+45}_{-26}$&$14^{+12}_{-8}$&$10^{+8}_{-6}$&$5.0^{+5.2}_{-3.3}$&$4.8^{+5.0}_{-3.2}$ \\ 
3.0&1000&0.06&0.05&0.08&3&$58^{+57}_{-32}$&$38^{+37}_{-21}$&$14^{+12}_{-8}$&$10^{+8}_{-6}$&$4.1^{+4.3}_{-2.7}$&$4.0^{+4.2}_{-2.6}$ \\ 
3.0&1313&0.06&0.07&0.10&3&$45^{+45}_{-25}$&$30^{+29}_{-17}$&$14^{+12}_{-8}$&$10^{+8}_{-6}$&$3.2^{+3.4}_{-2.1}$&$3.1^{+3.3}_{-2.1}$ \\ 
3.0&1724&0.06&0.09&0.13&3&$35^{+34}_{-19}$&$23^{+23}_{-13}$&$14^{+12}_{-8}$&$10^{+8}_{-6}$&$2.5^{+2.6}_{-1.6}$&$2.4^{+2.5}_{-1.6}$ \\ 
3.0&2264&0.06&0.11&0.17&3&$27^{+26}_{-15}$&$18^{+17}_{-10}$&$14^{+12}_{-8}$&$10^{+8}_{-6}$&$1.9^{+2.0}_{-1.3}$&$1.9^{+1.9}_{-1.2}$ \\ 
3.0&2972&0.06&0.14&0.21&3&$22^{+21}_{-12}$&$14^{+14}_{-8}$&$14^{+12}_{-8}$&$10^{+8}_{-6}$&$1.5^{+1.6}_{-1.0}$&$1.5^{+1.6}_{-1.0}$ \\ 
3.0&3903&0.06&0.17&0.26&3&$17^{+17}_{-10}$&$12^{+11}_{-6}$&$14^{+12}_{-8}$&$10^{+8}_{-6}$&$1.2^{+1.3}_{-0.8}$&$1.2^{+1.3}_{-0.8}$ \\ 
3.0&5125&0.06&0.20&0.30&3&$15^{+15}_{-8}$&$10^{+10}_{-5}$&$14^{+12}_{-8}$&$10^{+8}_{-6}$&$1.1^{+1.1}_{-0.7}$&$1.0^{+1.1}_{-0.7}$ \\ 
3.0&6729&0.06&0.23&0.34&3&$13^{+13}_{-7}$&$9^{+9}_{-5}$&$14^{+12}_{-8}$&$10^{+8}_{-6}$&$0.9^{+1.0}_{-0.6}$&$0.9^{+1.0}_{-0.6}$ \\ 
3.0&8835&0.06&0.25&0.38&6&$24^{+14}_{-9}$&$16^{+9}_{-6}$&$14^{+12}_{-8}$&$10^{+8}_{-6}$&$1.7^{+1.2}_{-0.9}$&$1.7^{+1.2}_{-0.9}$ \\ 
3.0&11601&0.06&0.27&0.40&6&$22^{+13}_{-9}$&$15^{+9}_{-6}$&$14^{+12}_{-8}$&$10^{+8}_{-6}$&$1.6^{+1.1}_{-0.9}$&$1.6^{+1.1}_{-0.8}$ \\ 
3.0&15232&0.06&0.29&0.44&7&$24^{+13}_{-9}$&$16^{+9}_{-6}$&$14^{+12}_{-8}$&$10^{+8}_{-6}$&$1.7^{+1.1}_{-0.9}$&$1.7^{+1.1}_{-0.9}$ \\ 
3.0&20000&0.06&0.32&0.46&7&$22^{+12}_{-8}$&$15^{+8}_{-6}$&$14^{+12}_{-8}$&$10^{+8}_{-6}$&$1.6^{+1.0}_{-0.8}$&$1.6^{+1.0}_{-0.8}$ \\ 
\hline1.0&1000&0.06&0.01&0.02&0&$0^{+132}_{-0}$&$0^{+91}_{-0}$&$14^{+12}_{-8}$&$10^{+8}_{-6}$&$0.0^{+9.4}_{-0.0}$&$0.0^{+9.5}_{-0.0}$ \\ 
1.4&1000&0.06&0.03&0.04&0&$0^{+73}_{-0}$&$0^{+47}_{-0}$&$14^{+12}_{-8}$&$10^{+8}_{-6}$&$0.0^{+5.2}_{-0.0}$&$0.0^{+4.9}_{-0.0}$ \\ 
2.1&1000&0.06&0.04&0.06&0&$0^{+48}_{-0}$&$0^{+32}_{-0}$&$14^{+12}_{-8}$&$10^{+8}_{-6}$&$0.0^{+3.4}_{-0.0}$&$0.0^{+3.3}_{-0.0}$ \\ 
3.0&1000&0.06&0.05&0.08&3&$58^{+57}_{-32}$&$38^{+37}_{-21}$&$14^{+12}_{-8}$&$10^{+8}_{-6}$&$4.1^{+4.3}_{-2.7}$&$4.0^{+4.2}_{-2.6}$ \\ 
4.1&1000&0.06&0.05&0.08&3&$56^{+55}_{-31}$&$37^{+36}_{-20}$&$14^{+12}_{-8}$&$10^{+8}_{-6}$&$4.0^{+4.2}_{-2.6}$&$3.9^{+4.0}_{-2.6}$ \\ 
5.7&1000&0.06&0.05&0.08&3&$56^{+55}_{-31}$&$37^{+36}_{-20}$&$14^{+12}_{-8}$&$10^{+8}_{-6}$&$4.0^{+4.1}_{-2.6}$&$3.8^{+4.0}_{-2.5}$ \\ 
7.8&1000&0.06&0.06&0.09&3&$48^{+47}_{-27}$&$32^{+32}_{-18}$&$14^{+12}_{-8}$&$10^{+8}_{-6}$&$3.4^{+3.6}_{-2.3}$&$3.4^{+3.5}_{-2.2}$ \\ 
10.7&1000&0.06&0.07&0.11&3&$40^{+39}_{-22}$&$27^{+26}_{-15}$&$14^{+12}_{-8}$&$10^{+8}_{-6}$&$2.8^{+3.0}_{-1.9}$&$2.8^{+2.9}_{-1.8}$ \\ 
14.8&1000&0.06&0.09&0.13&4&$45^{+36}_{-22}$&$31^{+24}_{-15}$&$14^{+12}_{-8}$&$10^{+8}_{-6}$&$3.2^{+2.8}_{-1.9}$&$3.2^{+2.8}_{-1.9}$ \\ 
20.3&1000&0.06&0.10&0.15&4&$41^{+32}_{-20}$&$28^{+22}_{-13}$&$14^{+12}_{-8}$&$10^{+8}_{-6}$&$2.9^{+2.5}_{-1.8}$&$2.9^{+2.5}_{-1.7}$ \\ 
27.9&1000&0.06&0.11&0.16&4&$37^{+29}_{-18}$&$25^{+20}_{-12}$&$14^{+12}_{-8}$&$10^{+8}_{-6}$&$2.6^{+2.3}_{-1.6}$&$2.6^{+2.3}_{-1.6}$ \\ 
38.4&1000&0.06&0.13&0.20&4&$30^{+24}_{-14}$&$20^{+16}_{-10}$&$14^{+12}_{-8}$&$10^{+8}_{-6}$&$2.1^{+1.9}_{-1.3}$&$2.1^{+1.8}_{-1.3}$ \\ 
52.9&1000&0.06&0.16&0.24&4&$25^{+20}_{-12}$&$17^{+13}_{-8}$&$14^{+12}_{-8}$&$10^{+8}_{-6}$&$1.8^{+1.5}_{-1.1}$&$1.7^{+1.5}_{-1.1}$ \\ 
72.7&1000&0.06&0.19&0.28&4&$21^{+17}_{-10}$&$14^{+11}_{-7}$&$14^{+12}_{-8}$&$10^{+8}_{-6}$&$1.5^{+1.3}_{-0.9}$&$1.5^{+1.3}_{-0.9}$ \\ 
100.0&1000&0.06&0.25&0.37&7&$28^{+15}_{-11}$&$19^{+10}_{-7}$&$14^{+12}_{-8}$&$10^{+8}_{-6}$&$2.0^{+1.3}_{-1.0}$&$2.0^{+1.3}_{-1.0}$ \\ 
\hline3.0&1000&0.03&0.05&0.08&3&$58^{+57}_{-32}$&$38^{+37}_{-21}$&$15^{+12}_{-7}$&$10^{+8}_{-5}$&$3.8^{+4.5}_{-3.2}$&$3.7^{+4.3}_{-3.1}$ \\ 
3.0&1000&0.04&0.05&0.08&3&$58^{+57}_{-32}$&$38^{+37}_{-21}$&$15^{+12}_{-7}$&$10^{+8}_{-5}$&$3.8^{+4.5}_{-3.2}$&$3.7^{+4.3}_{-3.1}$ \\ 
3.0&1000&0.04&0.05&0.08&3&$58^{+57}_{-32}$&$38^{+37}_{-21}$&$15^{+12}_{-7}$&$10^{+8}_{-5}$&$3.8^{+4.5}_{-3.2}$&$3.7^{+4.3}_{-3.1}$ \\ 
3.0&1000&0.05&0.05&0.08&3&$58^{+57}_{-32}$&$38^{+37}_{-21}$&$15^{+12}_{-7}$&$10^{+8}_{-5}$&$3.8^{+4.5}_{-3.2}$&$3.7^{+4.3}_{-3.1}$ \\ 
3.0&1000&0.05&0.05&0.08&3&$58^{+57}_{-32}$&$38^{+37}_{-21}$&$15^{+12}_{-7}$&$10^{+8}_{-5}$&$3.8^{+4.5}_{-3.2}$&$3.7^{+4.3}_{-3.1}$ \\ 
3.0&1000&0.06&0.05&0.08&3&$58^{+57}_{-32}$&$38^{+37}_{-21}$&$15^{+12}_{-7}$&$10^{+8}_{-5}$&$3.8^{+4.5}_{-3.2}$&$3.7^{+4.3}_{-3.1}$ \\ 
3.0&1000&0.07&0.05&0.08&2&$39^{+51}_{-25}$&$25^{+34}_{-17}$&$15^{+12}_{-7}$&$10^{+8}_{-5}$&$2.5^{+3.8}_{-2.3}$&$2.5^{+3.6}_{-2.3}$ \\ 
3.0&1000&0.08&0.05&0.08&2&$39^{+51}_{-25}$&$25^{+34}_{-17}$&$15^{+12}_{-7}$&$10^{+8}_{-5}$&$2.5^{+3.8}_{-2.3}$&$2.5^{+3.6}_{-2.3}$ \\ 
3.0&1000&0.09&0.05&0.08&2&$39^{+51}_{-25}$&$25^{+34}_{-17}$&$15^{+12}_{-7}$&$10^{+8}_{-5}$&$2.5^{+3.8}_{-2.3}$&$2.5^{+3.6}_{-2.3}$ \\ 
3.0&1000&0.10&0.05&0.08&2&$39^{+51}_{-25}$&$25^{+34}_{-17}$&$15^{+12}_{-7}$&$10^{+8}_{-5}$&$2.5^{+3.8}_{-2.3}$&$2.5^{+3.6}_{-2.3}$ \\ 
3.0&1000&0.11&0.05&0.08&1&$19^{+45}_{-17}$&$13^{+29}_{-11}$&$15^{+12}_{-7}$&$10^{+8}_{-5}$&$1.3^{+3.1}_{-1.3}$&$1.2^{+3.0}_{-1.2}$ \\ 
3.0&1000&0.12&0.05&0.08&1&$19^{+45}_{-17}$&$13^{+29}_{-11}$&$11^{+11}_{-6}$&$8^{+8}_{-4}$&$1.7^{+4.2}_{-1.7}$&$1.6^{+4.0}_{-1.6}$ \\ 
3.0&1000&0.14&0.05&0.08&1&$19^{+45}_{-17}$&$13^{+29}_{-11}$&$8^{+10}_{-5}$&$5^{+7}_{-3}$&$2.5^{+6.4}_{-2.5}$&$2.5^{+6.2}_{-2.5}$ \\ 
3.0&1000&0.16&0.05&0.08&1&$19^{+45}_{-17}$&$13^{+29}_{-11}$&$8^{+10}_{-5}$&$5^{+7}_{-3}$&$2.5^{+6.4}_{-2.5}$&$2.5^{+6.2}_{-2.5}$ \\ 
3.0&1000&0.18&0.05&0.08&1&$19^{+45}_{-17}$&$13^{+29}_{-11}$&$4^{+9}_{-3}$&$3^{+6}_{-2}$&$5.1^{+14.3}_{-5.1}$&$4.9^{+13.8}_{-4.9}$ \\ 

\hline
\end{tabular}
    
  \end{minipage}
\begin{minipage}{0.84\textwidth}
(1) Maximum transverse separation between cluster-pair axes and the Q1410 sightline (see \Cref{sec:dndz_method}).
(2) Maximum velocity difference to any cluster-pair within $\Delta d$ from the Q1410 sightline (see \Cref{sec:dndz_method}).
(3) Minimum rest-frame equivalent width (see \Cref{sec:dndz_method}).
(4) Redshift path (see \Cref{sec:dndz_method} and \Cref{sec:dz}).
(5) Absorption distance (see \Cref{sec:dz}).
(6) Total number of absorption components having $W_{\rm r} \ge W_{\rm r}^{\rm min}$ within $\Delta d$ and $\Delta v$ from cluster-pairs (see \Cref{sec:dndz_method}).
(7) Redshift number density associated to absorption components having $W_{\rm r} \ge W_{\rm r}^{\rm min}$ within $\Delta d$ and $\Delta v$ from cluster-pairs (see \Cref{sec:dndz_method}).
(8) Absorption distance number density to absorption components having $W_{\rm r} \ge W_{\rm r}^{\rm min}$ within $\Delta d$ and $\Delta v$ from cluster-pairs (analogous to $dN/dz$ but see also \Cref{sec:dz})
(9) Field expectation for the redshift number density (see \Cref{sec:dndz_field}).
(10) Field expectation for the absorption distance number density (see \Cref{sec:dndz_field}).
(11) Excess of the measured redshift number density compared to the field expectation (see \Cref{sec:excesses}).
(12) Excess of the measured absorption distance number density compared to the field expectation (see \Cref{sec:excesses}).

\end{minipage}
\end{table*}

    \begin{table*}
    \centering
    \scriptsize
    \begin{minipage}{0.84\textwidth}
    \centering
    \caption{Summary of relevant quantities for our sample of narrow \hi\ ($b < 50$\kms) absorption lines.}\label{tab:nla_summary}
    \begin{tabular}{@{}cccccccccccc@{}}

\hline                            
$\Delta d$ & $\Delta v$ & $W_r^{\rm min}$ &$\Delta z$ & $\Delta X$&  $N$  & $\frac{dN}{dz}$ & $\frac{dN}{dX}$ & $\frac{dN}{dz}|_{\rm field}$ & $\frac{dN}{dX}|_{\rm field}$ &$\frac{dN}{dz}/\frac{dN}{dz}|_{\rm field}$ & $\frac{dN}{dX}/\frac{dN}{dX}|_{\rm field}$ \\ 

(\mpc)     &  (\kms)    &      (\AA)        &            &              &       &                    &                      &                                      &                                      &                                                        &                                                       \\ 

(1)         &  (2)        &      (3)           &      (4)   & (5)          &  (6)  &    (7)             & (8)                  &     (9)                              & (10)                                  &   (11)                                                  &   (12)                                                  \\ 

\hline 

3.0&500&0.04&0.03&0.05&7&$228^{+123}_{-85}$&$151^{+82}_{-56}$&$106^{+23}_{-20}$&$75^{+16}_{-14}$&$2.2^{+1.2}_{-0.8}$&$2.0^{+1.1}_{-0.8}$ \\ 
3.0&630&0.04&0.04&0.06&7&$186^{+101}_{-69}$&$124^{+67}_{-46}$&$106^{+23}_{-20}$&$75^{+16}_{-14}$&$1.8^{+1.0}_{-0.7}$&$1.7^{+0.9}_{-0.6}$ \\ 
3.0&794&0.04&0.05&0.07&8&$174^{+86}_{-61}$&$116^{+57}_{-40}$&$106^{+23}_{-20}$&$75^{+16}_{-14}$&$1.6^{+0.8}_{-0.6}$&$1.6^{+0.8}_{-0.6}$ \\ 
3.0&1000&0.04&0.06&0.08&10&$179^{+77}_{-56}$&$120^{+51}_{-37}$&$106^{+23}_{-20}$&$75^{+16}_{-14}$&$1.7^{+0.7}_{-0.6}$&$1.6^{+0.7}_{-0.5}$ \\ 
3.0&1313&0.04&0.07&0.11&11&$154^{+62}_{-46}$&$103^{+41}_{-31}$&$106^{+23}_{-20}$&$75^{+16}_{-14}$&$1.5^{+0.6}_{-0.5}$&$1.4^{+0.6}_{-0.4}$ \\ 
3.0&1724&0.04&0.09&0.13&11&$121^{+49}_{-36}$&$82^{+33}_{-24}$&$106^{+23}_{-20}$&$75^{+16}_{-14}$&$1.2^{+0.5}_{-0.4}$&$1.1^{+0.5}_{-0.3}$ \\ 
3.0&2264&0.04&0.12&0.17&11&$95^{+38}_{-28}$&$64^{+26}_{-19}$&$106^{+23}_{-20}$&$75^{+16}_{-14}$&$0.9^{+0.4}_{-0.3}$&$0.9^{+0.4}_{-0.3}$ \\ 
3.0&2972&0.04&0.14&0.21&14&$101^{+35}_{-27}$&$68^{+23}_{-18}$&$106^{+23}_{-20}$&$75^{+16}_{-14}$&$1.0^{+0.3}_{-0.3}$&$0.9^{+0.3}_{-0.3}$ \\ 
3.0&3903&0.04&0.16&0.24&17&$104^{+32}_{-25}$&$70^{+21}_{-17}$&$106^{+23}_{-20}$&$75^{+16}_{-14}$&$1.0^{+0.3}_{-0.3}$&$0.9^{+0.3}_{-0.2}$ \\ 
3.0&5125&0.04&0.18&0.27&18&$99^{+29}_{-23}$&$67^{+20}_{-16}$&$106^{+23}_{-20}$&$75^{+16}_{-14}$&$0.9^{+0.3}_{-0.2}$&$0.9^{+0.3}_{-0.2}$ \\ 
3.0&6729&0.04&0.19&0.28&18&$93^{+28}_{-22}$&$63^{+19}_{-15}$&$106^{+23}_{-20}$&$75^{+16}_{-14}$&$0.9^{+0.3}_{-0.2}$&$0.8^{+0.3}_{-0.2}$ \\ 
3.0&8835&0.04&0.22&0.32&24&$110^{+27}_{-22}$&$76^{+19}_{-15}$&$106^{+23}_{-20}$&$75^{+16}_{-14}$&$1.0^{+0.3}_{-0.2}$&$1.0^{+0.3}_{-0.2}$ \\ 
3.0&11601&0.04&0.25&0.36&29&$115^{+26}_{-21}$&$80^{+18}_{-15}$&$106^{+23}_{-20}$&$75^{+16}_{-14}$&$1.1^{+0.3}_{-0.2}$&$1.1^{+0.3}_{-0.2}$ \\ 
3.0&15232&0.04&0.30&0.42&33&$112^{+23}_{-19}$&$79^{+16}_{-14}$&$106^{+23}_{-20}$&$75^{+16}_{-14}$&$1.1^{+0.2}_{-0.2}$&$1.1^{+0.2}_{-0.2}$ \\ 
3.0&20000&0.04&0.33&0.47&37&$112^{+22}_{-18}$&$79^{+15}_{-13}$&$106^{+23}_{-20}$&$75^{+16}_{-14}$&$1.1^{+0.2}_{-0.2}$&$1.1^{+0.2}_{-0.2}$ \\ 
\hline1.0&1000&0.04&0.01&0.02&1&$68^{+157}_{-59}$&$47^{+108}_{-40}$&$106^{+23}_{-20}$&$75^{+16}_{-14}$&$0.6^{+1.5}_{-0.6}$&$0.6^{+1.5}_{-0.5}$ \\ 
1.4&1000&0.04&0.03&0.04&2&$77^{+102}_{-51}$&$50^{+67}_{-33}$&$106^{+23}_{-20}$&$75^{+16}_{-14}$&$0.7^{+1.0}_{-0.5}$&$0.7^{+0.9}_{-0.4}$ \\ 
2.1&1000&0.04&0.04&0.06&3&$69^{+68}_{-38}$&$46^{+45}_{-26}$&$106^{+23}_{-20}$&$75^{+16}_{-14}$&$0.7^{+0.6}_{-0.4}$&$0.6^{+0.6}_{-0.4}$ \\ 
3.0&1000&0.04&0.06&0.08&10&$179^{+77}_{-56}$&$120^{+51}_{-37}$&$106^{+23}_{-20}$&$75^{+16}_{-14}$&$1.7^{+0.7}_{-0.6}$&$1.6^{+0.7}_{-0.5}$ \\ 
4.1&1000&0.04&0.06&0.09&10&$175^{+75}_{-55}$&$117^{+50}_{-36}$&$106^{+23}_{-20}$&$75^{+16}_{-14}$&$1.7^{+0.7}_{-0.5}$&$1.6^{+0.7}_{-0.5}$ \\ 
5.7&1000&0.04&0.06&0.09&10&$174^{+74}_{-54}$&$116^{+50}_{-36}$&$106^{+23}_{-20}$&$75^{+16}_{-14}$&$1.6^{+0.7}_{-0.5}$&$1.6^{+0.7}_{-0.5}$ \\ 
7.8&1000&0.04&0.07&0.10&10&$151^{+65}_{-47}$&$103^{+44}_{-32}$&$106^{+23}_{-20}$&$75^{+16}_{-14}$&$1.4^{+0.6}_{-0.5}$&$1.4^{+0.6}_{-0.5}$ \\ 
10.7&1000&0.04&0.08&0.12&11&$141^{+57}_{-42}$&$95^{+38}_{-28}$&$106^{+23}_{-20}$&$75^{+16}_{-14}$&$1.3^{+0.6}_{-0.4}$&$1.3^{+0.5}_{-0.4}$ \\ 
14.8&1000&0.04&0.09&0.14&12&$129^{+49}_{-37}$&$88^{+34}_{-25}$&$106^{+23}_{-20}$&$75^{+16}_{-14}$&$1.2^{+0.5}_{-0.4}$&$1.2^{+0.5}_{-0.4}$ \\ 
20.3&1000&0.04&0.11&0.16&14&$129^{+45}_{-34}$&$88^{+30}_{-23}$&$106^{+23}_{-20}$&$75^{+16}_{-14}$&$1.2^{+0.4}_{-0.3}$&$1.2^{+0.4}_{-0.3}$ \\ 
27.9&1000&0.04&0.12&0.17&16&$135^{+43}_{-34}$&$93^{+30}_{-23}$&$106^{+23}_{-20}$&$75^{+16}_{-14}$&$1.3^{+0.4}_{-0.3}$&$1.2^{+0.4}_{-0.3}$ \\ 
38.4&1000&0.04&0.15&0.22&19&$127^{+36}_{-29}$&$87^{+25}_{-20}$&$106^{+23}_{-20}$&$75^{+16}_{-14}$&$1.2^{+0.4}_{-0.3}$&$1.2^{+0.4}_{-0.3}$ \\ 
52.9&1000&0.04&0.17&0.24&19&$113^{+32}_{-26}$&$78^{+22}_{-18}$&$106^{+23}_{-20}$&$75^{+16}_{-14}$&$1.1^{+0.3}_{-0.3}$&$1.1^{+0.3}_{-0.3}$ \\ 
72.7&1000&0.04&0.19&0.28&21&$109^{+29}_{-24}$&$76^{+21}_{-16}$&$106^{+23}_{-20}$&$75^{+16}_{-14}$&$1.0^{+0.3}_{-0.2}$&$1.0^{+0.3}_{-0.2}$ \\ 
100.0&1000&0.04&0.24&0.35&29&$119^{+26}_{-22}$&$84^{+19}_{-15}$&$106^{+23}_{-20}$&$75^{+16}_{-14}$&$1.1^{+0.3}_{-0.2}$&$1.1^{+0.3}_{-0.2}$ \\ 
\hline3.0&1000&0.03&0.06&0.09&10&$176^{+75}_{-55}$&$118^{+50}_{-37}$&$122^{+25}_{-21}$&$87^{+18}_{-15}$&$1.5^{+0.7}_{-0.5}$&$1.4^{+0.6}_{-0.5}$ \\ 
3.0&1000&0.04&0.06&0.09&10&$176^{+75}_{-55}$&$118^{+50}_{-37}$&$122^{+25}_{-21}$&$87^{+18}_{-15}$&$1.5^{+0.7}_{-0.5}$&$1.4^{+0.6}_{-0.5}$ \\ 
3.0&1000&0.04&0.06&0.09&10&$176^{+75}_{-55}$&$118^{+50}_{-37}$&$118^{+24}_{-20}$&$84^{+17}_{-15}$&$1.5^{+0.7}_{-0.5}$&$1.4^{+0.7}_{-0.5}$ \\ 
3.0&1000&0.05&0.06&0.09&10&$176^{+75}_{-55}$&$118^{+50}_{-37}$&$114^{+24}_{-20}$&$82^{+17}_{-14}$&$1.5^{+0.7}_{-0.6}$&$1.4^{+0.7}_{-0.5}$ \\ 
3.0&1000&0.05&0.06&0.09&10&$176^{+75}_{-55}$&$118^{+50}_{-37}$&$107^{+23}_{-19}$&$77^{+17}_{-14}$&$1.6^{+0.8}_{-0.6}$&$1.5^{+0.7}_{-0.6}$ \\ 
3.0&1000&0.06&0.06&0.09&9&$159^{+73}_{-52}$&$106^{+48}_{-35}$&$107^{+23}_{-19}$&$77^{+17}_{-14}$&$1.5^{+0.7}_{-0.6}$&$1.4^{+0.7}_{-0.5}$ \\ 
3.0&1000&0.07&0.06&0.09&9&$159^{+73}_{-52}$&$106^{+48}_{-35}$&$104^{+23}_{-19}$&$74^{+16}_{-14}$&$1.5^{+0.8}_{-0.6}$&$1.4^{+0.7}_{-0.6}$ \\ 
3.0&1000&0.08&0.06&0.09&9&$159^{+73}_{-52}$&$106^{+48}_{-35}$&$104^{+23}_{-19}$&$74^{+16}_{-14}$&$1.5^{+0.8}_{-0.6}$&$1.4^{+0.7}_{-0.6}$ \\ 
3.0&1000&0.09&0.06&0.09&8&$141^{+70}_{-49}$&$94^{+47}_{-33}$&$96^{+22}_{-18}$&$69^{+16}_{-13}$&$1.5^{+0.8}_{-0.6}$&$1.4^{+0.7}_{-0.6}$ \\ 
3.0&1000&0.10&0.06&0.09&7&$123^{+67}_{-46}$&$82^{+44}_{-31}$&$82^{+21}_{-17}$&$59^{+15}_{-12}$&$1.5^{+0.9}_{-0.7}$&$1.4^{+0.8}_{-0.6}$ \\ 
3.0&1000&0.11&0.06&0.09&7&$123^{+67}_{-46}$&$82^{+44}_{-31}$&$79^{+21}_{-17}$&$56^{+15}_{-12}$&$1.6^{+0.9}_{-0.7}$&$1.5^{+0.9}_{-0.6}$ \\ 
3.0&1000&0.12&0.06&0.09&7&$123^{+67}_{-46}$&$82^{+44}_{-31}$&$71^{+20}_{-16}$&$51^{+14}_{-11}$&$1.7^{+1.0}_{-0.8}$&$1.6^{+1.0}_{-0.7}$ \\ 
3.0&1000&0.14&0.06&0.09&7&$123^{+67}_{-46}$&$82^{+44}_{-31}$&$68^{+19}_{-15}$&$48^{+14}_{-11}$&$1.8^{+1.1}_{-0.8}$&$1.7^{+1.0}_{-0.8}$ \\ 
3.0&1000&0.16&0.06&0.09&7&$123^{+67}_{-46}$&$82^{+44}_{-31}$&$57^{+18}_{-14}$&$41^{+13}_{-10}$&$2.2^{+1.3}_{-1.0}$&$2.0^{+1.2}_{-0.9}$ \\ 
3.0&1000&0.18&0.06&0.09&6&$106^{+63}_{-42}$&$71^{+42}_{-28}$&$46^{+17}_{-13}$&$33^{+12}_{-9}$&$2.3^{+1.5}_{-1.2}$&$2.1^{+1.4}_{-1.1}$ \\ 

\hline
\end{tabular}
    
  \end{minipage}
\begin{minipage}{0.84\textwidth}
(1) Maximum transverse separation between cluster-pair axes and the Q1410 sightline (see \Cref{sec:dndz_method}).
(2) Maximum velocity difference to any cluster-pair within $\Delta d$ from the Q1410 sightline (see \Cref{sec:dndz_method}).
(3) Minimum rest-frame equivalent width (see \Cref{sec:dndz_method}).
(4) Redshift path (see \Cref{sec:dndz_method} and \Cref{sec:dz}).
(5) Absorption distance (see \Cref{sec:dz}).
(6) Total number of absorption components having $W_{\rm r} \ge W_{\rm r}^{\rm min}$ within $\Delta d$ and $\Delta v$ from cluster-pairs (see \Cref{sec:dndz_method}).
(7) Redshift number density associated to absorption components having $W_{\rm r} \ge W_{\rm r}^{\rm min}$ within $\Delta d$ and $\Delta v$ from cluster-pairs (see \Cref{sec:dndz_method}).
(8) Absorption distance number density to absorption components having $W_{\rm r} \ge W_{\rm r}^{\rm min}$ within $\Delta d$ and $\Delta v$ from cluster-pairs (analogous to $dN/dz$ but see also \Cref{sec:dz})
(9) Field expectation for the redshift number density (see \Cref{sec:dndz_field}).
(10) Field expectation for the absorption distance number density (see \Cref{sec:dndz_field}).
(11) Excess of the measured redshift number density compared to the field expectation (see \Cref{sec:excesses}).
(12) Excess of the measured absorption distance number density compared to the field expectation (see \Cref{sec:excesses}).

\end{minipage}
\end{table*}

    \begin{table*}
    \centering
    \scriptsize
    \begin{minipage}{0.84\textwidth}
    \centering
    \caption{Summary of relevant quantities for our sample of broad \hi\ ($b \ge 50$\kms) absorption lines.}\label{tab:bla_summary}
    \begin{tabular}{@{}cccccccccccc@{}}

\hline                            
$\Delta d$ & $\Delta v$ & $W_r^{\rm min}$ &$\Delta z$ & $\Delta X$&  $N$  & $\frac{dN}{dz}$ & $\frac{dN}{dX}$ & $\frac{dN}{dz}|_{\rm field}$ & $\frac{dN}{dX}|_{\rm field}$ &$\frac{dN}{dz}/\frac{dN}{dz}|_{\rm field}$ & $\frac{dN}{dX}/\frac{dN}{dX}|_{\rm field}$ \\ 

(\mpc)     &  (\kms)    &      (\AA)        &            &              &       &                    &                      &                                      &                                      &                                                        &                                                       \\ 

(1)         &  (2)        &      (3)           &      (4)   & (5)          &  (6)  &    (7)             & (8)                  &     (9)                              & (10)                                  &   (11)                                                  &   (12)                                                  \\ 

\hline 

3.0&500&0.04&0.03&0.05&4&$130^{+103}_{-63}$&$86^{+69}_{-42}$&$18^{+13}_{-10}$&$12^{+9}_{-7}$&$7.4^{+6.4}_{-4.4}$&$7.0^{+6.0}_{-4.1}$ \\ 
3.0&630&0.04&0.04&0.06&4&$106^{+85}_{-52}$&$71^{+56}_{-34}$&$18^{+13}_{-10}$&$12^{+9}_{-7}$&$6.1^{+5.2}_{-3.6}$&$5.7^{+4.9}_{-3.4}$ \\ 
3.0&794&0.04&0.05&0.07&6&$131^{+78}_{-52}$&$87^{+52}_{-35}$&$18^{+13}_{-10}$&$12^{+9}_{-7}$&$7.5^{+5.1}_{-3.9}$&$7.0^{+4.8}_{-3.7}$ \\ 
3.0&1000&0.04&0.06&0.08&6&$108^{+65}_{-43}$&$72^{+43}_{-29}$&$18^{+13}_{-10}$&$12^{+9}_{-7}$&$6.1^{+4.2}_{-3.2}$&$5.8^{+4.0}_{-3.0}$ \\ 
3.0&1313&0.04&0.07&0.11&6&$84^{+50}_{-33}$&$56^{+34}_{-22}$&$18^{+13}_{-10}$&$12^{+9}_{-7}$&$4.8^{+3.3}_{-2.5}$&$4.5^{+3.1}_{-2.4}$ \\ 
3.0&1724&0.04&0.09&0.13&6&$66^{+40}_{-26}$&$44^{+27}_{-18}$&$18^{+13}_{-10}$&$12^{+9}_{-7}$&$3.8^{+2.6}_{-2.0}$&$3.6^{+2.5}_{-1.9}$ \\ 
3.0&2264&0.04&0.12&0.17&7&$61^{+33}_{-23}$&$41^{+22}_{-15}$&$18^{+13}_{-10}$&$12^{+9}_{-7}$&$3.5^{+2.2}_{-1.7}$&$3.3^{+2.1}_{-1.7}$ \\ 
3.0&2972&0.04&0.14&0.21&7&$50^{+27}_{-19}$&$34^{+18}_{-13}$&$18^{+13}_{-10}$&$12^{+9}_{-7}$&$2.9^{+1.8}_{-1.4}$&$2.7^{+1.7}_{-1.4}$ \\ 
3.0&3903&0.04&0.16&0.24&8&$49^{+24}_{-17}$&$33^{+16}_{-11}$&$18^{+13}_{-10}$&$12^{+9}_{-7}$&$2.8^{+1.7}_{-1.3}$&$2.6^{+1.6}_{-1.3}$ \\ 
3.0&5125&0.04&0.18&0.27&8&$44^{+22}_{-15}$&$30^{+15}_{-10}$&$18^{+13}_{-10}$&$12^{+9}_{-7}$&$2.5^{+1.5}_{-1.2}$&$2.4^{+1.4}_{-1.2}$ \\ 
3.0&6729&0.04&0.19&0.28&8&$41^{+20}_{-14}$&$28^{+14}_{-10}$&$18^{+13}_{-10}$&$12^{+9}_{-7}$&$2.4^{+1.4}_{-1.1}$&$2.3^{+1.4}_{-1.1}$ \\ 
3.0&8835&0.04&0.22&0.32&8&$37^{+18}_{-13}$&$25^{+13}_{-9}$&$18^{+13}_{-10}$&$12^{+9}_{-7}$&$2.1^{+1.3}_{-1.0}$&$2.0^{+1.2}_{-1.0}$ \\ 
3.0&11601&0.04&0.25&0.36&9&$36^{+16}_{-12}$&$25^{+11}_{-8}$&$18^{+13}_{-10}$&$12^{+9}_{-7}$&$2.0^{+1.2}_{-1.0}$&$2.0^{+1.1}_{-0.9}$ \\ 
3.0&15232&0.04&0.30&0.42&10&$34^{+14}_{-11}$&$24^{+10}_{-7}$&$18^{+13}_{-10}$&$12^{+9}_{-7}$&$1.9^{+1.1}_{-0.9}$&$1.9^{+1.0}_{-0.9}$ \\ 
3.0&20000&0.04&0.33&0.47&11&$33^{+13}_{-10}$&$24^{+9}_{-7}$&$18^{+13}_{-10}$&$12^{+9}_{-7}$&$1.9^{+1.0}_{-0.9}$&$1.9^{+1.0}_{-0.9}$ \\ 
\hline1.0&1000&0.04&0.01&0.02&2&$135^{+180}_{-89}$&$93^{+124}_{-62}$&$18^{+13}_{-10}$&$12^{+9}_{-7}$&$7.7^{+10.6}_{-5.7}$&$7.5^{+10.3}_{-5.6}$ \\ 
1.4&1000&0.04&0.03&0.04&4&$154^{+122}_{-75}$&$100^{+80}_{-48}$&$18^{+13}_{-10}$&$12^{+9}_{-7}$&$8.8^{+7.6}_{-5.2}$&$8.1^{+7.0}_{-4.8}$ \\ 
2.1&1000&0.04&0.04&0.06&5&$115^{+78}_{-50}$&$77^{+53}_{-34}$&$18^{+13}_{-10}$&$12^{+9}_{-7}$&$6.6^{+5.0}_{-3.6}$&$6.3^{+4.7}_{-3.4}$ \\ 
3.0&1000&0.04&0.06&0.08&6&$108^{+65}_{-43}$&$72^{+43}_{-29}$&$18^{+13}_{-10}$&$12^{+9}_{-7}$&$6.1^{+4.2}_{-3.2}$&$5.8^{+4.0}_{-3.0}$ \\ 
4.1&1000&0.04&0.06&0.09&6&$105^{+63}_{-42}$&$70^{+42}_{-28}$&$18^{+13}_{-10}$&$12^{+9}_{-7}$&$6.0^{+4.1}_{-3.1}$&$5.7^{+3.9}_{-3.0}$ \\ 
5.7&1000&0.04&0.06&0.09&6&$104^{+63}_{-42}$&$70^{+42}_{-28}$&$18^{+13}_{-10}$&$12^{+9}_{-7}$&$6.0^{+4.1}_{-3.1}$&$5.6^{+3.9}_{-2.9}$ \\ 
7.8&1000&0.04&0.07&0.10&6&$91^{+54}_{-36}$&$62^{+37}_{-25}$&$18^{+13}_{-10}$&$12^{+9}_{-7}$&$5.2^{+3.6}_{-2.7}$&$5.0^{+3.4}_{-2.6}$ \\ 
10.7&1000&0.04&0.08&0.12&6&$77^{+46}_{-31}$&$52^{+31}_{-21}$&$18^{+13}_{-10}$&$12^{+9}_{-7}$&$4.4^{+3.0}_{-2.3}$&$4.2^{+2.9}_{-2.2}$ \\ 
14.8&1000&0.04&0.09&0.14&7&$76^{+41}_{-28}$&$52^{+28}_{-19}$&$18^{+13}_{-10}$&$12^{+9}_{-7}$&$4.3^{+2.7}_{-2.2}$&$4.2^{+2.7}_{-2.1}$ \\ 
20.3&1000&0.04&0.11&0.16&7&$64^{+35}_{-24}$&$44^{+24}_{-16}$&$18^{+13}_{-10}$&$12^{+9}_{-7}$&$3.7^{+2.3}_{-1.8}$&$3.6^{+2.3}_{-1.8}$ \\ 
27.9&1000&0.04&0.12&0.17&7&$59^{+32}_{-22}$&$41^{+22}_{-15}$&$18^{+13}_{-10}$&$12^{+9}_{-7}$&$3.4^{+2.1}_{-1.7}$&$3.3^{+2.1}_{-1.6}$ \\ 
38.4&1000&0.04&0.15&0.22&8&$53^{+26}_{-19}$&$37^{+18}_{-13}$&$18^{+13}_{-10}$&$12^{+9}_{-7}$&$3.1^{+1.8}_{-1.5}$&$3.0^{+1.8}_{-1.4}$ \\ 
52.9&1000&0.04&0.17&0.24&8&$47^{+23}_{-16}$&$33^{+16}_{-11}$&$18^{+13}_{-10}$&$12^{+9}_{-7}$&$2.7^{+1.6}_{-1.3}$&$2.7^{+1.6}_{-1.3}$ \\ 
72.7&1000&0.04&0.19&0.28&8&$41^{+21}_{-14}$&$29^{+14}_{-10}$&$18^{+13}_{-10}$&$12^{+9}_{-7}$&$2.4^{+1.4}_{-1.1}$&$2.3^{+1.4}_{-1.1}$ \\ 
100.0&1000&0.04&0.24&0.35&10&$41^{+18}_{-13}$&$29^{+12}_{-9}$&$18^{+13}_{-10}$&$12^{+9}_{-7}$&$2.3^{+1.3}_{-1.1}$&$2.3^{+1.3}_{-1.1}$ \\ 
\hline3.0&1000&0.03&0.06&0.09&6&$106^{+63}_{-42}$&$71^{+42}_{-28}$&$18^{+12}_{-8}$&$13^{+9}_{-6}$&$5.9^{+4.9}_{-4.1}$&$5.5^{+4.5}_{-3.8}$ \\ 
3.0&1000&0.04&0.06&0.09&6&$106^{+63}_{-42}$&$71^{+42}_{-28}$&$18^{+12}_{-8}$&$13^{+9}_{-6}$&$5.9^{+4.9}_{-4.1}$&$5.5^{+4.5}_{-3.8}$ \\ 
3.0&1000&0.04&0.06&0.09&6&$106^{+63}_{-42}$&$71^{+42}_{-28}$&$18^{+12}_{-8}$&$13^{+9}_{-6}$&$5.9^{+4.9}_{-4.1}$&$5.5^{+4.5}_{-3.8}$ \\ 
3.0&1000&0.05&0.06&0.09&6&$106^{+63}_{-42}$&$71^{+42}_{-28}$&$18^{+12}_{-8}$&$13^{+9}_{-6}$&$5.9^{+4.9}_{-4.1}$&$5.5^{+4.5}_{-3.8}$ \\ 
3.0&1000&0.05&0.06&0.09&6&$106^{+63}_{-42}$&$71^{+42}_{-28}$&$18^{+12}_{-8}$&$13^{+9}_{-6}$&$5.9^{+4.9}_{-4.1}$&$5.5^{+4.5}_{-3.8}$ \\ 
3.0&1000&0.06&0.06&0.09&6&$106^{+63}_{-42}$&$71^{+42}_{-28}$&$18^{+12}_{-8}$&$13^{+9}_{-6}$&$5.9^{+4.9}_{-4.1}$&$5.5^{+4.5}_{-3.8}$ \\ 
3.0&1000&0.07&0.06&0.09&6&$106^{+63}_{-42}$&$71^{+42}_{-28}$&$18^{+12}_{-8}$&$13^{+9}_{-6}$&$5.9^{+4.9}_{-4.1}$&$5.5^{+4.5}_{-3.8}$ \\ 
3.0&1000&0.08&0.06&0.09&6&$106^{+63}_{-42}$&$71^{+42}_{-28}$&$18^{+12}_{-8}$&$13^{+9}_{-6}$&$5.9^{+4.9}_{-4.1}$&$5.5^{+4.5}_{-3.8}$ \\ 
3.0&1000&0.09&0.06&0.09&6&$106^{+63}_{-42}$&$71^{+42}_{-28}$&$18^{+12}_{-8}$&$13^{+9}_{-6}$&$5.9^{+4.9}_{-4.1}$&$5.5^{+4.5}_{-3.8}$ \\ 
3.0&1000&0.10&0.06&0.09&5&$88^{+60}_{-38}$&$59^{+40}_{-26}$&$14^{+11}_{-7}$&$10^{+8}_{-5}$&$6.2^{+5.8}_{-4.8}$&$5.8^{+5.4}_{-4.5}$ \\ 
3.0&1000&0.11&0.06&0.09&5&$88^{+60}_{-38}$&$59^{+40}_{-26}$&$11^{+10}_{-6}$&$8^{+7}_{-4}$&$8.2^{+8.4}_{-7.3}$&$7.7^{+7.9}_{-6.8}$ \\ 
3.0&1000&0.12&0.06&0.09&4&$71^{+56}_{-34}$&$47^{+37}_{-23}$&$11^{+10}_{-6}$&$8^{+7}_{-4}$&$6.6^{+7.3}_{-6.0}$&$6.1^{+6.8}_{-5.6}$ \\ 
3.0&1000&0.14&0.06&0.09&4&$71^{+56}_{-34}$&$47^{+37}_{-23}$&$7^{+9}_{-5}$&$5^{+7}_{-3}$&$9.9^{+12.6}_{-9.9}$&$9.2^{+11.7}_{-9.2}$ \\ 
3.0&1000&0.16&0.06&0.09&2&$35^{+47}_{-23}$&$24^{+31}_{-16}$&$4^{+8}_{-3}$&$3^{+6}_{-2}$&$9.9^{+20.5}_{-9.9}$&$9.2^{+19.1}_{-9.2}$ \\ 
3.0&1000&0.18&0.06&0.09&2&$35^{+47}_{-23}$&$24^{+31}_{-16}$&$4^{+8}_{-3}$&$3^{+6}_{-2}$&$9.9^{+20.5}_{-9.9}$&$9.2^{+19.1}_{-9.2}$ \\ 

\hline
\end{tabular}
    
  \end{minipage}
\begin{minipage}{0.84\textwidth}
(1) Maximum transverse separation between cluster-pair axes and the Q1410 sightline (see \Cref{sec:dndz_method}).
(2) Maximum velocity difference to any cluster-pair within $\Delta d$ from the Q1410 sightline (see \Cref{sec:dndz_method}).
(3) Minimum rest-frame equivalent width (see \Cref{sec:dndz_method}).
(4) Redshift path (see \Cref{sec:dndz_method} and \Cref{sec:dz}).
(5) Absorption distance (see \Cref{sec:dz}).
(6) Total number of absorption components having $W_{\rm r} \ge W_{\rm r}^{\rm min}$ within $\Delta d$ and $\Delta v$ from cluster-pairs (see \Cref{sec:dndz_method}).
(7) Redshift number density associated to absorption components having $W_{\rm r} \ge W_{\rm r}^{\rm min}$ within $\Delta d$ and $\Delta v$ from cluster-pairs (see \Cref{sec:dndz_method}).
(8) Absorption distance number density to absorption components having $W_{\rm r} \ge W_{\rm r}^{\rm min}$ within $\Delta d$ and $\Delta v$ from cluster-pairs (analogous to $dN/dz$ but see also \Cref{sec:dz})
(9) Field expectation for the redshift number density (see \Cref{sec:dndz_field}).
(10) Field expectation for the absorption distance number density (see \Cref{sec:dndz_field}).
(11) Excess of the measured redshift number density compared to the field expectation (see \Cref{sec:excesses}).
(12) Excess of the measured absorption distance number density compared to the field expectation (see \Cref{sec:excesses}).

\end{minipage}
\end{table*}

\bsp

\label{lastpage}
\end{document}